\documentclass[reprint,amsmath,amssymb,aps,floatfix]{revtex4-2}
\usepackage[utf8]{inputenc} 
\usepackage[T1]{fontenc}    
\usepackage[pdfusetitle]{hyperref}
\usepackage{graphicx,url,qrcode,bm,tikz,physics,dsfont,slashed,mathtools,booktabs}
\usepackage{subfig}
\usetikzlibrary{calc}
\usepackage{cleveref}
\usepackage{comment}
\usepackage{cancel}
\crefname{section}{Sec.}{Secs.}
\Crefname{section}{Section}{Sections}
\crefname{figure}{Fig.}{Figs.}
\crefname{equation}{Eq.}{Eqs.}
\crefname{appendix}{Appendix}{Appendices}
\newcommand{\im}[0]{\mathrm{i}}
\DeclareMathOperator\eu{e}
\renewcommand{\vec}[1]{\bm{#1}}

\newcommand{\Heavi}[0]{\Theta }

\newcommand{\diagram}[1]{\vcenter{\hbox{\begin{tikzpicture} #1 \end{tikzpicture}}}}

\newcommand{\dTwo}[0]{1.8cm}
\newcommand{\dThree}[0]{2.7cm}

\newcommand{\len}[0]{1.0}
\usetikzlibrary{calc}
\tikzset{
  gluon/.style={thick, decorate, decoration={coil, amplitude=.08cm, segment length=.15cm, aspect=.75}}}
\tikzset{pion/.style={
	dashed,
	postaction={decorate},
	decoration={markings,mark=at position .6 with {\arrow{>}}}
	}}
\tikzset{nucleon/.style={
	postaction={decorate},
	decoration={markings,mark=at position .6 with {\arrow{>}}}
	}}
\tikzset{inmedium/.style={
	very thick,
	}}
    \tikzset{inmediumarrow1/.style={
	very thick,
	postaction={decorate},
	decoration={markings,mark=at position .63 with {\arrow{>}}}
	}}
    \tikzset{inmediumarrow2/.style={
	very thick,
	postaction={decorate},
	decoration={markings,mark=at position .7 with {\arrow{>}}}
	}}
\tikzset{source/.style={
	decorate,
	decoration={snake,amplitude=0.8mm}
	}}
\newcommand{\vertex}[2]{
	\fill[black!5] (#1) circle (.2);
	\draw (#1) circle (.2);
	\node at (#1) {\scalebox{0.8}{$#2$}};
}

\tikzset{>=latex}
\usetikzlibrary{datavisualization}
\usetikzlibrary{datavisualization.formats.functions}
\usetikzlibrary{patterns}    	             
\usetikzlibrary{decorations.markings}        
\usetikzlibrary{shapes.geometric}            
\usetikzlibrary{decorations.pathmorphing}    
\usetikzlibrary{decorations}

\newcommand{\SelfEnergyTypeTwo}[0]{
  \coordinate (A) at (0,0);
  \coordinate (B) at ($(A)+(\len,0)$);
  \coordinate (C) at ($(B)+(\len,0)$);
  \coordinate (D) at ($(B)+(0,-0.6)$);
  \draw[pion] (A) -- (B);
  \draw[pion] (B) -- (C);
  \draw[inmedium] ($(B)!0.5!(D)$) circle (.3cm);
  \node at (D) {\scalebox{0.7}{$\blacktriangleright$}};
}

\newcommand{\SelfEnergyTypeThree}[0]{
  \coordinate (A) at (0,0);
  \coordinate (B) at ($(A) + (\len,0)$);
  \coordinate (C) at ($(B) + (\len,0)$);
  \coordinate (E) at ($(B) + (-\len*0.7,-\len)$);
  \coordinate (F) at ($(B) + (\len*0.7,-\len)$);
  \draw[pion] (A) -- (B);
  \draw[pion] (B) -- (C);
  \draw[pion] (B) to[in=90,out=220] (E);
  \draw[pion] (F) to[in=320,out=90] (B);
  \draw[inmedium] (E) to[in=135,out=45] (F);
  \draw[inmedium] (E) to[in=225,out=-45] (F);
  \node at ($(E)!0.5!(F) + (0,0.3)$) {\scalebox{0.7}{$\blacktriangleright$}};
  \node at ($(E)!0.5!(F) + (0,-0.3)$) {\scalebox{0.7}{$\blacktriangleleft$}};
}

\newcommand{\SelfEnergyTypeFour}[0]{
  \coordinate (A) at (0,0);
  \coordinate (B) at ($(A)+(\len,0)$);
  \coordinate (C) at ($(B)+(\len,0)$);
  \coordinate (D) at ($(C)+(\len,0)$);
  \draw[pion] (A) -- (B);
  \draw[pion] (B) -- (C);
  \draw[pion] (C) -- (D);
  \draw[inmedium] ($(B)!0.5!(C)$) circle (.5cm);
  \node at ($(B)!0.5!(C) + (0,0.5)$) {\scalebox{0.7}{$\blacktriangleright$}};
  \node at ($(B)!0.5!(C) - (0,0.5)$) {\scalebox{0.7}{$\blacktriangleleft$}};
}

\newcommand{\SelfEnergyTypeFive}[0]{
  \coordinate (A) at (0,0);
  \coordinate (B) at ($(A)+(\len,0)$);
  \coordinate (C) at ($(B)+(\len,0)$);
  \coordinate (D) at ($(B)+(0,-\len)$);
  \draw[pion] (A) -- (B);
  \draw[pion] (B) -- (C);
  \draw[pion] (D) -- (B);
  \draw[inmedium] ($(B)!0.5!(D)$) circle (\len/2);
  \node at ($(B)!0.5!(D) + (\len/2,0)$) {\scalebox{0.7}{$\blacktriangledown$}};
  \node at ($(B)!0.5!(D) - (\len/2,0)$) {\scalebox{0.7}{$\blacktriangle$}};
}

\newcommand{\SelfEnergyFirstA}[0]{
  \coordinate (A) at (0,0);
  \coordinate (B) at ($(A)+(\len,0)$);
  \coordinate (C) at ($(B)+(\len,0)$);
  \coordinate (D) at ($(C)+(\len,0)$);
  \draw[pion] (A) -- (B);
  \draw[pion] (C) -- (D);
  \draw ($(B)!0.5!(C)$) ++(180:.5) arc (180:0:.5);
  \draw[inmedium] ($(B)!0.5!(C)$) ++(180:.5) arc (180:360:.5);
  \node at ($(B)!0.5!(C)+(0,.5)$) {\scalebox{0.7}{$\blacktriangleright$}};
  \node at ($(B)!0.5!(C)-(0,.5)$) {\scalebox{0.7}{$\blacktriangleleft$}};
  \vertex{B}{1};
  \vertex{C}{1};
}

\newcommand{\SelfEnergyFirstB}[0]{
  \coordinate (A) at (0,0);
  \coordinate (B) at ($(A)+(\len,0)$);
  \coordinate (C) at ($(B)+(\len,0)$);
  \coordinate (D) at ($(C)+(\len,0)$);
  \draw[pion] (A) -- (B);
  \draw[pion] (C) -- (D);
  \draw[inmedium] ($(B)!0.5!(C)$) ++(180:.5) arc (180:0:.5);
  \draw ($(B)!0.5!(C)$) ++(180:.5) arc (180:360:.5);
  \node at ($(B)!0.5!(C)+(0,.5)$) {\scalebox{0.7}{$\blacktriangleright$}};
  \node at ($(B)!0.5!(C)-(0,.5)$) {\scalebox{0.7}{$\blacktriangleleft$}};
  \vertex{B}{1};
  \vertex{C}{1};
}

\newcommand{\DecayTree}[0]{
  \coordinate (A) at (0,0);
  \coordinate (B) at ($(A)+(\len,0)$);
  \coordinate (C) at ($(B)+(\len,0)$);
  \draw[pion] (A) -- (B);
  \draw[source] (B) -- (C);
  }

\newcommand{\DecayTypeTwo}[0]{
  \coordinate (A) at (0,0);
  \coordinate (B) at ($(A)+(\len,0)$);
  \coordinate (C) at ($(B)+(\len,0)$);
  \coordinate (D) at ($(B)+(0,-0.6)$);
  \draw[pion] (A) -- (B);
  \draw[source] (B) -- (C);
  \draw[inmedium] ($(B)!0.5!(D)$) circle (.3cm);
  \node at (D) {\scalebox{0.7}{$\blacktriangleright$}};
}

\newcommand{\DecayTypeThree}[0]{
  \coordinate (A) at (0,0);
  \coordinate (B) at ($(A) + (\len,0)$);
  \coordinate (C) at ($(B) + (\len,0)$);
  \coordinate (E) at ($(B) + (-\len*0.7,-\len)$);
  \coordinate (F) at ($(B) + (\len*0.7,-\len)$);
  \draw[pion] (A) -- (B);
  \draw[source] (B) -- (C);
  \draw[pion] (B) to[in=90,out=220] (E);
  \draw[pion] (F) to[in=320,out=90] (B);
  \draw[inmedium] (E) to[in=135,out=45] (F);
  \draw[inmedium] (E) to[in=225,out=-45] (F);
  \node at ($(E)!0.5!(F) + (0,0.3)$) {\scalebox{0.7}{$\blacktriangleright$}};
  \node at ($(E)!0.5!(F) + (0,-0.3)$) {\scalebox{0.7}{$\blacktriangleleft$}};
}

\newcommand{\DecayTypeFour}[0]{
  \coordinate (A) at (0,0);
  \coordinate (B) at ($(A)+(\len,0)$);
  \coordinate (C) at ($(B)+(\len,0)$);
  \coordinate (D) at ($(C)+(\len,0)$);
  \draw[pion] (A) -- (B);
  \draw[pion] (B) -- (C);
  \draw[source] (C) -- (D);
  \draw[inmedium] ($(B)!0.5!(C)$) circle (.5cm);
  \node at ($(B)!0.5!(C) + (0,0.5)$) {\scalebox{0.7}{$\blacktriangleright$}};
  \node at ($(B)!0.5!(C) - (0,0.5)$) {\scalebox{0.7}{$\blacktriangleleft$}};
}

\newcommand{\DecayTypeFive}[0]{
  \coordinate (A) at (0,0);
  \coordinate (B) at ($(A)+(\len,0)$);
  \coordinate (C) at ($(B)+(\len,0)$);
  \coordinate (D) at ($(B)+(0,-\len)$);
  \draw[pion] (A) -- (B);
  \draw[source] (B) -- (C);
  \draw[pion] (D) -- (B);
  \draw[inmedium] ($(B)!0.5!(D)$) circle (\len/2);
  \node at ($(B)!0.5!(D) + (\len/2,0)$) {\scalebox{0.7}{$\blacktriangledown$}};
  \node at ($(B)!0.5!(D) - (\len/2,0)$) {\scalebox{0.7}{$\blacktriangle$}};
}

\newcommand{\DecayConstantFirstA}[0]{
  \coordinate (A) at (0,0);
  \coordinate (B) at ($(A)+(\len,0)$);
  \coordinate (C) at ($(B)+(\len,0)$);
  \coordinate (D) at ($(C)+(\len,0)$);
  \draw[pion] (A) -- (B);
  \draw[source] (C) -- (D);
  \draw ($(B)!0.5!(C)$) ++(180:.5) arc (180:0:.5);
  \draw[inmedium] ($(B)!0.5!(C)$) ++(180:.5) arc (180:360:.5);
  \node at ($(B)!0.5!(C)+(0,.5)$) {\scalebox{0.7}{$\blacktriangleright$}};
  \node at ($(B)!0.5!(C)-(0,.5)$) {\scalebox{0.7}{$\blacktriangleleft$}};
  \vertex{B}{1};
  \vertex{C}{1};
}

\newcommand{\DecayConstantFirstB}[0]{
  \coordinate (A) at (0,0);
  \coordinate (B) at ($(A)+(\len,0)$);
  \coordinate (C) at ($(B)+(\len,0)$);
  \coordinate (D) at ($(C)+(\len,0)$);
  \draw[pion] (A) -- (B);
  \draw[source] (C) -- (D);
  \draw[inmedium] ($(B)!0.5!(C)$) ++(180:.5) arc (180:0:.5);
  \draw ($(B)!0.5!(C)$) ++(180:.5) arc (180:360:.5);
  \node at ($(B)!0.5!(C)+(0,.5)$) {\scalebox{0.7}{$\blacktriangleright$}};
  \node at ($(B)!0.5!(C)-(0,.5)$) {\scalebox{0.7}{$\blacktriangleleft$}};
  \vertex{B}{1};
  \vertex{C}{1};
}

\newcommand{\SelfEnergyTypeadditonalfirst}[0]{
  \coordinate (A) at (0,0);
  \coordinate (B) at ($(A)+(\len,0)$);
  \coordinate (C) at ($(B)+(\len,0)$);
  \coordinate (D) at ($(C)+(\len,0)$);
  \coordinate (E) at ($(C)+(-\len*0.5,+\len)$);
  \draw[pion] (A) -- (B);
  \draw[nucleon] (C) -- (B);
  \draw[pion] (C) -- (D);
  \draw[inmediumarrow1] (B) to[in=90,out=170] (E);
  \draw[inmediumarrow1] (E) to[in=40,out=90] (C);
  \draw[pion] (B) -- (E);
}

\newcommand{\SelfEnergyTypeadditonalsecond}[0]{
  \coordinate (A) at (0,0);
  \coordinate (B) at ($(A)+(\len,0)$);
  \coordinate (C) at ($(B)+(\len,0)$);
  \coordinate (D) at ($(C)+(\len,0)$);
  \coordinate (E) at ($(C)+(-\len*0.5,+\len)$);
  \draw[pion] (A) -- (B);
  \draw[nucleon] (B) -- (C);
  \draw[pion] (C) -- (D);
  \draw[inmediumarrow1] (E) to[in=170,out=90] (B);
  \draw[inmediumarrow1] (C) to[in=90,out=40] (E);
  \draw[pion] (B) -- (E);
}

\newcommand{\SelfEnergyTypeadditonalthird}[0]{
  \coordinate (A) at (0,0);
  \coordinate (B) at ($(A)+(\len,0)$);
  \coordinate (C) at ($(B)+(\len,0)$);
  \coordinate (D) at ($(C)+(\len,0)$);
  \coordinate (E) at ($(C)+(-\len*0.5,+\len)$);
  \draw[pion] (A) -- (B);
  \draw[nucleon] (C) -- (B);
  \draw[pion] (C) -- (D);
  \draw[inmediumarrow1] (B) to[in=90,out=170] (E);
  \draw[inmediumarrow1] (E) to[in=40,out=90] (C);
  \draw[pion] (C) -- (E);
}

\newcommand{\SelfEnergyTypeadditonalfourth}[0]{
  \coordinate (A) at (0,0);
  \coordinate (B) at ($(A)+(\len,0)$);
  \coordinate (C) at ($(B)+(\len,0)$);
  \coordinate (D) at ($(C)+(\len,0)$);
  \coordinate (E) at ($(C)+(-\len*0.5,+\len)$);
  \draw[pion] (A) -- (B);
  \draw[nucleon] (B) -- (C);
  \draw[pion] (C) -- (D);
  \draw[inmediumarrow1] (E) to[in=170,out=90] (B);
  \draw[inmediumarrow1] (C) to[in=90,out=40] (E);
  \draw[pion] (C) -- (E);
}

\newcommand{\SelfEnergyTypeadditonalfifth}[0]{
  \coordinate (A) at (0,0);
  \coordinate (B) at ($(A)+(\len,0)$);
  \coordinate (C) at ($(B)+(\len,0)$);
  \coordinate (D) at ($(C)+(\len,0)$);
  \coordinate (E) at ($(C)+(-\len*0.5,+\len)$);
  \draw[pion] (A) -- (B);
  \draw[inmediumarrow2] (C) -- (B);
  \draw[pion] (C) -- (D);
  \draw[inmediumarrow1] (B) to[in=90,out=170] (E);
  \draw[nucleon] (E) to[in=40,out=90] (C);
  \draw[pion] (B) -- (E);
}

\newcommand{\SelfEnergyTypeadditonalsixth}[0]{
  \coordinate (A) at (0,0);
  \coordinate (B) at ($(A)+(\len,0)$);
  \coordinate (C) at ($(B)+(\len,0)$);
  \coordinate (D) at ($(C)+(\len,0)$);
  \coordinate (E) at ($(C)+(-\len*0.5,+\len)$);
  \draw[pion] (A) -- (B);
  \draw[inmediumarrow2] (B) -- (C);
  \draw[pion] (C) -- (D);
  \draw[inmediumarrow1] (E) to[in=170,out=90] (B);
  \draw[nucleon] (C) to[in=90,out=40] (E);
  \draw[pion] (B) -- (E);
}

\newcommand{\SelfEnergyTypeadditonalnineth}[0]{
  \coordinate (A) at (0,0);
  \coordinate (B) at ($(A)+(\len,0)$);
  \coordinate (C) at ($(B)+(\len,0)$);
  \coordinate (D) at ($(C)+(\len,0)$);
  \coordinate (E) at ($(C)+(-\len*0.5,+\len)$);
  \draw[pion] (A) -- (B);
  \draw[inmediumarrow2] (C) -- (B);
  \draw[pion] (C) -- (D);
  \draw[nucleon] (B) to[in=90,out=170] (E);
  \draw[inmediumarrow1] (E) to[in=40,out=90] (C);
  \draw[pion] (C) -- (E);
}

\newcommand{\SelfEnergyTypeadditonaltenth}[0]{
  \coordinate (A) at (0,0);
  \coordinate (B) at ($(A)+(\len,0)$);
  \coordinate (C) at ($(B)+(\len,0)$);
  \coordinate (D) at ($(C)+(\len,0)$);
  \coordinate (E) at ($(C)+(-\len*0.5,+\len)$);
  \draw[pion] (A) -- (B);
  \draw[inmediumarrow2] (B) -- (C);
  \draw[pion] (C) -- (D);
  \draw[nucleon] (E) to[in=170,out=90] (B);
  \draw[inmediumarrow1] (C) to[in=90,out=40] (E);
  \draw[pion] (C) -- (E);
}

\newcommand{\Decayadditonalfirst}[0]{
  \coordinate (A) at (0,0);
  \coordinate (B) at ($(A)+(\len,0)$);
  \coordinate (C) at ($(B)+(\len,0)$);
  \coordinate (D) at ($(C)+(\len,0)$);
  \coordinate (E) at ($(C)+(-\len*0.5,+\len)$);
  \draw[pion] (A) -- (B);
  \draw[nucleon] (C) -- (B);
  \draw[source] (C) -- (D);
  \draw[inmediumarrow1] (B) to[in=90,out=170] (E);
  \draw[inmediumarrow1] (E) to[in=40,out=90] (C);
  \draw[pion] (B) -- (E);
}

\newcommand{\Decayadditonalsecond}[0]{
  \coordinate (A) at (0,0);
  \coordinate (B) at ($(A)+(\len,0)$);
  \coordinate (C) at ($(B)+(\len,0)$);
  \coordinate (D) at ($(C)+(\len,0)$);
  \coordinate (E) at ($(C)+(-\len*0.5,+\len)$);
  \draw[pion] (A) -- (B);
  \draw[nucleon] (B) -- (C);
  \draw[source] (C) -- (D);
  \draw[inmediumarrow1] (E) to[in=170,out=90] (B);
  \draw[inmediumarrow1] (C) to[in=90,out=40] (E);
  \draw[pion] (B) -- (E);
}

\newcommand{\Decayadditonalthird}[0]{
  \coordinate (A) at (0,0);
  \coordinate (B) at ($(A)+(\len,0)$);
  \coordinate (C) at ($(B)+(\len,0)$);
  \coordinate (D) at ($(C)+(\len,0)$);
  \coordinate (E) at ($(C)+(-\len*0.5,+\len)$);
  \draw[pion] (A) -- (B);
  \draw[nucleon] (C) -- (B);
  \draw[source] (C) -- (D);
  \draw[inmediumarrow1] (B) to[in=90,out=170] (E);
  \draw[inmediumarrow1] (E) to[in=40,out=90] (C);
  \draw[pion] (C) -- (E);
}

\newcommand{\Decayadditonalfourth}[0]{
  \coordinate (A) at (0,0);
  \coordinate (B) at ($(A)+(\len,0)$);
  \coordinate (C) at ($(B)+(\len,0)$);
  \coordinate (D) at ($(C)+(\len,0)$);
  \coordinate (E) at ($(C)+(-\len*0.5,+\len)$);
  \draw[pion] (A) -- (B);
  \draw[nucleon] (B) -- (C);
  \draw[source] (C) -- (D);
  \draw[inmediumarrow1] (E) to[in=170,out=90] (B);
  \draw[inmediumarrow1] (C) to[in=90,out=40] (E);
  \draw[pion] (C) -- (E);
}

\newcommand{\Decayadditonalfifth}[0]{
  \coordinate (A) at (0,0);
  \coordinate (B) at ($(A)+(\len,0)$);
  \coordinate (C) at ($(B)+(\len,0)$);
  \coordinate (D) at ($(C)+(\len,0)$);
  \coordinate (E) at ($(C)+(-\len*0.5,+\len)$);
  \draw[pion] (A) -- (B);
  \draw[inmediumarrow2] (C) -- (B);
  \draw[source] (C) -- (D);
  \draw[inmediumarrow1] (B) to[in=90,out=170] (E);
  \draw[nucleon] (E) to[in=40,out=90] (C);
  \draw[pion] (B) -- (E);
}

\newcommand{\Decayadditonalsixth}[0]{
  \coordinate (A) at (0,0);
  \coordinate (B) at ($(A)+(\len,0)$);
  \coordinate (C) at ($(B)+(\len,0)$);
  \coordinate (D) at ($(C)+(\len,0)$);
  \coordinate (E) at ($(C)+(-\len*0.5,+\len)$);
  \draw[pion] (A) -- (B);
  \draw[inmediumarrow2] (B) -- (C);
  \draw[source] (C) -- (D);
  \draw[inmediumarrow1] (E) to[in=170,out=90] (B);
  \draw[nucleon] (C) to[in=90,out=40] (E);
  \draw[pion] (B) -- (E);
}

\newcommand{\Decayadditonalnineth}[0]{
  \coordinate (A) at (0,0);
  \coordinate (B) at ($(A)+(\len,0)$);
  \coordinate (C) at ($(B)+(\len,0)$);
  \coordinate (D) at ($(C)+(\len,0)$);
  \coordinate (E) at ($(C)+(-\len*0.5,+\len)$);
  \draw[pion] (A) -- (B);
  \draw[inmediumarrow2] (C) -- (B);
  \draw[source] (C) -- (D);
  \draw[nucleon] (B) to[in=90,out=170] (E);
  \draw[inmediumarrow1] (E) to[in=40,out=90] (C);
  \draw[pion] (C) -- (E);
}

\newcommand{\Decayadditonaltenth}[0]{
  \coordinate (A) at (0,0);
  \coordinate (B) at ($(A)+(\len,0)$);
  \coordinate (C) at ($(B)+(\len,0)$);
  \coordinate (D) at ($(C)+(\len,0)$);
  \coordinate (E) at ($(C)+(-\len*0.5,+\len)$);
  \draw[pion] (A) -- (B);
  \draw[inmediumarrow2] (B) -- (C);
  \draw[source] (C) -- (D);
  \draw[nucleon] (E) to[in=170,out=90] (B);
  \draw[inmediumarrow1] (C) to[in=90,out=40] (E);
  \draw[pion] (C) -- (E);
}
\begin{document}
\title{Glashow-Weinberg relation in isospin-asymmetric nuclear matter}
\author{Kihong Kwon}
\email{kwon.k.297a@m.isct.ac.jp}
\author{Daisuke Jido}
\email{jido@phys.titech.ac.jp}
\author{Yamato Suda}
\email{basketball.yamato98@gmail.com}
\affiliation{Department of Physics, Institute of Science Tokyo, Meguro, Tokyo 152-8551, Japan}
\date{\today}
\begin{abstract}
The conservation of the axial-vector currents in nuclear matter with finite isospin asymmetry $\rho_n/\rho_p$ is explored in this paper. In explicit calculations in the in-medium chiral perturbation theory, a charged pion becomes massive in isospin-asymmetric nuclear matter. In the presence of the massive charged pion, two components of the axial-vector current are not conserved from the two-loop order when applying the massive in-medium pion state. As a result, the in-medium Glashow-Weinberg relation for the charged pions cannot be derived in isospin-asymmetric nuclear matter from operator relations in quantum chromodynamics. 
\end{abstract}
\maketitle

\section{Introduction}
Broken global and local symmetries reveal the spectra of low-energy physics, the mass generation, and dispersion relations of effective degrees of freedom \cite{salam1968elementary,weinberg1967model,weinberg1976implications,weinberg1972approximate,Glashow:1967rx,watanabe2012unified,hidaka2013counting,nielsen1976count}. Especially, spontaneously broken global symmetries imply the appearance of massless bosons, so-called the Nambu-Goldstone bosons \cite{Goldstone:1962es,goldstone1962broken}. The broken chiral symmetry explains the hadron spectrum in low-energy quantum chromodynamics (QCD). The chiral symmetry is supposed to be partially restored at finite densities or temperatures, which has been explored both theoretically and experimentally \cite{ikeno2011precision,ikeno2023pion,tani2021structure,kolomeitsev2003chiral,jido2008medium,birse1994chiral,hatsuda1999precursor,celenza1992partial,birse1994two,jido2007medium,marczenko2023reconciling,jido2013eta,hirenzaki1991d,toki1989structure,galain1991vector,fernandez2007pion,gubler2017novel,narayanan2006chiral,meisinger1996chiral, suzuki2004precision,friedman2004medium,nishi2023chiral,friedman2005elastic,nishi2018spectroscopy,geissel2002deeply,itahashi2000deeply,yamazaki1998effective,grion2005pi,palmese2016chiral}. Since the masses of hadrons are strongly related to the broken chiral symmetry and the hadron properties are constrained by the symmetry breaking, one can expect hadron properties to be influenced by background matter effects at finite densities. 

The spontaneous breaking of chiral symmetry is confirmed by a nonzero value of the quark condensate. This quantity can be obtained from properties of the pions such as the Gell-Mann-Oakes-Renner relation \cite{Gell-Mann:1968hlm}, which relates the pion mass and decay constant to the quark condensate. The quark condensate can also be estimated using another relation in terms of pion form factors without masses, and this relation, so-called the Glashow-Weinberg relation \cite{glashow1968breaking,jido2008medium}, is equivalent to the Gell-Mann-Oakes-Renner relation under the partially conserved axial-vector current relation. The Glashow-Weinberg relation is the connection among the decay constant $f_\pi$, pseudoscalar coupling constant $G_\pi$, and the quark condensate, which is more useful in the chiral limit. We will derive the in-medium version of the Glashow-Weinberg relation.

The pions are the Nambu-Goldstone bosons in the chiral limit produced by the spontaneously broken chiral symmetry, $SU(2)_L\times SU(2)_R\rightarrow SU(2)_V$. In our previous work \cite{kwon2025pion}, the pion properties exhibit different behaviors in isospin-asymmetric nuclear matter depending on the matter density $\rho$ and isospin asymmetry $\rho_n/\rho_p$ with nonzero quark masses, as asymmetric in-medium corrections such as the Weinberg-Tomozawa term are nonzero contributions in the asymmetric matter. In the presence of asymmetric contributions, the in-medium effects may affect the masses of the Nambu-Goldstone bosons in the chiral limit. The in-medium corrections for the pion masses will be examined with explicit calculations.

The chiral symmetry restoration in nuclear matter can be studied by pion properties using in-medium versions of the operator relations. In Refs.~\cite{jido2008medium,jido2007medium}, relations between the in-medium quark condensate and pion properties were derived with the conservation of the axial-vector current in isospin-symmetric nuclear matter. In this work, the conservation of the axial-vector currents in isospin-asymmetric nuclear matter is investigated with operator relations and explicit calculations of the pion properties. 

We focus mainly on arguments in the chiral limit and take this limit at the end of the procedures. This paper is organized as follows: In \cref{sec:pionproper}, in-medium pion properties are defined, and the correlation function approach is introduced to calculate in-medium corrections using relevant Feynman diagrams. The in-medium corrections will be evaluated in the next-to-leading order in the in-medium chiral perturbation theory, which is reviewed in \cref{sec:inmediumchiral} with the chiral Lagrangian up to the second chiral order. In \cref{sec:inmeGW}, the in-medium Glashow-Weinberg relation is derived from operator relations in QCD. The mass corrections of the pions are evaluated in \cref{sec:mass} with relevant Feynman diagrams. In \cref{sec:breaking}, the conservation of the axial-vector currents in isospin-asymmetric nuclear matter is discussed. Finally, \cref{sec:summary} is devoted to the summary. 
\section{Pion properties in nuclear matter}\label{sec:pionproper}
In this section, the pion properties in nuclear matter with general isospin asymmetry $r\equiv\rho_n/\rho_p$ are defined from the in-medium pion states and the correlation function approach. The charged and neutral pions are collectively denoted as $\pi$ or $\pi^{\pm,0}$. This section follows the arguments in Refs.~\cite{goda2014pion,jido2008medium,kwon2025pion,meissner2002medium}. Note that we assume the isotropic systems, and thus all pion properties are rotationally symmetric.
\subsection{Definitions of in-medium pion properties}
We first define the in-medium pion states $|\pi^*\rangle$ with the conventional normalization similar to the vacuum pion state $|\pi\rangle$ as
\begin{subequations}\label{eq:normalization}
\begin{align}
    &\langle\pi^{\pm,0}(q_{\pi^{\pm,0}})|{\pi^{\pm,0}}(p_{\pi^{\pm,0}})\rangle=(2\pi)^32E_{\pi^{\pm,0}}(\vec p)\delta^{(3)}(\vec q-\vec p),\\
    &\langle\pi^{*\pm,0}(q_{\pi^{\pm,0}})|{\pi^{*\pm,0}}(p_{\pi^{\pm,0}})\rangle=(2\pi)^32\omega_{\pi^{\pm,0}}(\vec p)\delta^{(3)}(\vec q-\vec p),
\end{align}
\end{subequations}
where $E_\pi(\vec p)$ and $\omega_\pi(\vec p)$ respectively denote energies of the in-vacuum and in-medium pions, and $p_\pi$ is on-shell pion momentum, which is expressed for the in-medium pions as $p_\pi=(\omega_\pi(\vec p),\vec p)$. Note that the in-vacuum pion energy satisfies the special relativistic dispersion relation with the vacuum pion mass $m_\pi$ in vacuum, $E_\pi(\vec p)^2=m_\pi^2+\vec p^2$, but the in-medium pion energy $\omega_\pi(\vec p)$ does not follow this simple dispersion relation due to in-medium effects. Since $\omega_\pi(\vec p)$ is the on-shell in-medium pion energy, one can define the in-medium pion mass $m_\pi^*$ as
\begin{align}\label{eq:deinme}
    m_{\pi^{\pm,0}}^*=\omega_{\pi^{\pm,0}}(\vec 0).
\end{align}
As shall be seen, even in the chiral limit, some of the pions can obtain a nonzero mass in isospin-asymmetric nuclear matter.

In this paper, we tacitly assume all the vacuum effects are properly renormalized. Thus, the renormalized pion fields have the matrix elements between the pure vacuum state $|0\rangle$ and one vacuum pion state in the form
\begin{align}
    \langle0|\pi^{\pm,0}(x)|\pi^{\pm,0}(p_{\pi^{\pm,0}})\rangle=\eu^{-\im p_{\pi^{\pm,0}}\cdot x}.
\end{align}
Then, the matrix elements of the pion fields between the asymptotic state of nuclear matter $|\Omega\rangle$ and one in-medium pion state can be written as
\begin{align}
    \langle\Omega|\pi^{\pm,0}(x)|\pi^{*\pm,0}(p_{\pi^{\pm,0}})\rangle=\sqrt{Z_{\pi^{\pm,0}}}\eu^{-\im p_{\pi^{\pm,0}}\cdot x},
\end{align}
where a constant wave function renormalization $Z_\pi$ has been introduced. We shall see how the wave function renormalization is calculated for the on-shell pions in the next section. Note that by taking into account the wave function renormalization, one can appropriately include in-medium effects on the pion fields, even though the in-medium effects have not been considered in the quantization process directly \cite{jido2008medium,meissner2002medium}.

Similarly, the matrix elements of composite operators can be defined. The pseudoscalar densities in QCD,
\begin{subequations}\label{eq:pesudoscq}
\begin{align}
    &\quad P^a=\im\bar q\gamma_5\tau^aq,\qquad a=1,2,3\\
    &P^\pm=\frac{1}{\sqrt2}(P^1\mp\im P^2),\quad P^0=P^3,
\end{align}
\end{subequations}
where $q$ and $\bar q$ are quark and antiquark fields, respectively, and $\tau^a$ is the Pauli matrix, have the matrix elements,
\begin{align}\label{eq:defpeu}
    \langle\Omega|P^{\pm,0}(x)|\pi^{*\pm,0}(p_{\pi^{\pm,0}})\rangle=G_{\pi^{\pm,0}}^*(p_{\pi^{\pm,0}})\eu^{-\im p_{\pi^{\pm,0}}\cdot x},
\end{align}
where the in-medium pseudoscalar form factor $G_\pi^*$ has been defined. Also, the matrix elements of the axial-vector currents,
\begin{subequations}\label{eq:axialquark}
\begin{align}
    &\quad A_\mu^a=\bar q\gamma_\mu\gamma_5\frac{\tau^a}{2}q,\quad a=1,2,3\\
    &A_\mu^\pm=\frac{1}{\sqrt 2}(A_\mu^1\mp\im A_\mu^2),\quad A_\mu^0=A_\mu^3,
\end{align}
\end{subequations}
which are conserved Noether currents of the chiral symmetry in vacuum QCD,
can be expressed from Lorentz covariance as
\begin{align}\label{eq:decayinme}
    &\langle\Omega|A_\mu^{\pm,0}(x)|\pi^{*\pm,0}(p_{\pi^{\pm,0}})\rangle\nonumber\\
    &\quad=\im\big[n_\mu N_{\pi^{\pm,0}}^*(p_{\pi^{\pm,0}})+p_\mu^{\pi^{\pm,0}}F_{\pi^{\pm,0}}^*(p_{\pi^{\pm,0}})\big]\eu^{-\im p_{\pi^{\pm,0}}\cdot x},
\end{align}
in which $n_\mu$ is the Lorentz frame vector of nuclear matter, which is taken as $n_\mu=(1,\vec 0)$ in the rest frame of nuclear matter, and the form factor $N_\pi^*$ has one more mass dimension than $F_\pi^*$. From these form factors, one can define in-medium pion decay constants \cite{jido2008medium,kwon2025pion}, but some of the decay constants seem to exhibit singularities in the soft limit of pion momenta depending on the definitions. Thus, observables must be carefully defined. As we are concerned with the chiral limit, we focus on the form factors $N_\pi^*$ and $F_\pi^*$ for the matrix elements of the axial-vector currents.

Note that the pseudoscalar densities and axial-vector currents satisfy the following operator relation in QCD:
\begin{align}\label{eq:PCAC}
    \partial^\mu A_\mu^{\pm,0}(x)=m_qP^{\pm,0}(x),
\end{align}
where we have assumed that the up- and down- quarks have the common mass, $m_u=m_d\equiv m_q$. This is the partially conserved axial-vector current (PCAC) relation, in which the right-hand side of \cref{eq:PCAC} is zero in the chiral limit, yielding the conserved axial-vector current relation. This confirms that the action and measure exhibit chiral symmetry in the chiral limit, but the chiral symmetry is spontaneously broken by quantum states, producing the Nambu-Goldstone bosons (the pions in the two-flavor case).
\subsection{Correlation function approach}
The in-medium pion properties defined in the previous section can be calculated in the correlation function approach, in which the pion properties are identified from the residues at the on-shell momentum pole positions of various correlation functions in momentum space. Let us first consider the two-point function of the pseudoscalar densities 
\begin{align}
    \Pi^{\pm\mp,00}(q)\equiv\int\dd[4]x\eu^{\im q\cdot x}\langle\Omega|\mathsf{T}\{P^{\pm,0}(x)P^{\mp,0}(0)\}|\Omega\rangle.
\end{align}
Using the complete sets of the in-medium pion states with the matrix elements \cref{eq:defpeu}, this correlation function can be expressed as
\begin{align}
    \Pi^{\pm\mp,00}(q)&=\frac{1}{2\omega_{\pi^{\pm,0}}(\vec q)}G_{\pi^{\pm,0}}^*\frac{\im}{q_0-\omega_{\pi^{\pm,0}}(\vec q)+\im\epsilon}\bar G^*_{\pi^{\pm,0}}\nonumber\\
    &-\frac{1}{2\omega_{\pi^{\mp,0}}(\vec q)}G_{\pi^{\mp,0}}^*\frac{\im}{q_0+\omega_{\pi^{\mp,0}}(\vec q)-\im\epsilon}\bar G^*_{\pi^{\mp,0}}+\cdots,
\end{align}
where $\cdots$ represents contributions from pion excited states. We are interested in the in-medium properties of pions and focus on the pion modes. On the other hand, the correlation function can be written in terms of vertex corrections and in-medium corrections of the pion propagators as
\begin{align}\label{eq:ppinme}
    \Pi^{\pm\mp,00}(q)=\hat G_{\pi^{\pm,0}}\frac{\im}{q^2-m_\pi^2-\Sigma_{\pi^{\pm,0}}(q)+\im\epsilon}\bar{\hat G}_{\pi^{\pm,0}}+\cdots,
\end{align}
where $\hat G_\pi$ is the vertex correction for the matrix element \cref{eq:defpeu}, and $\Sigma_\pi$ denotes the in-medium pion self-energy, which respectively include all one-particle-irreducible diagram contributions of in-medium effects for the corresponding matrix element and propagator. This expression can be expanded as
\begin{align}\label{eq:ppinva}
    &\Pi^{\pm\mp,00}(q)\nonumber\\
    &=\hat G_{\pi^{\pm,0}}\frac{\im}{2\mathcal{E}_{\pi^{\pm,0}}}\bigg(\frac{1}{q_0-\mathcal{E}_{\pi^{\pm,0}}+\im\epsilon}-\frac{1}{q_0+\mathcal{E}_{\pi^{\pm,0}}-\im\epsilon}\bigg)\bar{\hat G}_{\pi^{\pm,0}}\nonumber\\
    &\quad+\cdots,
\end{align}
where $\mathcal{E}_\pi\equiv\sqrt{m_\pi^2+\Sigma_\pi(q)+\vec q^2}$. By comparing \cref{eq:ppinme,eq:ppinva} at the on-shell momentum pole positions of the pions, one can find
\begin{align}
    \omega_{\pi^{\pm,0}}(\vec q)=\sqrt{m_\pi^2+\Sigma_{\pi^{\pm,0}}(q_{\pi^{\pm,0}})+\vec q^2},
\end{align}
from which, with \cref{eq:deinme}, the mass equations can be found as
\begin{align}\label{eq:inmemasseq}
    m^*_{\pi^{\pm,0}}=\sqrt{m_\pi^2+\Sigma_{\pi^{\pm,0}}(m^*_{\pi^{\pm,0}},\vec 0)}.
\end{align}
Each mass equation exhibits positive and negative solutions, which respectively correspond to particle and antiparticle mass solutions. As the neutral pion is its own antiparticle, its positive and negative mass solutions are the same. On the other hand, the charged pions are antiparticles of each other, and they generally have different masses in nuclear matter from isospin asymmetry effects, which come from the Weinberg-Tomozawa interaction at the one-loop order. Then, some of the pions might have nonzero mass solutions, which will be argued in more detail later.

Around the pion on-shell momenta, \cref{eq:ppinva} can be written as
\begin{align}\label{eq:ppcorewav}
    &\Pi^{\pm\mp,00}(q)\nonumber\\
    &=\frac{\im}{2\omega_{\pi^{\pm,0}}(\vec q)}\hat G_{\pi^{\pm,0}}\frac{Z_{\pi^{\pm,0}}}{q_0-\omega_{\pi^{\pm,0}}(\vec q)+\im\epsilon}\bar{\hat{G}}_{\pi^{\pm,0}}+\cdots,
\end{align}
where the wave function renormalization has been defined as
\begin{align}\label{eq:defwavefunction}
    Z_\pi\equiv\bigg(1-\frac{1}{2\omega_{\pi}(\vec q)}\frac{\partial\Sigma_\pi(q)}{\partial q_0}\bigg|_{q_0=\omega_\pi(\vec q)}\bigg)^{-1}.
\end{align}
Then, the pseudoscalar form factor of the in-medium pion at on-shell momenta can be calculated as
\begin{align}
    G_{\pi^{\pm,0}}^*=\sqrt{Z_{\pi^{\pm,0}}}\hat G_{\pi^{\pm,0}}.
\end{align}
As mentioned earlier, one can properly include in-medium effects on the pion fields by using the wave function renormalization. The importance of the wave function renormalization was already emphasized in Refs.~\cite{jido2008medium,hatsuda1999precursor,meissner2002medium}.

In similar ways, the two-point function of the axial-vector currents can also be discussed,
\begin{align}
    \Pi_{\mu\nu}^{\pm\mp,00}(q)\equiv\int\dd[4]x\eu^{\im q\cdot x}\langle\Omega|\mathsf{T}\{A_\mu^{\pm,0}(x)A_\nu^{\mp,0}(0)\}|\Omega\rangle.
\end{align}
It can be expanded from the complete sets of the in-medium pion states as
\begin{align}
    &\Pi_{\mu\nu}^{{\pm\mp,00}}(q)\nonumber\\
    &=\im\frac{\big(n_\mu N_{\pi^{\pm,0}}^*+q_\mu^{\pi^{\pm,0}}(\vec q)F_{\pi^{\pm,0}}^*\big)\big(n_\nu\bar N_{\pi^{\pm,0}}^*+q_\nu^{\pi^{\pm,0}}(\vec q)\bar F_{\pi^{\pm,0}}^*\big)}{2\omega_{\pi^{\pm,0}}(\vec q)(q^0-\omega_{\pi^{\pm,0}}(\vec q)+\im\epsilon)}\nonumber\\
    &-\im\frac{\big(n_\mu \bar N_{\pi^{\mp,0}}^*+q_\mu^{\pi^{\mp,0}}(-\vec q)\bar F_{\pi^{\mp,0}}^*\big)\big(n_\nu N_{\pi^{\mp,0}}^*+q_\nu^{\pi^{\mp,0}}(-\vec q) F_{\pi^{\mp,0}}^*\big)}{2\omega_{\pi^{\mp,0}}(\vec q)(q^0+\omega_{\pi^{\mp,0}}(\vec q)-\im\epsilon)}\nonumber\\
    &\quad+\cdots,
\end{align}
where $q^\pi$ and $q$ denote on-shell and off-shell in-medium pion momenta, respectively, and $q^\pi(\pm \vec q)\equiv(\omega_\pi(\vec q),\pm\vec q)$. This correlation function can also be expressed in terms of in-medium corrections from vacuum quantities as
\begin{align}
    &\Pi_{\mu\nu}^{\pm\mp,00}(q)\nonumber\\
    &\quad=\big(\im\hat \Gamma_\mu^{\pi^{\pm,0}}
    \big)\frac{\im}{q^2-m_\pi^2-\Sigma_{\pi^{\pm,0}}(q)+\im\epsilon}\big(-\im\bar{\hat \Gamma}_\nu^{\pi^{\pm,0}}
    \big)+\cdots,
\end{align}
where $\im\hat \Gamma_\mu^\pi$ is the vertex correction for the matrix element \cref{eq:decayinme}, including all one-particle-irreducible diagram contributions. By expanding this expression around an on-shell pion momentum similar to \cref{eq:ppcorewav}, one obtains
\begin{align}\label{eq:axial-vector form}
    n_\mu N_\pi^*+q_\mu^\pi(\vec q)F_\pi^*=\sqrt{Z_\pi}\hat{\Gamma}_\mu^\pi.
\end{align}
Note that the form factors and self-energies are rotationally symmetric, so that they are invariant under $\vec q\leftrightarrow-\vec q$ in isotropic nuclear matter as well as the in-medium pion energies.

\section{In-medium chiral perturbation theory}\label{sec:inmediumchiral}
In this work, the in-medium chiral perturbation theory with the $SU(2)$ chiral symmetry is used to evaluate the in-medium pion properties. In this section, the in-medium chiral perturbation theory is briefly introduced with the chiral Lagrangian of the nucleon and pion up to the second order in chiral power, as in-medium effects from the $\mathcal{O}(\rho^2)$ order are ignored.
\subsection{The generating functional}
The in-medium chiral perturbation theory is derived in the path integral formalism assuming the asymptotic state of nuclear matter as Fermi seas of nucleons \cite{oller2002chiral,meissner2002medium}:
\begin{align}
    |\Omega\rangle=\prod_i^{|\vec p_i|\le k_F^{i}}a^\dagger(\vec p_i)|0\rangle,
\end{align}
where $i$ denotes all the quantum numbers of a nucleon, such as isospin, and $a^\dagger(\vec p_i)$ is a nucleon creation operator. The nucleons are included in the matter up to the Fermi momenta of the protons and neutrons, which can be written in terms of the densities as
\begin{align}
    k_F^{p,n}=(3\pi^2\rho_{p,n})^{1/3},
\end{align}
where $p$ and $n$ indices represent proton and neutron, respectively. Note that the isospin asymmetry of matter can be parametrized by the proton and neutron densities as $r\equiv\rho_n/\rho_p$.

The generating functional $Z[J]$ in the path integral formalism is defined as the vacuum-to-vacuum transition amplitude. In nuclear matter, the generating function can be expressed as
\begin{align}
    Z[J]&=\int\mathcal{D}U\mathcal{D}\bar N\mathcal{D}N\langle\Omega_{\text{out}}|N(t\rightarrow\infty)\rangle\nonumber\\
    &\times\eu^{\im\int\dd[4]x[\mathcal{L}_{\pi}+\mathcal{L}_{\pi N}]}\langle N(t\rightarrow-\infty)|\Omega_{\text{in}}\rangle,
\end{align}
where $U$ and $N$ indicate the chiral and nucleon fields, respectively. Also, the full Lagrangian contains pion-nucleon interaction terms as well as separate pion and nucleon terms. According to Refs.~\cite{oller2019medium,oller2010chiral,alarcon2023nuclear}, nucleon-nucleon contact interactions are important from the $\mathcal{O}(\rho^2)$ order, which is beyond our present scope. Therefore, the nucleon fields appear only in the bilinear form, and then the generating functional can be integrated out, yielding the expression \cite{oller2002chiral},
\begin{widetext}
\begin{align}\label{eq:generating}
    Z[J]&=\int\mathcal{D}U\exp\bigg\{\im\int\dd[4]x\bigg[\mathcal{L}_\pi-\int\dd\Tilde{p}\mathcal{F}\text{Tr}\big(\im\Gamma(x,y)(\cancel{p}+m_N)n(\vec p)\big)\nonumber\\
    &-\frac{\im}{2}\int\dd\Tilde{p}\dd\Tilde{q}\mathcal{F}\text{Tr}\big(\im\Gamma(x,x')(\cancel{q}+m_N)n(\vec q)\im\Gamma(y',y)(\cancel{p}+m_N)n(\vec p)\big)+\cdots\bigg]\bigg\},
\end{align}
\end{widetext}
where $d\Tilde{p}\equiv\dd[3]\vec p(2\pi)^{-3}[2E(\vec p)]^{-1}$ is the Lorentz invariant three-momentum integration measure, $E(\vec p)=(\vec p^2+m_N^2)^{1/2}$ denotes the on-shell nucleon energy, $\mathcal{F}$ indicates a Fourier transformation of spacetime variables except for $x$, and $\Gamma(x,y)$ is a nonlocal vertex defined with interaction operator $A$ and the vacuum nucleon propagator $D_0^{-1}$ as
\begin{align}
    \im\Gamma\equiv A\big[\mathds{1}_4-D_0^{-1}A\big]^{-1}.
\end{align}
The matrix $n(\vec p)$ restricts the momentum integral of an in-medium nucleon propagator up to the Fermi momenta, defined with the Heaviside step function $\Theta$ in a diagonal form as
\begin{align}
	n(\vec p) &\equiv \begin{pmatrix}
		\Heavi(k_F^p - |\vec p|) & 0 \\
		0 & \Heavi(k_F^n - |\vec p|) 
	\end{pmatrix}.
\end{align}
The Heaviside step functions for the Fermi momenta are often abbreviated as $\Theta_{\vec p}^{p,n}\equiv\Theta(k_F^{p,n}-|\vec p|)$.

The expression \cref{eq:generating} exhibits three independent expansion schemes. The first is the expansion in Fermi sea insertions, which is explicitly represented in \cref{eq:generating}. Next, the nonlocal vertex can be expanded by the interaction operator $A$ as $\im\Gamma=A+AD_0^{-1}A+AD_0^{-1}AD_0^{-1}A+\cdots$. The final scheme is the chiral power expansion. The pion-nucleon interaction operator can be expanded as $A=A^{(1)}+A^{(2)}+\cdots$, in which number indices denote the small-momentum powers in the chiral perturbation theory. In our perturbative scheme, we include up to the next-to-leading order in each expansion, excluding any contribution from the $\mathcal{O}(\rho^2)$ order. 
\subsection{The chiral Lagrangian}
In this work, the chiral Lagrangian is considered in the second order of the chiral power counting. The lowest contribution of the pion Lagrangian is given by \cite{gasser1985chiral}
\begin{align}
    \mathcal{L}_\pi^{(2)}=\frac{f^2}{4}\text{Tr}\big(D_\mu U^\dagger D^\mu U+\chi^\dagger U+\chi U^\dagger \big),
\end{align}
where $D_\mu$ denotes the covariant derivative for the chiral field $U$, and $\chi$ is a linear combination of the scalar and pseudoscalar external currents. This Lagrangian contains the lowest-order pion-pion interaction terms as well as the pion kinetic term. Note that we employ the parametrization for the chiral field given by
\begin{align}
    U=\exp{\im\pi^a\tau^a\frac{y(\pi^2)}{2\sqrt{\pi^2}}},\quad(\pi^2=\pi^a\pi^a)
\end{align}
where the function $y(\pi^2)$ satisfies
\begin{align}
    y(\pi^2)-\sin[y(\pi^2)]=\frac{4}{3}\bigg(\frac{\pi^2}{f^2}\bigg)^{3/2}.
\end{align}
This parametrization ensures chiral symmetry in the path integral formalism, enabling a naive treatment of perturbative calculations \cite{charap1970closed,charap1971closed,gerstein1971chiral}.

The nucleon terms are written as
\begin{align}\label{eq:nuclLag}
    \mathcal{L}_{\pi N}=\bar N\bigg(\im\cancel{\partial}-m_N-\sum_{j=1}A^{(j)}\bigg)N. 
\end{align}
The interaction operators $A^{(j)}$ of the first and second orders are given by \cite{gasser1988nucleons,fettes2001annals,gasser1988nucleons}
\begin{align}\label{eq:interaction}
    A^{(1)}&=-\im\gamma^\mu\Gamma_\mu-\im g_A\gamma^\mu\gamma^5\Delta_\mu,\nonumber\\
    A^{(2)}&=-c_1\text{Tr}\big(\chi_+\big)+\frac{2c_2}{2m_N^2}\text{Tr}\big(u^\mu u^\nu\big)D^N_\mu D^N_\nu\nonumber\\
    &\qquad-\frac{c_3}{2}\text{Tr}\big(u^\mu u_\mu\big)+\cdots,
\end{align}
with
\begin{align}
    D_\mu^N&=\partial_\mu+\Gamma_\mu,\nonumber\\
    \Gamma_\mu&=\frac{1}{2}[u^\dagger,\partial_\mu u]-\frac{\im}{2}u^\dagger(v_\mu+a_\mu)u-\frac{\im}{2}u(v_\mu-a_\mu)u^\dagger,\nonumber\\
    u&=\sqrt U\nonumber\\
    \Delta_\mu&=\frac{1}{2}\big[u^\dagger\{\partial_\mu-\im(v_\mu+a_\mu)\}u-u\{\partial_\mu-\im(v_\mu-a_\mu)\}u^\dagger\big],\nonumber\\
    u_\mu&=2\im\Delta_\mu,\nonumber\\
    \chi_{\pm}&=u^\dagger\chi u^\dagger\pm u\chi^\dagger u.\nonumber
\end{align}
Explicit expressions of the interaction terms are listed in Appendix~\ref{app:interaction}.

As all the vacuum effects are assumed to be renormalized by counterterms in the chiral Lagrangian, physical values are used for the in-vacuum quantities; the in-vacuum pion decay constant $f=92.4$ MeV, the axial-vector coupling $g_A=1.26$, and the in-vacuum nucleon mass $m_N=939$ MeV. Also, the low-energy constants in $A^{(2)}$ are given in units of $10^{-3}\text{MeV}^{-1}$ by $c_1=0.59$, $c_2=3.30$, and $c_3=-4.43$ \cite{gasser1991sigma,alarcon2012chiral,alarcon2014strangeness,baru2011precision,goda2014pion}. Note that in-medium corrections proportional to $c_4$ are in higher orders, and thus the interaction proportional to $c_4$ is ignored in \cref{eq:interaction}.

\section{In-medium Glashow-Weinberg relation}\label{sec:inmeGW}
The Glashow-Weinberg (GW) relation is the representation of the chiral Ward identity near the physical pion pole positions \cite{jido2008medium,glashow1968breaking}, which shows the connection between the quark condensate and pion form factors. This relation can be written as the well-known Gell-Mann-Oakes-Renner relation \cite{Gell-Mann:1968hlm,hatsuda1985soft,thorsson1995s} with a nonzero  quark mass, which is expressed in terms of observable quantities such as the pion mass and decay constant. In the chiral limit, the GW relation is more useful to estimate the extent of the spontaneous breaking of chiral symmetry, which is written as 
\begin{align}
    f_\pi G_\pi=-\langle0|\bar qq|0\rangle,
\end{align}
where $|0\rangle$ denotes the QCD vacuum. The in-medium version of the GW relation can be derived using the operator relations in QCD by taking into account in-medium pion states. To achieve this, we maintain the averaged quark mass $m_q$ in operator relations and set it to zero at the end. Let us first consider the following correlation function,
\begin{align}
    \int\dd[4]x\eu^{\im q\cdot x}\partial^\mu\langle\Omega|\mathsf{T}\{A_\mu^{\pm,0}(x)P^{\mp,0}(0)\}|\Omega\rangle,
\end{align}
which can be written using the PCAC relation, \cref{eq:PCAC}, and integration by parts as
\begin{align}\label{eq:chrialwardbe}
    &\im q^\mu\Pi_{5\mu}^{\pm\mp,00}(q)+m_q\Pi^{\pm\mp,00}(q)\nonumber\\
    &=-\int\dd[4]x\eu^{\im q\cdot x}\delta(x_0)\langle\Omega|[A_0^{\pm,0}(x),P^{\mp,0}(0)]|\Omega\rangle,
\end{align}
where
\begin{align}
    \Pi_{5\mu}^{\pm\mp,00}(q)\equiv\int\dd[4]x\eu^{\im q\cdot x}\langle\Omega|\mathsf{T}\{A_\mu^{\pm,0}(x)P^{\mp,0}(0)\}|\Omega\rangle.
\end{align}
The right-hand side of \cref{eq:chrialwardbe} can be calculated by writing the operators in terms of the quark fields:
\begin{subequations}
\begin{align}
    &\int\dd[4]x\eu^{\im q\cdot x}\delta(x_0)[A_0^{\pm}(x),P^{\mp}(0)]\nonumber\\
    &=\frac{1}{2}\int\dd[3]\vec x\eu^{-\im\vec q\cdot\vec x}\nonumber\\
    &\qquad\times\bigg[q^\dagger(0,\vec x)\frac{\tau^1\mp\im\tau^2}{2}\gamma_5q(0,\vec x),\im q^\dagger(0)\gamma^0\gamma_5\big(\tau^1\pm\im\tau^2\big)q(0)\bigg]\nonumber\\
     &=-\im\int\dd[3]\vec x\eu^{-\im \vec q\cdot \vec x}\delta^{(3)}(\vec x)\bar q(0,\vec x)q(0),\\
    &\int\dd[4]x\eu^{\im q\cdot x}\delta(x_0)[A_0^{0}(x),P^{0}(0)]\nonumber\\
    &=\int\dd[3]\vec x\eu^{-\im\vec q\cdot\vec x}\bigg[q^\dagger(0,\vec x)\frac{\tau^3}{2}\gamma_5q(0,\vec x),\im q^\dagger(0)\gamma^0\gamma_5\tau^3q(0)\bigg]\nonumber\\
    &=-\im\int\dd[3]\vec x\eu^{-\im \vec q\cdot \vec x}\delta^{(3)}(\vec x)\bar q(0,\vec x)q(0).
\end{align}
\end{subequations}
Then, one arrives at the chiral Ward identity:
\begin{align}\label{eq:chiralward}
    \im q^\mu\Pi_{5\mu}^{\pm\mp,00}(q)+m_q\Pi^{\pm\mp,00}(q)=\im\langle\Omega|\bar qq|\Omega\rangle\equiv\im\langle\bar qq\rangle^*.
\end{align}
Note that the axial-vector currents and pseudoscalar densities are treated as local functions of the quark fields, which provides the Dirac delta function in coordinate space; that is, the internal structures of the operators in terms of the quark fields have not been considered. In other words, we are treating long-distance scales in this argument compared to the inner structures, so that this identity is valid for small $q$. As shall be shown, Adler's zero confirms that the left-hand side exhibits the $q$ dependence only through the corresponding form factors at on-shell momenta for a Nambu-Goldstone boson. As in the vacuum case, the in-medium form factors are independent of the on-shell pion momenta unless the matter provides a momentum change to an external pion, which is not considered in our work.

As derived in Appendix \ref{app:decorr}, the correlation function on the left-hand side of the chiral Ward identity can be expressed in terms of the in-medium pion properties as
\begin{widetext}
\begin{subequations}
\begin{align}
    &\im q^\mu\Pi_{5\mu}^{\pm\mp,00}(q)=-\im q^\mu\sum_{n}\frac{\big[n_\mu N_{\pi_n^{\pm,0}}^*+p_\mu^{\pi_n^{\pm,0}}(\vec q)F_{\pi_n^{\pm,0}}^*\big]\bar G_{\pi_
    {n}^{\pm,0}}^*}{2\omega_{\pi_n^{\pm,0}}(\vec q)(q^0-\omega_{\pi_n^{\pm,0}}(\vec q)+\im\epsilon)}-\im q^\mu\sum_{m}\frac{\big[n_\mu \bar N_{\pi_m^{\mp,0}}^*+p_\mu^{\pi_m^{\mp,0}}(-\vec q)\bar F_{\pi_m^{\mp,0}}^*\big]G_{\pi_
    {m}^{\mp,0}}^*}{2\omega_{\pi_m^{\mp,0}}(\vec q)(q^0+\omega_{\pi_m^{\mp,0}}(\vec q)-\im\epsilon)},\\
    &m_q\Pi^{\pm\mp,00}(q)=\im m_q\sum_{n}\frac{G_{\pi_
    {n}^{\pm,0}}^*\bar G_{\pi_
    {n}^{\pm,0}}^*}{2\omega_{\pi_n^{\pm,0}}(\vec q)(q^0-\omega_{\pi_n^{\pm,0}}(\vec q)+\im\epsilon)}-\im m_q\sum_{m}\frac{G_{\pi_
    {m}^{\mp,0}}^*\bar G_{\pi_
    {m}^{\mp,0}}^*}{2\omega_{\pi_m^{\mp,0}}(\vec q)(q^0+\omega_{\pi_m^{\mp,0}}(\vec q)-\im\epsilon)},
\end{align}
\end{subequations}
in which the on-shell momentum dependence of the form factors has been dropped. Using the PCAC relation in terms of the in-medium pion properties, one finds the following relation from \cref{eq:defpeu,eq:decayinme,eq:PCAC}
\begin{align}
    \big[\omega_{\pi_n^{\pm,0}}(\vec p)^2-\vec p\cdot \vec p\big]F_{\pi_n^{\pm,0}}^*+\omega_{\pi_n^{\pm,0}}(\vec p)N_{\pi_n^{\pm,0}}^*=m_qG_{\pi_n^{\pm,0}}^*,
\end{align}
so that
\begin{subequations}
\begin{align}
    \im q^\mu\Pi_{5\mu}^{\pm\mp,00}(q)&=-\im q^0 \sum_{n} \frac{( N_{\pi_n^{\pm,0}}^*+\omega_{\pi_n^{\pm,0}}(\vec q) F^*_{\pi_n^{\pm,0}}) \bar G^*_{\pi^{\pm,0}_n}}{2\omega_{\pi_n^{\pm,0}}(\vec q)\big(q^0 -\omega_{\pi_n^{\pm,0}}(\vec q) + \im \epsilon\big)} -\im q^0 \sum_{n} \frac{( \bar N_{\pi_n^{\mp,0}}^*+\omega_{\pi_n^{\mp,0}}(\vec q) \bar F^*_{\pi_n^{\mp,0}}) G^*_{\pi^{\mp,0}_n}}{2\omega_{\pi_n^{\mp,0}}(\vec q)\big(q^0 + \omega_{\pi_n^{\mp,0}}(\vec q) - \im \epsilon\big)}\nonumber\\
    &+\im \vec q^2 \sum_{n} \frac{ F^*_{\pi_n^{\pm,0}} \bar G^*_{\pi^{\pm,0}_n}}{2\omega_{\pi_n^{\pm,0}}(\vec q)\big(q^0 -\omega_{\pi_n^{\pm,0}}(\vec q) + \im \epsilon\big)} -\im \vec q^2 \sum_{n} \frac{\bar F^*_{\pi_n^{\mp,0}} G^*_{\pi^{\mp,0}_n}}{2\omega_{\pi_n^{\mp,0}}(\vec q)\big(q^0 + \omega_{\pi_n^{\mp,0}}(\vec q) - \im \epsilon\big)},\\
    m_q\Pi^{\pm\mp,00}(q)&=\im\sum_{n} \frac{\omega_{\pi_n^{\pm,0}}(\vec q)(\omega_{\pi_n^{{\pm,0}}}(\vec q)F_{\pi_n^{\pm,0}}^{*}+N_{\pi_n^{\pm,0}}^{*}) \bar G^*_{\pi^{\pm,0}_n}}{2\omega_{\pi_n^{\pm,0}}(\vec q)(q^0 -\omega_{\pi_n^{\pm,0}}(\vec q) + \im \epsilon)} - \im \sum_{n} \frac{\omega_{\pi_n^{\mp,0}}(\vec q)(\omega_{\pi_n^{\mp,0}}(\vec q)\bar F_{\pi_n^{\mp,0}}^{*}+\bar N_{\pi_n^{\mp,0}}^{*}) G^{*}_{\pi^{\mp,0}_n}}{2\omega_{\pi_n^{\mp,0}}(\vec q)(q^0 + \omega_{\pi_n^{\mp,0}}(\vec q) - \im \epsilon)}\nonumber\\
    &-\im\vec q^2\sum_{n} \frac{F_{\pi_n^{\pm,0}}^{*} \bar G^*_{\pi^{\pm,0}_n}}{2\omega_{\pi_n^{\pm,0}}(\vec q)(q^0 -\omega_{\pi_n^{\pm,0}}(\vec q) + \im \epsilon)} + \im \vec q^2\sum_{n} \frac{\bar F_{\pi_n^{\mp,0}}^{*} G^{*}_{\pi^{\mp,0}_n}}{2\omega_{\pi_n^{\mp,0}}(\vec q)(q^0 + \omega_{\pi_n^{\mp,0}}(\vec q) - \im \epsilon)}.
\end{align}
\end{subequations}
Then, the in-medium Glashow-Weinberg relation can be derived:
\begin{subequations}\label{eq:inmeGW}
\begin{align}
    \frac{1}{2} \bigg[\sum\limits_{n}  \bigg(F_{\pi_n^+}^*\big(p_{\pi_n^+}\big)+\frac{N_{\pi_n^+}^*\big(p_{\pi_n^+}\big)}{\omega_{\pi_n^+}(\vec p)}\bigg) \bar G^{*}_{\pi^+_n}\big(p_{\pi_n^+}\big)+\sum\limits_{n}\bigg(\bar F_{\pi_n^-}^*\big(p_{\pi_n^-}\big)+\frac{\bar N_{\pi_n^-}^*\big(p_{\pi_n^-}\big)}{\omega_{\pi_n^-}(\vec p)}\bigg) G^{*}_{\pi^-_n}\big(p_{\pi_n^-}\big)  \bigg]&=-\langle\bar qq\rangle^*,\label{eq:inmeGWch}\\
    \sum\limits_{n}\bigg(\bar F_{\pi^0_n}^*\big(p_{\pi_n^0}\big)+\frac{\bar N_{\pi_n^0}^*\big(p_{\pi_n^0}\big)}{\omega_{\pi_n^0}(\vec p)}\bigg) G^{*}_{\pi^0_n}\big(p_{\pi_n^0}\big)&=-\langle\bar qq\rangle^*,\label{eq:inmeGWne}
\end{align}
\end{subequations}
\end{widetext}
which are valid for sufficiently small momenta to treat the pions as point-like particles. From the chiral Ward identity, \cref{eq:chiralward}, one additional relation can be written, which is the complex conjugate of \cref{eq:inmeGWch}. Since we are not considering the inner structure and decay of the pions, the form factors are assumed to be real, so that the additional relation is the same as \cref{eq:inmeGWch}.
Therefore, if the operator relations in QCD, e.g., the PCAC relation, \cref{eq:PCAC}, are also valid in nuclear matter, the partial restoration of the chiral symmetry in nuclear matter with general isospin asymmetry can be estimated from in-medium pion properties using the in-medium Glashow-Weinberg relation, \cref{eq:inmeGW}. 

Note that in the vacuum, there is no contribution for $N_{\pi_n}^*$ due to the absence of the Lorentz frame vector of the matter. From the PCAC relation, \cref{eq:PCAC}, the decay constants must be zero for pion excited states in the chiral limit; thus, the chiral symmetry forces that only Nambu-Goldstone modes contribute to the Glashow-Weinberg relation in the vacuum. More detailed arguments can be read off in Ref.~\cite{jido2008medium}.

As shall be shown, however, one of the charged pions becomes massive in isospin-asymmetric nuclear matter. In this case, the PCAC relation, \cref{eq:PCAC}, may not be valid in isospin-asymmetric nuclear matter. To inspect this, we will explore the in-medium corrections in isospin-asymmetric nuclear matter.

\section{Mass generation of a charged pion}\label{sec:mass}
Since the pion self-energies at the pion rest frame are functions of the square of pion energy, $q_0^2$, the pions are massless in the chiral limit in vacuum and isospin-symmetric nuclear matter. However, in-medium quantities generally depend on linear in-medium pion energy $q_0$. Therefore, it may occur a mass of an in-medium pion, which was already pointed out in Ref.~\cite{cohen1995pseudo} in dense nuclear matter. In this section, we examine the mass generation of a pion with explicit perturbative calculations. The Feynman diagrams will be considered up to the two Fermi-sea insertions. Note that we do not consider sub-diagrams which can be contracted into pure vacuum corrections, as all vacuum quantities are assumed to be renormalized already \cite{goda2014pion,goda2013chiral}.

The self-energies contain the following Feynman diagram contributions up to the next-to-leading order:
\begin{align}\label{eq:self-energy diagrams}
	-\im\Sigma_{\pi}(q) &=\resizebox{\dThree}{!}{$\diagram{
		\SelfEnergyFirstA;
	}$} + \resizebox{\dThree}{!}{$\diagram{
		\SelfEnergyFirstB;
	}$}\nonumber\\
    &+ \resizebox{\dTwo}{!}{$\diagram{
		\SelfEnergyTypeTwo;
		\vertex{B}{1};
	}$} + \resizebox{\dTwo}{!}{$\diagram{
		\SelfEnergyTypeTwo;
		\vertex{B}{2};
	}$}\nonumber\\
    &+ \resizebox{\dTwo}{!}{$\diagram{
		\SelfEnergyTypeThree;
		\vertex{B}{2};
		\vertex{E}{1};
		\vertex{F}{1};}
	$}
        + \resizebox{\dThree}{!}{$\diagram{
		\SelfEnergyTypeFour;
		\vertex{B}{1};
            \vertex{C}{1}
	}$}\nonumber\\
    &+\resizebox{\dThree}{!}{$\diagram{
		\SelfEnergyTypeFour;
		\vertex{B}{1};
            \vertex{C}{2}
	}$} +\resizebox{\dThree}{!}{$\diagram{
		\SelfEnergyTypeFour;
		\vertex{B}{2};
            \vertex{C}{1}
	}$}\nonumber\\
    &+\resizebox{\dThree}{!}{$\diagram{
		\SelfEnergyTypeFour;
		\vertex{B}{2};
            \vertex{C}{2}
	}$} + \resizebox{\dTwo}{!}{$\diagram{
		\SelfEnergyTypeFive;
		\vertex{B}{1};
		\vertex{D}{1};
	}$}\nonumber
\end{align}
\begin{align}
     &+\resizebox{\dThree}{!}{$\diagram{
		\SelfEnergyTypeadditonalfifth;
		\vertex{B}{1};
        \vertex{C}{1};
        \vertex{E}{1}
	}$}+\resizebox{\dThree}{!}{$\diagram{
		\SelfEnergyTypeadditonalsixth;
		\vertex{B}{1};
        \vertex{C}{1};
        \vertex{E}{1}
	}$}\nonumber\\
    &+\resizebox{\dThree}{!}{$\diagram{
		\SelfEnergyTypeadditonalfifth;
		\vertex{B}{2};
        \vertex{C}{1};
        \vertex{E}{1}
	}$}+\resizebox{\dThree}{!}{$\diagram{
		\SelfEnergyTypeadditonalsixth;
		\vertex{B}{2};
        \vertex{C}{1};
        \vertex{E}{1}
	}$}\nonumber\\
    &+\resizebox{\dThree}{!}{$\diagram{
		\SelfEnergyTypeadditonalnineth;
		\vertex{B}{1};
        \vertex{C}{1};
        \vertex{E}{1}
	}$}+\resizebox{\dThree}{!}{$\diagram{
		\SelfEnergyTypeadditonaltenth;
		\vertex{B}{1};
        \vertex{C}{1};
        \vertex{E}{1}
	}$}\nonumber\\
    &+\resizebox{\dThree}{!}{$\diagram{
		\SelfEnergyTypeadditonalnineth;
		\vertex{B}{1};
        \vertex{C}{2};
        \vertex{E}{1}
	}$}+\resizebox{\dThree}{!}{$\diagram{
		\SelfEnergyTypeadditonaltenth;
		\vertex{B}{1};
        \vertex{C}{2};
        \vertex{E}{1}
	}$}\nonumber
\end{align}
\begin{align}
    &\ +\resizebox{\dThree}{!}{$\diagram{
		\SelfEnergyTypeadditonalfirst;
		\vertex{B}{1};
        \vertex{C}{1};
        \vertex{E}{1}
	}$}+\resizebox{\dThree}{!}{$\diagram{
		\SelfEnergyTypeadditonalfirst;
		\vertex{B}{2};
        \vertex{C}{1};
        \vertex{E}{1}
	}$}\nonumber\\
    &+\resizebox{\dThree}{!}{$\diagram{
		\SelfEnergyTypeadditonalsecond;
		\vertex{B}{1};
        \vertex{C}{1};
        \vertex{E}{1}
	}$}+\resizebox{\dThree}{!}{$\diagram{
		\SelfEnergyTypeadditonalsecond;
		\vertex{B}{2};
        \vertex{C}{1};
        \vertex{E}{1}
	}$}\nonumber\\
    &+\resizebox{\dThree}{!}{$\diagram{
		\SelfEnergyTypeadditonalthird;
		\vertex{B}{1};
        \vertex{C}{1};
        \vertex{E}{1}
	}$}+\resizebox{\dThree}{!}{$\diagram{
		\SelfEnergyTypeadditonalthird;
		\vertex{B}{1};
        \vertex{C}{2};
        \vertex{E}{1}
	}$}\nonumber\\
    &+\resizebox{\dThree}{!}{$\diagram{
		\SelfEnergyTypeadditonalfourth;
		\vertex{B}{1};
        \vertex{C}{1};
        \vertex{E}{1}
	}$}+\resizebox{\dThree}{!}{$\diagram{
		\SelfEnergyTypeadditonalfourth;
		\vertex{B}{1};
        \vertex{C}{2};
        \vertex{E}{1}
	}$}
\end{align}
where the dashed lines denote the in-vacuum pion propagators, thin and thick lines respectively indicate the in-vacuum and in-medium nucleon propagators. The number indices in the above diagrams represent the chiral orders of the corresponding interaction at the vertices. Detailed expressions of the self-energies corresponding to the Feynman diagrams are shown in Appendix~\ref{app:explicit}. Note that additional diagrams have been included in \cref{eq:self-energy diagrams} than those in Ref \cite{kwon2025pion}, which contain two in-medium nucleon propagators with one more in-vacuum propagator. Since we aim for the mass solutions from the self-energies, the spatial momentum $\vec q$ dependence of the self-energies is ignored.

The in-medium pion masses are obtained by solving the mass equation \cref{eq:inmemasseq}. With a nonzero vacuum pion mass, the in-medium pion masses can be evaluated by inserting the vacuum pion mass in the self-energies below the $\mathcal{O}(\rho^2)$ order due to proper perturbation arguments \cite{goda2014pion,meissner2002medium} in the density expansion. However, this will yield only massless pions in the chiral limit, as the interactions of the Nambu-Goldstone bosons in the chiral perturbation theory vanish at the soft limit according to the Adler zero \cite{adler1965consistency}. In other words, since there is no constant term in the self-energies for $q_0$ in the chiral limit, $q_0=0$ provides $m_\pi^*=0$. 

\begin{figure*}%
\subfloat[]{{\includegraphics[width=0.4\textwidth ]{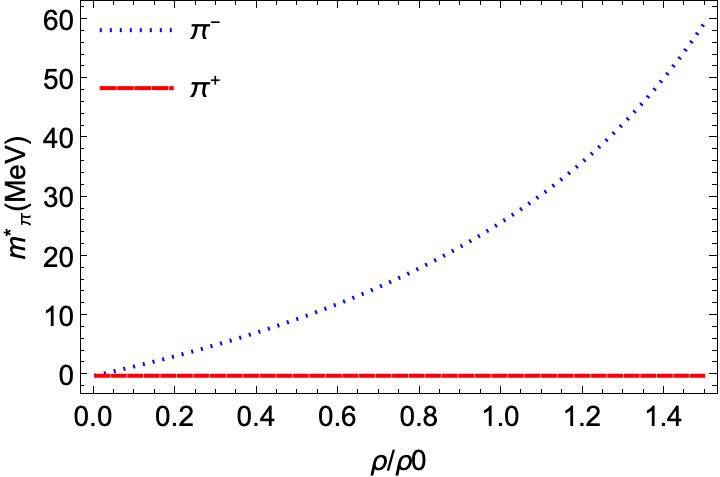} }}%
\subfloat[]{{\includegraphics[width=0.4\textwidth ]{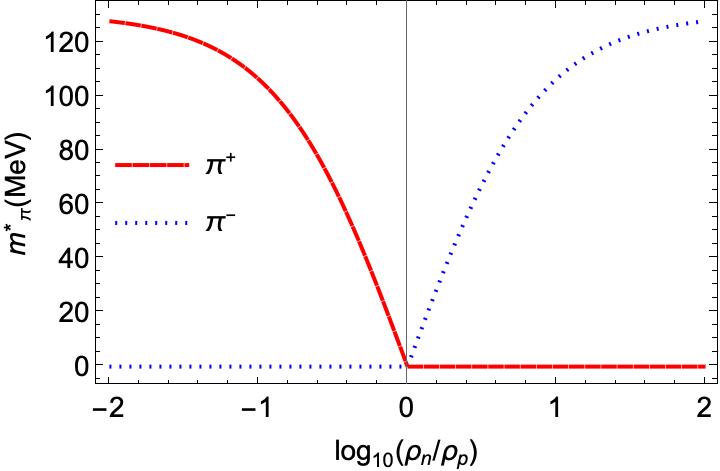} }}%
\caption{The behaviors of the in-medium masses of the charged pions in the chiral limit. (a) and (b) respectively show the density dependence of the in-medium masses at $\rho_n/\rho_p=1.5$ and the ratio dependence at $\rho=\rho_0$.}%
\label{fig:chmass}%
\end{figure*}
To show the qualitative picture of the mass generation of a pion, the mass equation is solved in the linear-density order for simplicity. Although the self-energies include only terms in linear density order, the nonzero solution of the self-consistent mass equation contains higher-order density terms. This process is similar to the iteration of a potential in the Lippmann-Schwinger equation in chiral unitary approaches \cite{kaiser1995chiral1,kaiser1995chiral2,meissner2000chiral,oset1998non}.
The behaviors of in-medium masses of the charged pions are shown in~\cref{fig:chmass} in terms of the normal nuclear density $\rho_0\approx0.17\ \text{fm}^{-3}$. Figure \ref{fig:chmass} (a) represents the density dependence at $\rho_n/\rho_p=1.5$. In neutron-rich matter, the negative pion becomes massive, but the positive pion remains massless.  Figure \ref{fig:chmass} (b) shows the ratio dependence at $\rho=\rho_0$ and indicates that the behaviors are flipped in proton-rich matter. Note that the neutral pion is always zero at all densities and ratios. The dominant contribution of the mass generation of a charged pion comes from the Weinberg-Tomozawa interaction,
\begin{align}
    \mathcal{L}_{WT}=-\frac{1}{4f^2}\bar N\gamma^\mu\pi^a\partial_\mu\pi^b\epsilon^{abc}\tau^cN.
\end{align}
This contribution must vanish due to the isovector nature of the nucleon bispinor in the symmetric nuclear matter, while for the asymmetric nuclear matter, this term contributes to the self-energies with $\epsilon^{ab3}$ and linear $q_0$;
\begin{align}\label{eq:selfarbi}
    \Sigma_{\pi^c}(q)=A(\rho,r)q_0+B(\rho,r)q_0^2+\cdots,
\end{align}
from which the mass equation in the chiral limit can be written as
\begin{align}
   A(\rho,r)q_0+B(\rho,r)q_0^{2}-q_0^2+\cdots=0, 
\end{align}
where the odd powers of $q_0$ vanish in vacuum and isospin-symmetric nuclear matter; thus, all pions are massless in these two cases. To show this explicitly, we truncate higher-order terms than the second order of $q_0$ for simplicity, and then the mass solutions are
\begin{align}
    m_\pi^*=0,\qquad m_\pi^*=\frac{A(\rho,r)}{1-B(\rho,r)},
\end{align}
where, if $A$ has no contribution, the in-medium mass of the pions is zero, but $A$ is not generally zero in isospin-asymmetric nuclear matter.
Thus, the odd powers cause a nonzero solution in the mass equation and provide asymmetry between the positive and negative mass solutions. In this way, the presence of the Weinberg-Tomozawa interaction makes the charged pions have different masses, with one of them being massless and the other becoming massive. Due to the isospin factor $\epsilon^{ab3}$, the Weinberg-Tomozawa interaction does not contribute to the neutral pion, and hence $\pi^0$ retains zero mass even in the asymmetric nuclear matter. The mass generation of the negative pion in neutron matter was analyzed in Ref.~\cite{cohen1995pseudo} using the isovector chemical potential.

\section{Nonzero corrections of the partially conserved axial-vector current relation in isospin-asymmetric nuclear matter}\label{sec:breaking}
We have seen a charged pion become massive in isospin-asymmetric nuclear matter according to the asymmetry $\rho_p/\rho_n\ne1$. Then, the matrix element of the divergence of an axial-vector current between the asymptotic state and one in-medium charged pion at the rest frame can be written as
\begin{align}\label{eq:cacpion}
    \langle\Omega|\partial^\mu A^c_\mu|\pi^{*c}(m_{\pi^c}^*,\vec 0)\rangle=m_{\pi^c}^*\sqrt{Z_{\pi^c}}\hat{\Gamma}_0^{\pi^c},
\end{align}
where $c$ indicates the component for the massive charged pion. Due to a nonzero mass, it seems that the axial-vector current is not proportional to the $SU(2)$ averaged quark mass with the massive in-medium pion state. In this section, the conservation of the axial-vector current is taken into account at one- and two-loop orders. The vertex correction can be obtained perturbatively as
\begin{align}
    \hat\Gamma^\pi_0=q_0f+\hat \Gamma_{0,\ \text{1-loop}}^\pi+\hat\Gamma_{0,\ \text{2-loop}}^\pi+\cdots,
\end{align}
which are given by the following Feynman diagram contributions:
\begin{align}\label{axial 1PI diagrams}
	\im \hat \Gamma^{\pi}_0 &=\resizebox{\dTwo}{!}{$\diagram{
		\DecayTree;
		\vertex{B}{2};
	}$} + \resizebox{\dTwo}{!}{$\diagram{
		\DecayTypeTwo;
		\vertex{B}{2};
	}$}\nonumber\\
    &+ \resizebox{\dThree}{!}{$\diagram{
		\DecayConstantFirstA;
	}$} + \resizebox{\dThree}{!}{$\diagram{
		\DecayConstantFirstB;
	}$}\nonumber\\
    &+ \resizebox{\dTwo}{!}{$\diagram{
		\DecayTypeThree;
		\vertex{B}{2};
		\vertex{E}{1};
		\vertex{F}{1};}
	$} + \resizebox{\dThree}{!}{$\diagram{
		\DecayTypeFour;
		\vertex{B}{1};
		\vertex{C}{1};
	}$}\nonumber\\
    &+ \resizebox{\dThree}{!}{$\diagram{
		\DecayTypeFour;
		\vertex{B}{1};
		\vertex{C}{2};
	}$} + \resizebox{\dThree}{!}{$\diagram{
		\DecayTypeFour;
		\vertex{B}{2};
		\vertex{C}{1};
	}$}\nonumber\\
    &+ \resizebox{\dThree}{!}{$\diagram{
		\DecayTypeFour;
		\vertex{B}{2};
		\vertex{C}{2};
	}$} + \resizebox{\dTwo}{!}{$\diagram{
		\DecayTypeFive;
		\vertex{B}{1};
		\vertex{D}{1};
	}$}\nonumber
\end{align}
\begin{align}
     &\ +\resizebox{\dThree}{!}{$\diagram{
		\Decayadditonalfifth;
		\vertex{B}{1};
        \vertex{C}{1};
        \vertex{E}{1}
	}$}+\resizebox{\dThree}{!}{$\diagram{
		\Decayadditonalsixth;
		\vertex{B}{1};
        \vertex{C}{1};
        \vertex{E}{1}
	}$}\nonumber\\
    &+\resizebox{\dThree}{!}{$\diagram{
		\Decayadditonalfifth;
		\vertex{B}{2};
        \vertex{C}{1};
        \vertex{E}{1}
	}$}+\resizebox{\dThree}{!}{$\diagram{
		\Decayadditonalsixth;
		\vertex{B}{2};
        \vertex{C}{1};
        \vertex{E}{1}
	}$}\nonumber\\
    &+\resizebox{\dThree}{!}{$\diagram{
		\Decayadditonalnineth;
		\vertex{B}{1};
        \vertex{C}{1};
        \vertex{E}{1}
	}$}+\resizebox{\dThree}{!}{$\diagram{
		\Decayadditonaltenth;
		\vertex{B}{1};
        \vertex{C}{1};
        \vertex{E}{1}
	}$}\nonumber\\
    &+\resizebox{\dThree}{!}{$\diagram{
		\Decayadditonalnineth;
		\vertex{B}{1};
        \vertex{C}{2};
        \vertex{E}{1}
	}$}+\resizebox{\dThree}{!}{$\diagram{
		\Decayadditonaltenth;
		\vertex{B}{1};
        \vertex{C}{2};
        \vertex{E}{1}
	}$}\nonumber
\end{align}
\begin{align}
    &\ +\resizebox{\dThree}{!}{$\diagram{
		\Decayadditonalfirst;
		\vertex{B}{1};
        \vertex{C}{1};
        \vertex{E}{1}
	}$}+\resizebox{\dThree}{!}{$\diagram{
		\Decayadditonalfirst;
		\vertex{B}{2};
        \vertex{C}{1};
        \vertex{E}{1}
	}$}\nonumber\\
    &+\resizebox{\dThree}{!}{$\diagram{
		\Decayadditonalsecond;
		\vertex{B}{1};
        \vertex{C}{1};
        \vertex{E}{1}
	}$}+\resizebox{\dThree}{!}{$\diagram{
		\Decayadditonalsecond;
		\vertex{B}{2};
        \vertex{C}{1};
        \vertex{E}{1}
	}$}\nonumber\\
    &+\resizebox{\dThree}{!}{$\diagram{
		\Decayadditonalthird;
		\vertex{B}{1};
        \vertex{C}{1};
        \vertex{E}{1}
	}$}+\resizebox{\dThree}{!}{$\diagram{
		\Decayadditonalthird;
		\vertex{B}{1};
        \vertex{C}{2};
        \vertex{E}{1}
	}$}\nonumber\\
    &+\resizebox{\dThree}{!}{$\diagram{
		\Decayadditonalfourth;
		\vertex{B}{1};
        \vertex{C}{1};
        \vertex{E}{1}
	}$}+\resizebox{\dThree}{!}{$\diagram{
		\Decayadditonalfourth;
		\vertex{B}{1};
        \vertex{C}{2};
        \vertex{E}{1}
	}$}
\end{align}
where the additional wavy line denotes the axial-vector currents. The explicit forms are given in Appendix~\ref{app:explicit}.
\subsection{One-loop calculation}
Let us consider the case in neutron-rich matter, where the negative pion becomes massive. To conserve the axial-vector current, the vertex correction may vanish at the on-shell pion rest frame. The mass equation in the chiral limit is given by \cref{eq:inmemasseq},
\begin{align}
    q_0^2-\Sigma_{\pi^-}^{1\text{-loop}}(q^0)=0.
\end{align}
Comparing the one-loop expressions of the self-energy and vertex correction in Appendix~\ref{app:explicit}, one can find the following relation,
\begin{align}
    q_0^2f+q_0\hat \Gamma_{0,\ \text{1-loop}}^{\pi^-}(q_0)=f\big(q_0^2-\Sigma_{\pi^-}^{\text{1-loop}}(q_0)\big).
\end{align}
Therefore, the nonzero mass solution makes the divergence of the axial vector current, \cref{eq:cacpion}, zero at the one-loop order. Then, all components of the axial-vector current are conserved in the chiral limit. In proton-rich matter, the vertex correction of the positive pion is zero at the non-zero mass solution.
\subsection{Two-loop calculation}
Now, let us take into account the two-loop calculations with $m_\pi=0$. Unlike the one-loop calculation, the vertex correction exhibits a different structure from the mass equation, so that
\begin{align}
    q_0^2f+q_0\hat \Gamma_{0,\ \text{1-loop}}^{\pi^c}(q_0)&+q_0\hat \Gamma_{0,\ \text{2-loop}}^{\pi^c}(q_0)\nonumber\\
    &\ne f\big(q_0^2-\Sigma_{\pi^c}^{\text{1-loop}}(q_0)-\Sigma_{\pi^c}^{\text{2-loop}}(q_0)\big).
\end{align}
The different structures are produced by the following diagrams,
\begin{align}
    -\im\Sigma_{\text{diff}}^{\pi^c}=&\resizebox{\dTwo}{!}{$\diagram{
		\SelfEnergyTypeThree;
		\vertex{B}{2};
		\vertex{E}{1};
		\vertex{F}{1};}
	$}+ \resizebox{\dTwo}{!}{$\diagram{
		\SelfEnergyTypeFive;
		\vertex{B}{1};
		\vertex{D}{1};
	}$},\\
    \im\hat\Gamma_{0,\ \text{diff}}^{\pi^c}=&\resizebox{\dTwo}{!}{$\diagram{
		\DecayTypeThree;
		\vertex{B}{2};
		\vertex{E}{1};
		\vertex{F}{1};}
	$}+ \resizebox{\dTwo}{!}{$\diagram{
		\DecayTypeFive;
		\vertex{B}{1};
		\vertex{D}{1};
	}$}.
\end{align}
These diagram contributions for the charged pions are explicitly expressed as
\begin{widetext}
\begin{align}
    &\Sigma^{\pi^+}_{\text{diff}}(q_0)
    = \frac{m_N^2g_A^2}{5(2\pi f)^4}\int \dd |\vec p|\,\dd |\vec k|\,\dd \cos \theta\frac{\vec p^2\vec k^2}{p_0k_0}\frac{[q_0^2(\Theta_{\boldsymbol{k}}^p\Theta_{\boldsymbol{p}}^p+\Theta_{\boldsymbol{k}}^n\Theta_{\boldsymbol{p}}^n-\Theta_{\boldsymbol{k}}^p\Theta_{\boldsymbol{p}}^n-\Theta_{\vec k}^n\Theta_{\vec p}^p)+5q^0(p_0- k_0)(\Theta_{\boldsymbol{k}}^n\Theta_{\boldsymbol{p}}^p-\Theta_{\vec k}^p\Theta_{\vec p}^n)]}{(2m_N^2-2p_0k_0+2|\vec p||\vec k|\cos\theta)},\\
    &\Sigma^{\pi^-}_{\text{diff}}(q_0)
    = \frac{m_N^2g_A^2}{5(2\pi f)^4}\int \dd |\vec p|\,\dd |\vec k|\,\dd \cos \theta\frac{\vec p^2\vec k^2}{p_0k_0}\frac{[q_0^2(\Theta_{\boldsymbol{k}}^p\Theta_{\boldsymbol{p}}^p+\Theta_{\boldsymbol{k}}^n\Theta_{\boldsymbol{p}}^n-\Theta_{\boldsymbol{k}}^p\Theta_{\boldsymbol{p}}^n-\Theta_{\vec k}^n\Theta_{\vec p}^p)-5q^0(p_0- k_0)(\Theta_{\boldsymbol{k}}^n\Theta_{\boldsymbol{p}}^p-\Theta_{\vec k}^p\Theta_{\vec p}^n)]}{(2m_N^2-2p_0k_0+2|\vec p||\vec k|\cos\theta)},\\
    &\hat{\Gamma}_{0,\ \mathrm{\text{diff}}}^{\pi^+}(q_0)
    =\frac{g_A^2m_N^2}{10(2\pi)^4f^3}\int \dd |\vec p|\,\dd |\vec k|\,\dd \cos \theta\frac{\vec p^2\vec k^2}{p_0k_0}\frac{\bigg(5(p_0-k_0)\big[(\Theta_{\boldsymbol{k}}^p\Theta^n_{\boldsymbol{p}}-\Theta_{\boldsymbol{k}}^n\Theta^p_{\boldsymbol{p}})\big]-q_0\big[4(\Theta^p_{\boldsymbol{p}}\Theta^n_{\boldsymbol{k}}+\Theta^n_{\boldsymbol{p}}\Theta^p_{\boldsymbol{k}})+6(\Theta^p_{\boldsymbol{p}}\Theta^p_{\boldsymbol{k}}+\Theta^n_{\boldsymbol{p}}\Theta^n_{\boldsymbol{k}})\big]\bigg)}{(2m_N^2-2p_0k_0+2|\vec p||\vec k|\cos{\theta})},\\
    &\hat{\Gamma}_{0,\ \mathrm{\text{diff}}}^{\pi^-}(q_0)
    =\frac{g_A^2m_N^2}{10(2\pi)^4f^3}\int \dd |\vec p|\,\dd |\vec k|\,\dd \cos \theta\frac{\vec p^2\vec k^2}{p_0k_0}\frac{\bigg(5(k_0-p_0)\big[(\Theta_{\boldsymbol{k}}^p\Theta^n_{\boldsymbol{p}}-\Theta_{\boldsymbol{k}}^n\Theta^p_{\boldsymbol{p}})\big]-q_0\big[4(\Theta^p_{\boldsymbol{p}}\Theta^n_{\boldsymbol{k}}+\Theta^n_{\boldsymbol{p}}\Theta^p_{\boldsymbol{k}})+6(\Theta^p_{\boldsymbol{p}}\Theta^p_{\boldsymbol{k}}+\Theta^n_{\boldsymbol{p}}\Theta^n_{\boldsymbol{k}})\big]\bigg)}{(2m_N^2-2p_0k_0+2|\vec p||\vec k|\cos{\theta})}.
\end{align}
\end{widetext}
Then, the matrix element \cref{eq:cacpion} for the massive negatively charged pion and its complex conjugate are nonzero,
\begin{align}
    \langle\Omega|\partial^\mu A^-_\mu|\pi^{*-}(m_{\pi^-}^*,\vec 0)\rangle\ne 0,\nonumber\\
    \langle\pi^{*-}(m_{\pi^-}^*,\vec 0)|\partial^\mu A^+_\mu|\Omega\rangle\ne 0,\nonumber
\end{align}
so that the divergence of the axial-vector currents for the charged pions receives nonzero corrections with the massive charged pion states in isospin-asymmetric nuclear matter, yielding only one Nambu-Goldstone boson, the neutral pion. This is consistent with the counting rule of the number of Nambu-Goldstone bosons without Lorentz symmetry in Ref. \cite{watanabe2012unified} as the number of independent spontaneously broken symmetries is one, and  the neutral pion has a linear dispersion in isospin-asymmetric nuclear matter. 

In deriving the in-medium Glashow-Weinberg relation, the PCAC relation was used, but we found that this relation receives nonzero in-medium corrections for the charged pions. Thus, the relation \cref{eq:inmeGWch} cannot be used to estimate the in-medium quark condensate. On the other hand, the neutral pion is still a Nambu-Goldstone boson and satisfies the PCAC relation.
According to the Adler zero \cite{adler1965consistency}, the matrix element of the axial-vector current between the asymptotic state and one neutral pion state must vanish in the soft limit,
\begin{align}
    \lim_{p_{\pi^0}\rightarrow0}\langle\Omega|A_\mu^0(x)|\pi^{*0}(p_{\pi^0})\rangle=0.
\end{align}
To satisfy this condition, the matrix element \cref{eq:decayinme} can be written as \cite{jido2008medium}
\begin{align}
    &\langle\Omega|A_\mu^{0}(x)|\pi^{*0}(p_{\pi^{0}})\rangle\nonumber\\
    &\quad=\im\big[n_\mu (n_\nu p_{\pi^0}^\nu) N_{\pi^{0}}^*(p_{\pi^{0}})+p_\mu^{\pi^{0}}F_{\pi^{0}}^*(p_{\pi^{0}})\big]\eu^{-\im p_{\pi^{0}}\cdot x}.
\end{align}
Then, the in-medium Glashow-Weinberg relation at the nucleon matter rest frame for the neutral pion can be written as
\begin{align}\label{eq:NGGWrelation}
    \mathrm{Re}\bigg[\sum\limits_{n}\bigg(\bar F_{\pi^0_n}^*\big(p_{\pi_n^0}\big)+\bar N_{\pi_n^0}^*\big(p_{\pi_n^0}\big)\bigg) G^{*}_{\pi^0_n}\big(p_{\pi_n^0}\big) \bigg]&=-\langle\bar qq\rangle^*.
\end{align}
Therefore, the quark condensate in isospin-asymmetric nuclear matter can be obtained from the neutral pion properties with \cref{eq:NGGWrelation}.

\subsection{Discussion}
Due to the mass generation of a charged pion in isospin-asymmetric nuclear matter, the absolute values of the on-shell in-medium pion energies of the charged pions are different. This states one cannot combine the positive energy and negative energy pole parts in correlation functions. For a clear argument, let us consider the two-point function of the pseudoscalar densities:
\begin{align}
    \Pi^{\pm\mp,00}(q)&=\frac{1}{2\omega_{\pi^{\pm,0}}(\vec q)}G_{\pi^{\pm,0}}^*\frac{\im}{q_0-\omega_{\pi^{\pm,0}}(\vec q)+\im\epsilon}\bar G^*_{\pi^{\pm,0}}\nonumber\\
    &-\frac{1}{2\omega_{\pi^{\mp,0}}(\vec q)}G_{\pi^{\mp,0}}^*\frac{\im}{q_0+\omega_{\pi^{\mp,0}}(\vec q)-\im\epsilon}\bar G^*_{\pi^{\mp,0}}+\cdots.
\end{align}
For the neutral pion, the positive and negative energy solutions exhibit the same absolute value at a fixed spatial momentum; thus, the correlation function can be expressed as
\begin{align}
    \Pi^{00}(q)&=G_{\pi^{0}}^*\frac{\im}{q^2_0-\omega^2_{\pi^{0}}(\vec q)+\im\epsilon}\bar G^*_{\pi^{0}}+\cdots,
\end{align}
so that near the physical pole position, the form factor can be nonzero even in the soft pion limit. On the other hand, since the absolute values of the physical pole positions of the charged pions at a given spatial momentum are different, the two expressions cannot be combined. Therefore, even if the residue near the pole position is finite, due to the on-shell energy in the denominator, the form factor of the charged pion is zero at the rest frame, $\vec q=\vec 0$, in the standard normalization convention \cref{eq:normalization}.

This can be shown in explicit calculations through the wave function renormalization. From the expression of the self-energy \cref{eq:selfarbi}, the wave function renormalization for a charged pion can be expressed as
\begin{align}
    Z_{\pi^c}=\bigg(1-\frac{A(\rho,r)}{2\omega_{\pi^c}(\vec q)}-B(\rho,r)+\cdots\bigg)^{-1}.
\end{align}
Then, near $\vec q=\vec0$, the wave function renormalization goes to zero, so that general form factors are zero at the rest frame unless the corresponding vertex correction exhibits a singularity. The self-energy of the neutral pion does not have the term proportional to linear $q^0$; its wave function renormalization is not zero in the soft limit.

One additional remark is that the in-medium correction terms of the PCAC relation for the massive charged pion can be estimated from the following expression:
\begin{align}
    \langle\Omega|\partial^\mu A^c_\mu|\pi^{*c}(m_{\pi^c}^*,\vec 0)\rangle=m_{\pi^c}^*N_{\pi^{c}}^*(m^*_{\pi^{c}})+m^{*2}_{\pi^{c}}F_{\pi^{c}}^*(m^*_{\pi^{c}}),
\end{align}
where 
\begin{align}
    m_{\pi^c}^*\approx\mathcal{O}(\rho),\qquad N_{\pi^c}^*\approx\mathcal{O}(\rho),\qquad F_{\pi^c}^*\approx\mathcal{O}(\rho^0).
\end{align}
From this expression, one can confirm that the in-medium correction terms begin with $\mathcal O(\rho^2)$. Thus, at lower orders than $\mathcal O(\rho^2)$, the chiral symmetry can also be estimated using the charged pions with \cref{eq:inmeGWch}.

\section{Summary}\label{sec:summary}
The mass generation of a charged pion in isospin-asymmetric nuclear matter has been investigated in the chiral limit with explicit calculations in the in-medium chiral perturbation theory. Even though the vacuum pion mass is zero, one of the charged pions becomes massive from asymmetric contributions such as the Weinberg-Tomozawa term; the positive pion becomes massive in proton-rich matter, and the negative pion becomes massive in neutron-rich matter. As a result, the remaining massless modes are only two pions at finite $\rho_n/\rho_p\ne1$.

Based on the mass generation of a charged pion, the conservation of the axial-vector currents in isospin-asymmetric nuclear matter has been explored.
At the one-loop calculation, the axial-vector form factor $\hat \Gamma_0^\pi$ for the massive charged pion vanishes, so that all axial-vector currents are conserved. On the other hand, the vertex correction does not vanish for all pions from the two-loop order, and thus, the PCAC relation receives finite in-medium corrections.

The in-medium Glashow-Weinberg relation is derived from the operator relations, such as the PCAC relation. As the axial-vector currents are not conserved in the chiral limit, however, the PCAC relation is not satisfied for the charged pions with the massive charged pion states. Therefore, the partial restoration of the chiral symmetry in isospin-asymmetric nuclear matter must be estimated from the neutral pion.

\section*{Acknowledgements}
The work of D.\,J.~was partly supported by Grants-in-Aid for Scientific Research from JSPS (17K05449, 21K03530, 25K07315).

\appendix
\begin{widetext}
\section{Derivation of the expressions for the correlation functions}\label{app:decorr}
In this appendix, the expressions for the correlation functions in terms of the in-medium quantities are derived. We specify the two-point function of the axial vector currents,
\begin{align}
    \Pi_{\mu\nu}^{\pm\mp,00}(q)=\int\dd[4]x\eu^{\im q\cdot x}\langle\Omega|\mathsf{T}\{A_\mu^{\pm,0}(x)A_\nu^{\mp,0}(0)\}|\Omega\rangle,
\end{align}
and for the other correlation functions, the derivation processes are the same. The two-point function of the axial vector currents can be written using the complete set of the in-medium pion states in the coordinate space as
\begin{align}
    \langle\Omega|\mathsf{T}\{A_\mu^{\pm,0}(x)A_\nu^{\mp,0}(0)\}|\Omega\rangle&=\sum_{n}\int\frac{\dd[3]\vec p_{\pi_{n}^{\pm,0}}}{(2\pi)^32\omega_{\pi_n^{\pm,0}}}\big[\Theta(x_0)\langle\Omega|A_\mu^{\pm,0}(x)|\pi_n^{*\pm,0}(p_{\pi_n^{\pm,0}})\rangle\langle \pi_n^{*\pm,0}(p_{\pi_n^{\pm,0}})|A_\nu^{\mp,0}(0)|\Omega\rangle\big]\nonumber\\
    &\qquad+\sum_m\int\frac{\dd[3]\vec p_{\pi_m^{\mp,0}}}{(2\pi)^32\omega_m}\big[\Theta(-x_0)\langle\Omega|A_\nu^{\mp,0}(0)\}|\pi_m^{*\mp,0}(p_{\pi_m^{\mp,0}})\rangle\langle \pi_m^{*\mp,0}(p_{\pi_m^{\mp,0}})|A_\mu^{\pm,0}(x)|\Omega\rangle\big],
\end{align}
where all pion excited states are included in the complete sets. Using the definition of the matrix element from \cref{eq:decayinme}, this can be written as
\begin{align}
     &\langle\Omega|\mathsf{T}\{A_\mu^{\pm,0}(x)A_\nu^{\mp,0}(0)|\Omega\rangle\nonumber\\
     &=\sum_{n}\int\frac{\dd[3]\vec p_{\pi_{n}^{\pm,0}}}{(2\pi)^3}\frac{\big[n_\mu N_{\pi_n^{\pm,0}}^*(p_{\pi_n^{\pm,0}})+p_\mu^{\pi_n^{\pm,0}}F_{\pi_n^{\pm,0}}^*(p_{\pi_n^{\pm,0}})\big]\big[n_\nu \bar N_{\pi_n^{\pm,0}}^*(p_{\pi_n^{\pm,0}})+p_\nu^{\pi_n^{\pm,0}}\bar F_{\pi_n^{\pm,0}}^*(p_{\pi_n^{\pm,0}})\big]\eu^{-\im p_{\pi_n^{\pm,0}}\cdot x}}{2\omega_{\pi_n^{\pm,0}}}\Theta(x_0)\nonumber\\
     &\qquad+\sum_{m}\int\frac{\dd[3]\vec p_{\pi_{m}^{\pm,0}}}{(2\pi)^3}\frac{\big[n_\mu \bar N_{\pi_m^{\mp,0}}^*(p_{\pi_m^{\mp,0}})+p_\mu^{\pi_m^{\mp,0}}\bar F_{\pi_m^{\mp,0}}^*(p_{\pi_m^{\mp,0}})\big]\big[n_\nu N_{\pi_m^{\mp,0}}^*(p_{\pi_m^{\mp,0}})+p_\nu^{\pi_m^{\mp,0}} F_{\pi_m^{\mp,0}}^*(p_{\pi_m^{\mp,0}})\big]\eu^{\im p_{\pi_m^{\mp,0}}\cdot x}}{2\omega_{\pi_m^{\mp,0}}}\Theta(-x_0)
\end{align}
Then, the mathematical identity \cite{weinberg1995quantum}
\begin{align}
    \Theta(\pm t)\eu^{\mp\im\omega t}=\mp\frac{\im}{2\pi}\int^{+\infty}_{-\infty}\dd s\frac{\eu^{\im st}}{s\pm\omega\mp\im\epsilon},
\end{align}
enables us to write the two-point function in the Fourier space as
\begin{align}
    &\Pi_{\mu\nu}^{\pm\mp,00}(q)\nonumber\\
    &=-\im\sum_{n}\int\dd[4]x\eu^{\im q\cdot x}\frac{\dd s\dd[3]\vec p_{\pi_{n}^{\pm,0}}}{(2\pi)^4}\frac{\big[n_\mu N_{\pi_n^{\pm,0}}^*(p_{\pi_n^{\pm,0}})+p_\mu^{\pi_n^{\pm,0}}F_{\pi_n^{\pm,0}}^*(p_{\pi_n^{\pm,0}})\big]\big[n_\nu \bar N_{\pi_n^{\pm,0}}^*(p_{\pi_n^{\pm,0}})+p_\nu^{\pi_n^{\pm,0}}\bar F_{\pi_n^{\pm,0}}^*(p_{\pi_n^{\pm,0}})\big]\eu^{\im st+\im \vec p_{\pi_n^{\pm,0}}\cdot \vec x}}{2\omega_{\pi_n^{\pm,0}}(s+\omega_{\pi_n^{\pm,0}}-\im\epsilon)}\nonumber\\
     &\qquad+\im\sum_{m}\int\dd[4]x\eu^{\im q\cdot x}\frac{\dd s\dd[3]\vec p_{\pi_{m}^{\pm,0}}}{(2\pi)^4}\frac{\big[n_\mu \bar N_{\pi_m^{\mp,0}}^*(p_{\pi_m^{\mp,0}})+p_\mu^{\pi_m^{\mp,0}}\bar F_{\pi_m^{\mp,0}}^*(p_{\pi_m^{\mp,0}})\big]\big[n_\nu N_{\pi_m^{\mp,0}}^*(p_{\pi_m^{\mp,0}})+p_\nu^{\pi_m^{\mp,0}} F_{\pi_m^{\mp,0}}^*(p_{\pi_m^{\mp,0}})\big]\eu^{\im st-\im \vec p_{\pi_m^{\mp,0}}\cdot \vec x}}{2\omega_{\pi_m^{\mp,0}}(s-\omega_{\pi_m^{\mp,0}}+\im\epsilon)}.
\end{align}
The integration over the four-dimensional coordinate space yields
the Dirac-delta functions in this expression, and therefore, the final expression is
\begin{align}
     &\Pi_{\mu\nu}^{\pm\mp,00}(q)\nonumber\\
    &=-\im\sum_{n}\int\dd s\dd[3]\vec p_{\pi_{n}^{\pm,0}}\frac{\big[n_\mu N_{\pi_n^{\pm,0}}^*(p_{\pi_n^{\pm,0}})+p_\mu^{\pi_n^{\pm,0}}F_{\pi_n^{\pm,0}}^*(p_{\pi_n^{\pm,0}})\big]\big[n_\nu \bar N_{\pi_n^{\pm,0}}^*(p_{\pi_n^{\pm,0}})+p_\nu^{\pi_n^{\pm,0}}\bar F_{\pi_n^{\pm,0}}^*(p_{\pi_n^{\pm,0}})\big]\delta^{(3)}(\vec p_{\pi_n^{\pm,0}}-\vec q)\delta(s+q^0)}{2\omega_{\pi_n^{\pm,0}}(s+\omega_{\pi_n^{\pm,0}}-\im\epsilon)}\nonumber\\
     &\qquad+\im\sum_{m}\int\dd s\dd[3]\vec p_{\pi_{m}^{\pm,0}}\frac{\big[n_\mu \bar N_{\pi_m^{\mp,0}}^*(p_{\pi_m^{\mp,0}})+p_\mu^{\pi_m^{\mp,0}}\bar F_{\pi_m^{\mp,0}}^*(p_{\pi_m^{\mp,0}})\big]\big[n_\nu N_{\pi_m^{\mp,0}}^*(p_{\pi_m^{\mp,0}})+p_\nu^{\pi_m^{\mp,0}} F_{\pi_m^{\mp,0}}^*(p_{\pi_m^{\mp,0}})\big]\delta^{(3)}(\vec p_{\pi_m^{\mp,0}}+\vec q)\delta(s+q^0)}{2\omega_{\pi_m^{\mp,0}}(s-\omega_{\pi_m^{\mp,0}}+\im\epsilon)}\nonumber\\
     &=\im\sum_{n}\frac{\big[n_\mu N_{\pi_n^{\pm,0}}^*(p_{\pi_n^{\pm,0}})+p_\mu^{\pi_n^{\pm,0}}(\vec q)F_{\pi_n^{\pm,0}}^*(p_{\pi_n^{\pm,0}})\big]\big[n_\nu \bar N_{\pi_n^{\pm,0}}^*(p_{\pi_n^{\pm,0}})+p_\nu^{\pi_n^{\pm,0}}(\vec q)\bar F_{\pi_n^{\pm,0}}^*(p_{\pi_n^{\pm,0}})\big]}{2\omega_{\pi_n^{\pm,0}}(\vec q)(q^0-\omega_{\pi_n^{\pm,0}}(\vec q)+\im\epsilon)}\nonumber\\
     &\qquad-\im\sum_{m}\frac{\big[n_\mu \bar N_{\pi_m^{\mp,0}}^*(p_{\pi_m^{\mp,0}})+p_\mu^{\pi_m^{\mp,0}}(-\vec q)\bar F_{\pi_m^{\mp,0}}^*(p_{\pi_m^{\mp,0}})\big]\big[n_\nu N_{\pi_m^{\mp,0}}^*(p_{\pi_m^{\mp,0}})+p_\nu^{\pi_m^{\mp,0}}(-\vec q) F_{\pi_m^{\mp,0}}^*(p_{\pi_m^{\mp,0}})\big]}{2\omega_{\pi_m^{\mp,0}}(\vec q)(q^0+\omega_{\pi_m^{\mp,0}}(\vec q)-\im\epsilon)}.
\end{align}
In the same ways, one can derive the expressions for other correlation functions:
\begin{align}
    &\Pi^{\pm\mp,00}(q)=\im\sum_{n}\frac{G_{\pi_
    {n}^{\pm,0}}^*(p_{\pi_n^{\pm,0}})\bar G_{\pi_
    {n}^{\pm,0}}^*(p_{\pi_n^{\pm,0}})}{2\omega_{\pi_n^{\pm,0}}(\vec q)(q^0-\omega_{\pi_n^{\pm,0}}(\vec q)+\im\epsilon)}-\im\sum_{m}\frac{G_{\pi_
    {m}^{\mp,0}}^*(p_{\pi_m^{\mp,0}})\bar G_{\pi_
    {m}^{\mp,0}}^*(p_{\pi_m^{\mp,0}})}{2\omega_{\pi_m^{\mp,0}}(\vec q)(q^0+\omega_{\pi_m^{\mp,0}}(\vec q)-\im\epsilon)},\\
    &\Pi_{5\mu}^{\pm\mp,00}(q)=-\sum_{n}\frac{\big[n_\mu N_{\pi_n^{\pm,0}}^*(p_{\pi_n^{\pm,0}})+p_\mu^{\pi_n^{\pm,0}}(\vec q)F_{\pi_n^{\pm,0}}^*(p_{\pi_n^{\pm,0}})\big]\bar G_{\pi_
    {n}^{\pm,0}}^*(p_{\pi_n^{\pm,0}})}{2\omega_{\pi_n^{\pm,0}}(\vec q)(q^0-\omega_{\pi_n^{\pm,0}}(\vec q)+\im\epsilon)}\nonumber\\
    &\qquad\qquad\qquad\qquad-\sum_{m}\frac{\big[n_\mu \bar N_{\pi_m^{\mp,0}}^*(p_{\pi_m^{\mp,0}})+p_\mu^{\pi_m^{\mp,0}}(-\vec q)\bar F_{\pi_m^{\mp,0}}^*(p_{\pi_m^{\mp,0}})\big]G_{\pi_
    {m}^{\mp,0}}^*(p_{\pi_m^{\mp,0}})}{2\omega_{\pi_m^{\mp,0}}(\vec q)(q^0+\omega_{\pi_m^{\mp,0}}(\vec q)-\im\epsilon)}.
\end{align}
Note that as the form factors and on-shell energies of the in-medium pions are symmetric under $\vec p\leftrightarrow-\vec p$ in isotropic nuclear matter, the sign of the spatial momentum $\vec q$ is not important for them.
\end{widetext}
\section{Interaction terms in the chiral Lagrangian}\label{app:interaction}
In this appendix, the relevant nonzero interaction terms are listed in explicit forms. The interaction terms contain relevant external currents as well as pion and nucleon fields. First, the interaction terms in the pionic Lagrangian are written by
\begin{subequations}
\begin{align}
    \mathcal{L}_{\pi^4}^{(2)}&=\frac{1}{10f^2}\partial_\mu\pi^i\partial^\mu\pi^j\pi^k\pi^l(3\delta^{ik}\delta^{jl}-\delta^{ij}\delta^{kl})\nonumber\\
    &-\frac{m_qB_0}{20f^2}\pi^i\pi^i\pi^j\pi^j,\\
    \mathcal{L}_{\pi a}^{(2)}&=-f\partial^\mu\pi^ia_\mu^j\delta^{ij},\\
    \mathcal{L}_{\pi^3a}^{(2)}&=\frac{1}{5f}a_\mu^i\partial^\mu\pi^j\pi^k\pi^l(3\delta^{ij}\delta^{kl}-4\delta^{ik}\delta^{jl}).
\end{align}
\end{subequations}
The pion-nucleon interactions appear as $A^{(j)}$ in \cref{eq:nuclLag,eq:interaction}. At the lowest chiral order, the interactions are given by
\begin{subequations}
\begin{align}
    A_\pi^{(1)}&=\frac{g_A}{2f}\gamma^\mu\gamma^5\partial_\mu\pi^i\tau^i,\\
    A_{\pi\pi}^{(1)}&=\frac{1}{4f^2}\gamma^\mu\pi^i\partial_\mu\pi^j\epsilon^{ijk}\tau^k,\\
    A_{\pi^3}^{(1)}&=\frac{g_A}{20f^3}\gamma^\mu\gamma^5(3\pi^i\pi^j\partial_\mu\pi^j-\pi^j\pi^j\partial_\mu\pi^i)\tau^i,\\
    A_a^{(1)}&=-g_A\gamma^\mu\gamma^5a_\mu^i\frac{\tau^i}{2},\\
    A_{\pi a}^{(1)}&=-\frac{1}{2f}\gamma^\mu\pi^i a_\mu^j\epsilon^{ijk}\tau^k,\\
    A_{\pi^2a}^{(1)}&=\frac{g_A}{4f^2}\gamma^\mu\gamma^5\pi^a\pi^ba_\mu^c\tau^d(\delta^{ab}\delta^{cd}-\delta^{ac}\delta^{bd}).
\end{align}
\end{subequations}
At the second chiral order, the interactions are
\begin{subequations}
\begin{align}
    A_{\pi\pi}^{(2)}&=\frac{4B_0c_1m_q}{f^2}\pi^i\pi^i+\frac{c_2}{f^2m_N^2}\partial_\mu\pi^i\partial_\nu\pi^i\partial^\mu\partial^\nu\nonumber\\
    &-\frac{c_3}{f^2}\partial_\mu\pi^i\partial^\mu\pi^i+\cdots,\\
    A_{\pi a}^{(2)}&=-\frac{2c_2}{fm_N^2}\partial_\mu\pi^ia_\nu^i\partial^\mu\partial^\nu+\frac{2c_3}{f}\partial^\mu\pi^ia_\mu^i+\cdots.
\end{align}
Note that the interaction terms proportional to $c_4,\ c_5,\ c_6$, and $c_7$ are omitted, and light quark masses are set to zero in the chiral limit.
\end{subequations}
\begin{widetext}
\section{Explicit forms of the in-medium corrections}\label{app:explicit}
The self-energies and vertex corrections are respectively given by the Feynman diagram contributions in \cref{eq:self-energy diagrams,axial 1PI diagrams}. In this appendix, the explicit expressions of the in-medium corrections with the vacuum pion mass $m_\pi$ are provided. At the one-loop order, the self-energies are given by
\begin{subequations}\label{eq:exselfone}
\begin{align}
    \Sigma^{\mathrm{1\text{L}}}_{\pi^+}(q_0)
    &= \int \dd |\vec p| \frac{\vec p^2}{4\pi^2p_0f^2} \left[ \frac{q_0^2g_A^2p_0}{q_0+2p_0}\big[(\Theta^p_{\boldsymbol{p}}+\Theta^n_{\boldsymbol{p}})-(\Theta^p_{\boldsymbol{p}}-\Theta^n_{\boldsymbol{p}})\big]-\frac{q_0^2g_A^2p_0}{q_0-2p_0}\big[(\Theta^p_{\boldsymbol{p}}+\Theta^n_{\boldsymbol{p}})+(\Theta^p_{\boldsymbol{p}}-\Theta^n_{\boldsymbol{p}})\big]\right. \nonumber\\
    &\qquad +\left.  \frac{2q_0g_A^2\vec p^2}{q_0+2p_0}\big[(\Theta^p_{\boldsymbol{p}}+\Theta^n_{\boldsymbol{p}})-(\Theta^p_{\boldsymbol{p}}-\Theta^n_{\boldsymbol{p}})\big]+\frac{2q_0g_A^2\vec p^2}{q_0-2p_0}\big[(\Theta^p_{\boldsymbol{p}}+\Theta^n_{\boldsymbol{p}})+(\Theta^p_{\boldsymbol{p}}-\Theta^n_{\boldsymbol{p}})\big]\right.\nonumber\\
    &\qquad \left. -4m_N\left(8B_0m_qc_1+\frac{2c_2}{m_N^2}q_0^2p_0^2+2q_0^2c_3\right)(\Theta^p_{\boldsymbol{p}}+\Theta^n_{\boldsymbol{p}})+2q_0p_0(\Theta^p_{\boldsymbol{p}}-\Theta^n_{\boldsymbol{p}})\right] \nonumber\\
    &\approx\frac{q_0^2g_A^2m_N\rho}{f^2(4m_N^2-q_0^2)}+\frac{q_0^3g_A^2(1-r)\rho}{2f^2(1+r)(4m_N^2-q_0^2)}+\frac{8B_0m_qc_1-2q_0^2(c_2+c_3)\rho}{f^2}+\frac{q_0(1-r)\rho}{2f^2(1+r)},
\end{align}
\begin{align}
    \Sigma^{\mathrm{1\text{L}}}_{\pi^-}(q_0)
    &=\int \dd |\vec p| \frac{\vec p^2}{4\pi^2p_0f^2} \left[ \frac{q_0^2g_A^2p_0}{q_0+2p_0}\big[(\Theta^p_{\boldsymbol{p}}+\Theta^n_{\boldsymbol{p}})+(\Theta^p_{\boldsymbol{p}}-\Theta^n_{\boldsymbol{p}})\big]-\frac{q_0^2g_A^2p_0}{q_0-2p_0}\big[(\Theta^p_{\boldsymbol{p}}+\Theta^n_{\boldsymbol{p}})-(\Theta^p_{\boldsymbol{p}}-\Theta^n_{\boldsymbol{p}})\big]\right. \nonumber\\
    &\qquad +\left.  \frac{2q_0g_A^2\vec p^2}{q_0+2p_0}\big[(\Theta^p_{\boldsymbol{p}}+\Theta^n_{\boldsymbol{p}})+(\Theta^p_{\boldsymbol{p}}-\Theta^n_{\boldsymbol{p}})\big]+\frac{2q_0g_A^2\vec p^2}{q_0-2p_0}\big[(\Theta^p_{\boldsymbol{p}}+\Theta^n_{\boldsymbol{p}})-(\Theta^p_{\boldsymbol{p}}-\Theta^n_{\boldsymbol{p}})\big]\right.\nonumber\\
    &\qquad \left. -4m_N\left(8B_0m_qc_1+\frac{2c_2}{m_N^2}q_0^2p_0^2+2q_0^2c_3\right)(\Theta^p_{\boldsymbol{p}}+\Theta^n_{\boldsymbol{p}})-2q_0p_0(\Theta^p_{\boldsymbol{p}}-\Theta^n_{\boldsymbol{p}})\right]\nonumber\\
    &\approx\frac{q_0^2g_A^2m_N\rho}{f^2(4m_N^2-q_0^2)}-\frac{q_0^3g_A^2(1-r)\rho}{2f^2(1+r)(4m_N^2-q_0^2)}+\frac{8B_0m_qc_1-2q_0^2(c_2+c_3)\rho}{f^2}-\frac{q_0(1-r)\rho}{2f^2(1+r)},
\end{align}
\begin{align}
    \Sigma^{\mathrm{1\text{L}}}_{\pi^0}(q_0)
    &=\int \dd |\vec p|\frac{q_0^2g_A^2\vec p^2}{4\pi^2f^2p_0}\frac{4p_0^2-4\vec p^2}{4p_0^2-q_0^2}(\Theta^p_{\boldsymbol{p}}+\Theta^n_{\boldsymbol{p}})-\int \dd |\vec p|\frac{2m_N\vec p^2}{\pi^2p_0f^2}\left(4B_0m_qc_1+c_2\frac{q_0^2p_0^2}{m_N^2}+q_0^2c_3\right)(\Theta^p_{\boldsymbol{p}}+\Theta^n_{\boldsymbol{p}})\nonumber\\
    &\approx \frac{q_0^2g_A^2m_N\rho}{f^2(4m_N^2-q_0^2)}+\frac{8B_0m_qc_1-2q_0^2(c_2+c_3)\rho}{f^2}.
\end{align}
\end{subequations}
In the one-loop calculations, we also indicate a linear-density approximation on the right-hand side of $\approx$ for each pion. The two-loop contributions are written as
\begin{subequations}\label{eq:exselftwo}
\begin{align}
    \Sigma^{\mathrm{2\text{L}}}_{\pi^+}(q_0)
    &=\frac{m_N^2g_A^2}{5(2\pi f)^4}\int \dd |\vec p|\,\dd |\vec k|\,\dd \cos \theta\frac{\vec p^2\vec k^2}{p_0k_0}\frac{p_0k_0-|\vec p||\vec k|\cos\theta-m_N^2}{((p-k)^2-m_\pi^2)^2}\nonumber\\
    &\times[(20q_0p_0-20q_0k_0-4p_0k_0+4|\vec p||\vec k|\cos\theta+4m_N^2+2q_0^2-4m_\pi^2)\Theta_{\boldsymbol{k}}^p\Theta_{\boldsymbol{p}}^n\nonumber\\
    &+(-20q_0p_0+20q_0k_0-4p_0k_0+4|\vec p||\vec k|\cos\theta+4m_N^2+2q_0^2-4m_\pi^2)\Theta_{\boldsymbol{k}}^n\Theta_{\boldsymbol{p}}^p\nonumber\\
    &+(4p_0k_0-4|\vec p||\vec k|\cos\theta-4m_N^2-2q_0^2-m_\pi^2)(\Theta_{\boldsymbol{k}}^p\Theta_{\boldsymbol{p}}^p+\Theta_{\boldsymbol{k}}^n\Theta_{\boldsymbol{p}}^n)]+\frac{m_N^2g_A^2}{5(2\pi f)^4}\int \dd |\vec p|\,\dd |\vec k|\,\dd \cos \theta\frac{\vec p^2\vec k^2}{p_0k_0}\nonumber\\
    &\times\frac{1}{(p-k)^2-m_\pi^2}[(5q_0k_0-5q_0p_0+2m_N^2-2p_0k_0+2|\vec p||\vec k|\cos\theta)\Theta^p_{\boldsymbol{p}}\Theta_{\boldsymbol{k}}^n\nonumber\\
    &+(-5q_0k_0+5q_0p_0+2m_N^2-2p_0k_0+2|\vec p||\vec k|\cos\theta)\Theta^p_{\boldsymbol{k}}\Theta_{\boldsymbol{p}}^n+(2p_0k_0-2|\vec p||\vec k|\cos\theta-2m_N^2)(\Theta^p_{\boldsymbol{p}}\Theta^p_{\boldsymbol{k}}+\Theta^n_{\boldsymbol{p}}\Theta^n_{\boldsymbol{k}})]\nonumber\\
    &-\frac{q_0^2}{2(2\pi)^4f^4}\int \dd |\vec p|\,\dd |\vec k|\,\dd \cos \theta\frac{\vec p^2\vec k^2}{p_0k_0}\frac{p^0 k^0+|\vec p||\vec k|\cos\theta+m_N^2}{q_0^2+2q_0(p^0-k^0)+2m_N^2-2p_0k_0+2|\vec p||\vec k|\cos\theta-m_\pi^2}\nonumber\\
    &\times\big[(\Theta^p_{\boldsymbol{p}}\Theta^p_{\boldsymbol{k}}+\Theta^n_{\boldsymbol{k}}\Theta^n_{\boldsymbol{p}})+(\Theta^p_{\boldsymbol{p}}\Theta^n_{\boldsymbol{k}}+\Theta^n_{\boldsymbol{p}}\Theta^p_{\boldsymbol{k}})-(\Theta^p_{\boldsymbol{p}}\Theta^n_{\boldsymbol{k}}-\Theta^n_{\boldsymbol{p}}\Theta^p_{\boldsymbol{k}})\big]\nonumber\\
    &-\frac{2q_0^2}{(2\pi)^4}\int \dd |\vec p|\,\dd |\vec k|\,\dd \cos \theta\frac{\vec p^2\vec k^2}{p_0k_0}\frac{C_s^1C_s^2(p_0k_0-|\vec p||\vec k|\cos\theta+m_N^2)}{q_0^2+2q_0(p^0-k^0)+2m_N^2-2p_0k_0+2|\vec p||\vec k|\cos\theta-m_\pi^2}(\Theta^p_{\boldsymbol{p}}\Theta^p_{\boldsymbol{k}}+\Theta^n_{\boldsymbol{k}}\Theta^n_{\boldsymbol{p}})\nonumber\\
    &-\frac{m_Nq^2_0}{(2\pi)^4f^2}\int \dd |\vec p|\,\dd |\vec k|\,\dd \cos \theta\frac{\vec p^2\vec k^2}{p_0k_0}\frac{(p_0+k_0)}{q_0^2+2q_0(p^0-k^0)+2m_N^2-2p_0k_0+2|\vec p||\vec k|\cos\theta-m_\pi^2}(\Theta^p_{\boldsymbol{p}}\Theta^p_{\boldsymbol{k}}-\Theta^n_{\boldsymbol{k}}\Theta^n_{\boldsymbol{p}})(C_s^1+C_s^2)\nonumber
\end{align}
\begin{align}
    \qquad\qquad&+\frac{q_0 g_A^2}{(2\pi)^4f^4}\int\frac{\dd |\vec p|\dd|\vec k|\dd\cos\theta|\vec p|^2|\vec k|^2}{p_0k_0}\frac{\Theta^n_{\vec k}\Theta^p_{\vec p}+\Theta^n_{\vec k}\Theta^n_{\vec p}}{q_0^2-2q_0(p^0-k^0)+2m_N^2-2p_0k_0+2|\vec p||\vec k|\cos\theta-m_\pi^2}\nonumber\\
    &\times\frac{2k^0p^0q_0^2+q_0^2m_N^2-(q_0^2+2k^0q_0)(k\cdot p)-2p^0q_0m_N^2+4k_0^2p^0q_0+4k^0p^0m_N^2-4m_N^2(k\cdot p+k_0^2)+4m_N^4}{q_0+2k^0}\nonumber\\
    &-\frac{q_0 g_A^2}{(2\pi)^4f^4}\int\frac{\dd |\vec p|\dd|\vec k|\dd\cos\theta|\vec p|^2|\vec k|^2}{p_0k_0}\frac{\Theta^p_{\vec k}\Theta^n_{\vec p}+\Theta^p_{\vec k}\Theta^p_{\vec p}}{q_0^2+2q_0(p^0-k^0)+2m_N^2-2p_0k_0+2|\vec p||\vec k|\cos\theta-m_\pi^2}\nonumber\\
    &\times\frac{-q_0^2m_N^2-2k^0p^0q_0^2+(q_0^2-2k^0q_0)(k\cdot p)+4k_0^2p^0q_0-2p^0q_0m_N^2-4k^0p^0m_N^2+4m_N^2(k\cdot p+k_0^2)-4m_N^4}{q_0-2k^0}\nonumber\\
    &-\frac{q_0m_Ng_A^2}{(2\pi)^4f^2}\int\frac{\dd |\vec p|\dd|\vec k|\dd\cos\theta|\vec p|^2|\vec k|^2}{p_0k_0}\frac{\Theta_{\vec k}^n\Theta_{\vec p}^n}{q_0^2-2q_0(p^0-k^0)+2m_N^2-2p_0k_0+2|\vec p||\vec k|\cos\theta-m_\pi^2}C_s^3\nonumber\\
    &\times\frac{q_0^2p^0+q_0^2k^0+2p^0q_0k^0+2k_0^2q_0-2q_0m_N^2-(2q_0+4k^0)(k\cdot p)+4p^0m_N^2}{q_0+2k^0}\nonumber\\
    &-\frac{ q_0m_Ng_A^2}{(2\pi)^4f^2}\int\frac{\dd |\vec p|\dd|\vec k|\dd\cos\theta|\vec p|^2|\vec k|^2}{p_0k_0}\frac{\Theta_{\vec k}^p\Theta_{\vec p}^p}{q_0^2+2q_0(p^0-k^0)+2m_N^2-2p_0k_0+2|\vec p||\vec k|\cos\theta-m_\pi^2}C_s^4\nonumber\\
    &\times\frac{-q_0^2p^0-k^0q_0^2+2p^0k^0q_0+2q_0k_0^2-2q_0m_N^2-(2q_0-4k^0)(k\cdot p)-4p^0m_N^2}{q_0-2k^0}\nonumber\\
    &-\frac{q_0m_Ng_A^2}{(2\pi)^4f^2}\int\frac{\dd |\vec p|\dd|\vec k|\dd\cos\theta|\vec p|^2|\vec k|^2}{p_0k_0}\frac{\Theta_{\vec k}^n\Theta_{\vec p}^n}{q_0^2+2q_0(p^0-k^0)+2m_N^2-2p_0k_0+2|\vec p||\vec k|\cos\theta-m_\pi^2}C_s^5\nonumber\\
    &\times\frac{p^0q_0^2+k^0q_0^2+2k^0p^0q_0+2p_0^2q_0-2q_0m_N^2-2q_0(k\cdot p)-4p^0(k\cdot p)+4k^0m_N^2}{q_0+2 p^0}\nonumber\\
    &-\frac{q_0m_N^2g_A^2}{(2\pi)^4f^4}\int\frac{\dd |\vec p|\dd|\vec k|\dd\cos\theta|\vec p|^2|\vec k|^2}{p_0k_0}\frac{\Theta^p_{\vec k}\Theta^p_{\vec p}+\Theta^p_{\vec k}\Theta^n_{\vec p}}{(p-k)^2-m_\pi^2}\frac{3(k\cdot p)-3m_N^2}{q_0-2k^0}\nonumber\\
    &-\frac{q_0m_N^2g_A^2}{(2\pi)^4f^4}\int\frac{\dd |\vec p|\dd|\vec k|\dd\cos\theta|\vec p|^2|\vec k|^2}{p_0k_0}\frac{\Theta^n_{\vec k}\Theta^p_{\vec p}+\Theta^n_{\vec k}\Theta^n_{\vec p}}{(p-k)^2-m_\pi^2}\frac{3(k\cdot p)-3m_N^2}{q_0+2k^0}\nonumber\\
    &-\frac{q_0m_N^2g_A^2}{(2\pi)^4f^4}\int\frac{\dd |\vec p|\dd|\vec k|\dd\cos\theta|\vec p|^2|\vec k|^2}{p_0k_0}\frac{\Theta^n_{\vec k}\Theta^p_{\vec p}-\Theta^p_{\vec k}\Theta^n_{\vec p}}{(p-k)^2-m_\pi^2}(k^0-p^0)\nonumber\\
    &-\frac{q_0m_Ng_A^2}{(2\pi)^4f^2}\int\frac{\dd |\vec p|\dd|\vec k|\dd\cos\theta|\vec p|^2|\vec k|^2}{p_0k_0}\frac{\Theta_{\vec k}^p\Theta_{\vec p}^p}{q_0^2-2q_0(p^0-k^0)+2m_N^2-2p_0k_0+2|\vec p||\vec k|\cos\theta-m_\pi^2}C_s^6\nonumber\\
    &\times\frac{-k^0q_0^2-p_0q_0^2+2k^0p^0q_0-2q_0(k\cdot p)+2p_0^2q_0-2q_0m_N^2+4p^0(k\cdot p)-4k_0m_N^2}{q_0-2 p^0}\nonumber\\
      &-\frac{m_Ng_A^2}{(2\pi)^4f^2}\int\frac{\dd |\vec p|\dd|\vec k|\dd\cos\theta|\vec p|^2|\vec k|^2}{p_0k_0}\frac{\Theta_{\vec k}^p\Theta_{\vec p}^n}{(p-k)^2-m_\pi^2}\frac{2q_0m_N^2-2q_0(k\cdot p)-4k^0(k\cdot p)+4p^0m_N^2}{q_0-2k^0}C_s^7\nonumber\\
      &-\frac{m_Ng_A^2}{(2\pi)^4f^2}\int\frac{\dd |\vec p|\dd|\vec k|\dd\cos\theta|\vec p|^2|\vec k|^2}{p_0k_0}\frac{\Theta_{\vec k}^n\Theta_{\vec p}^p}{(p-k)^2-m_\pi^2}\frac{2q_0m_N^2-2q_0(k\cdot p)+4k^0(k\cdot p)-4p^0m_N^2}{q_0+2k^0}C_s^8\nonumber\\
      &-\frac{m_Ng_A^2}{(2\pi)^4f^2}\int\frac{\dd |\vec p|\dd|\vec k|\dd\cos\theta|\vec p|^2|\vec k|^2}{p_0k_0}\frac{\Theta_{\vec k}^n\Theta_{\vec p}^p}{(p-k)^2-m_\pi^2}\frac{2q_0m_N^2-2q_0(k\cdot p)-4p^0(k\cdot p)+4k^0m_N^2}{q_0-2p^0}C_s^9\nonumber\\
      &-\frac{m_Ng_A^2}{(2\pi)^4f^2}\int\frac{\dd |\vec p|\dd|\vec k|\dd\cos\theta|\vec p|^2|\vec k|^2}{p_0k_0}\frac{\Theta_{\vec k}^p\Theta_{\vec p}^n}{(p-k)^2-m_\pi^2}\frac{2q_0m_N^2-2q_0(k\cdot p)+4p^0(k\cdot p)-4k^0m_N^2}{q_0+2p^0}C_s^{10},
\end{align}
\begin{align}
    \Sigma^{\mathrm{2\text{L}}}_{\pi^-}(q_0)
    &=\frac{m_N^2g_A^2}{5(2\pi f)^4}\int \dd |\vec p|\,\dd |\vec k|\,\dd \cos \theta\frac{\vec p^2\vec k^2}{p_0k_0}\frac{p_0k_0-|\vec p||\vec k|\cos\theta-m_N^2}{((p-k)^2-m_\pi^2)^2}\nonumber\\
    &\times[(-20q_0p_0+20q_0k_0-4p_0k_0+4|\vec p||\vec k|\cos\theta+4m_N^2+2q_0^2-4m_\pi^2)\Theta_{\boldsymbol{k}}^p\Theta_{\boldsymbol{p}}^n\nonumber\\
    &+(20q_0p_0-20q_0k_0-4p_0k_0+4|\vec p||\vec k|\cos\theta+4m_N^2+2q_0^2-4m_\pi^2)\Theta_{\boldsymbol{k}}^n\Theta_{\boldsymbol{p}}^p\nonumber\\
    &+(4p_0k_0-4|\vec p||\vec k|\cos\theta-4m_N^2-2q_0^2-m_\pi^2)(\Theta_{\boldsymbol{k}}^p\Theta_{\boldsymbol{p}}^p+\Theta_{\boldsymbol{k}}^n\Theta_{\boldsymbol{p}}^n)]+\frac{m_N^2g_A^2}{5(2\pi f)^4}\int \dd |\vec p|\,\dd |\vec k|\,\dd \cos \theta\frac{\vec p^2\vec k^2}{p_0k_0}\nonumber
\end{align}
\begin{align}
    \qquad\qquad\ &\times\frac{1}{(p-k)^2-m_\pi^2}[(-5q_0k_0+5q_0p_0+2m_N^2-2p_0k_0+2|\vec p||\vec k|\cos\theta)\Theta^p_{\boldsymbol{p}}\Theta_{\boldsymbol{k}}^n\nonumber\\
    &+(5q_0k_0-5q_0p_0+2m_N^2-2p_0k_0+2|\vec p||\vec k|\cos\theta)\Theta^p_{\boldsymbol{k}}\Theta_{\boldsymbol{p}}^n+(2p_0k_0-2|\vec p||\vec k|\cos\theta-2m_N^2)(\Theta^p_{\boldsymbol{p}}\Theta^p_{\boldsymbol{k}}+\Theta^n_{\boldsymbol{p}}\Theta^n_{\boldsymbol{k}})]\nonumber\\
    &-\frac{q_0^2}{2(2\pi)^4f^4}\int \dd |\vec p|\,\dd |\vec k|\,\dd \cos \theta\frac{\vec p^2\vec k^2}{p_0k_0}\frac{p^0k^0+|\vec p||\vec k|\cos\theta+m_N^2}{q_0^2+2q_0(p^0-k^0)+2m_N^2-2p_0k_0+2|\vec p||\vec k|\cos\theta-m_\pi^2}\nonumber\\
    &\times\big[(\Theta^p_{\boldsymbol{p}}\Theta^p_{\boldsymbol{k}}+\Theta^n_{\boldsymbol{p}}\Theta^n_{\boldsymbol{k}})+(\Theta^p_{\boldsymbol{p}}\Theta^n_{\boldsymbol{k}}+\Theta^n_{\boldsymbol{p}}\Theta^p_{\boldsymbol{k}})+(\Theta^p_{\boldsymbol{p}}\Theta^n_{\boldsymbol{k}}-\Theta^n_{\boldsymbol{p}}\Theta^p_{\boldsymbol{k}})\big]\nonumber\\
    &-\frac{2q_0^2}{(2\pi)^4}\int \dd |\vec p|\,\dd |\vec k|\,\dd \cos \theta\frac{\vec p^2\vec k^2}{p_0k_0}\frac{C_s^1C_s^2(p_0k_0-|\vec p||\vec k|\cos\theta+m_N^2)}{q_0^2+2q_0(p^0-k^0)+2m_N^2-2p_0k_0+2|\vec p||\vec k|\cos\theta-m_\pi^2}(\Theta^p_{\boldsymbol{p}}\Theta^p_{\boldsymbol{k}}+\Theta^n_{\boldsymbol{k}}\Theta^n_{\boldsymbol{p}})\nonumber\\
    &+\frac{m_Nq^2_0}{(2\pi)^4f^2}\int \dd |\vec p|\,\dd |\vec k|\,\dd \cos \theta\frac{\vec p^2\vec k^2}{p_0k_0}\frac{(p_0+k_0)}{q_0^2+2q_0(p^0-k^0)+2m_N^2-2p_0k_0+2|\vec p||\vec k|\cos\theta-m_\pi^2}(\Theta^p_{\boldsymbol{p}}\Theta^p_{\boldsymbol{k}}-\Theta^n_{\boldsymbol{k}}\Theta^n_{\boldsymbol{p}})(C_s^1+C_s^2)\nonumber\\
    &+\frac{q_0 g_A^2}{(2\pi)^4f^4}\int\frac{\dd |\vec p|\dd|\vec k|\dd\cos\theta|\vec p|^2|\vec k|^2}{p_0k_0}\frac{\Theta^p_{\vec k}\Theta^n_{\vec p}+\Theta^p_{\vec k}\Theta^p_{\vec p}}{q_0^2-2q_0(p^0-k^0)+2m_N^2-2p_0k_0+2|\vec p||\vec k|\cos\theta-m_\pi^2}\nonumber\\
    &\times\frac{2k^0p^0q_0^2+q_0^2m_N^2-(q_0^2+2k^0q_0)(k\cdot p)-2p^0q_0m_N^2+4k_0^2p^0q_0+4k^0p^0m_N^2-4m_N^2(k\cdot p+k_0^2)+4m_N^4}{q_0+2k^0}\nonumber\\
    &-\frac{q_0g_A^2}{(2\pi)^4f^4}\int\frac{\dd |\vec p|\dd|\vec k|\dd\cos\theta|\vec p|^2|\vec k|^2}{p_0k_0}\frac{\Theta^n_{\vec k}\Theta^p_{\vec p}+\Theta^n_{\vec k}\Theta^n_{\vec p}}{q_0^2+2q_0(p^0-k^0)+2m_N^2-2p_0k_0+2|\vec p||\vec k|\cos\theta-m_\pi^2}\nonumber\\
    &\times\frac{-q_0^2m_N^2-2k^0p^0q_0^2+(q_0^2-2k^0q_0)(k\cdot p)+4k_0^2p^0q_0-2p^0q_0m_N^2-4k^0p^0m_N^2+4m_N^2(k\cdot p+k_0^2)-4m_N^4}{q_0-2k^0}\nonumber\\
    &-\frac{ q_0m_Ng_A^2}{(2\pi)^4f^2}\int\frac{\dd |\vec p|\dd|\vec k|\dd\cos\theta|\vec p|^2|\vec k|^2}{p_0k_0}\frac{\Theta_{\vec p}^p\Theta_{\vec k}^p}{q_0^2-2q_0(p^0-k^0)+2m_N^2-2p_0k_0+2|\vec p||\vec k|\cos\theta-m_\pi^2}C_s^3\nonumber\\
    &\times\frac{q_0^2p^0+q_0^2k^0+2p^0q_0k^0+2k_0^2q_0-2q_0m_N^2-(2q_0+4k^0)(k\cdot p)+4p^0m_N^2}{q_0+2k^0}\nonumber\\
    \qquad\qquad&-\frac{q_0m_N g_A^2}{(2\pi)^4f^2}\int\frac{\dd |\vec p|\dd|\vec k|\dd\cos\theta|\vec p|^2|\vec k|^2}{p_0k_0}\frac{\Theta_{\vec p}^n\Theta_{\vec k}^n}{q_0^2+2q_0(p^0-k^0)+2m_N^2-2p_0k_0+2|\vec p||\vec k|\cos\theta-m_\pi^2}C_s^4\nonumber\\
    &\times\frac{-q_0^2p^0-k^0q_0^2+2p^0k^0q_0+2q_0k_0^2-2q_0m_N^2-(2q_0-4k^0)(k\cdot p)-4p^0m_N^2}{q_0-2k^0}\nonumber\\
    &-\frac{q_0m_Ng_A^2}{(2\pi)^4f^2}\int\frac{\dd |\vec p|\dd|\vec k|\dd\cos\theta|\vec p|^2|\vec k|^2}{p_0k_0}\frac{\Theta_{\vec p}^p\Theta_{\vec k}^p}{q_0^2+2q_0(p^0-k^0)+2m_N^2-2p_0k_0+2|\vec p||\vec k|\cos\theta-m_\pi^2}C_s^5\nonumber\\
    &\times\frac{p^0q_0^2+k^0q_0^2+2k^0p^0q_0+2p_0^2q_0-2q_0m_N^2-2q_0(k\cdot p)-4p^0(k\cdot p)+4k^0m_N^2}{q_0+2 p^0}\nonumber\\
    &-\frac{q_0m_Ng_A^2}{(2\pi)^4f^2}\int\frac{\dd |\vec p|\dd|\vec k|\dd\cos\theta|\vec p|^2|\vec k|^2}{p_0k_0}\frac{\Theta_{\vec p}^n\Theta_{\vec k}^n}{q_0^2-2q_0(p^0-k^0)+2m_N^2-2p_0k_0+2|\vec p||\vec k|\cos\theta-m_\pi^2}C_s^6\nonumber\\
    &\times\frac{-k^0q_0^2-p_0q_0^2+2k^0p^0q_0-2q_0(k\cdot p)+2p_0^2q_0-2q_0m_N^2+4p^0(k\cdot p)-4k_0m_N^2}{q_0-2p^0}\nonumber\\
    &-\frac{q_0m_N^2g_A^2}{(2\pi)^4f^4}\int\frac{\dd |\vec p|\dd|\vec k|\dd\cos\theta|\vec p|^2|\vec k|^2}{p_0k_0}\frac{\Theta^n_{\vec k}\Theta^p_{\vec p}+\Theta^n_{\vec k}\Theta^n_{\vec p}}{(p-k)^2-m_\pi^2}\frac{3(k\cdot p)-3m_N^2}{q_0-2k^0}\nonumber\\
    &-\frac{q_0m_N^2g_A^2}{(2\pi)^4f^4}\int\frac{\dd |\vec p|\dd|\vec k|\dd\cos\theta|\vec p|^2|\vec k|^2}{p_0k_0}\frac{\Theta^p_{\vec k}\Theta^n_{\vec p}+\Theta^p_{\vec k}\Theta^p_{\vec p}}{(p-k)^2-m_\pi^2}\frac{3(k\cdot p)-3m_N^2}{q_0+2k^0}\nonumber\\
    &-\frac{q_0m_N^2g_A^2}{(2\pi)^4f^4}\int\frac{\dd |\vec p|\dd|\vec k|\dd\cos\theta|\vec p|^2|\vec k|^2}{p_0k_0}\frac{\Theta^p_{\vec k}\Theta^n_{\vec p}-\Theta^n_{\vec k}\Theta^p_{\vec p}}{(p-k)^2-m_\pi^2}(k^0-p^0)\nonumber\\
    &-\frac{m_Ng_A^2}{(2\pi)^4f^2}\int\frac{\dd |\vec p|\dd|\vec k|\dd\cos\theta|\vec p|^2|\vec k|^2}{p_0k_0}\frac{\Theta_{\vec k}^n\Theta_{\vec p}^p}{(p-k)^2-m_\pi^2}\frac{2q_0m_N^2-2q_0(k\cdot p)-4k^0(k\cdot p)+4p^0m_N^2}{q_0-2k^0}C_s^7\nonumber\\
    &-\frac{m_Ng_A^2}{(2\pi)^4f^2}\int\frac{\dd |\vec p|\dd|\vec k|\dd\cos\theta|\vec p|^2|\vec k|^2}{p_0k_0}\frac{\Theta_{\vec k}^p\Theta_{\vec p}^n}{(p-k)^2-m_\pi^2}\frac{2q_0m_N^2-2q_0(k\cdot p)+4k^0(k\cdot p)-4p^0m_N^2}{q_0+2k^0}C_s^8\nonumber\\
    &-\frac{m_Ng_A^2}{(2\pi)^4f^2}\int\frac{\dd |\vec p|\dd|\vec k|\dd\cos\theta|\vec p|^2|\vec k|^2}{p_0k_0}\frac{\Theta_{\vec k}^p\Theta_{\vec p}^n}{(p-k)^2-m_\pi^2}\frac{2q_0m_N^2-2q_0(k\cdot p)-4p^0(k\cdot p)+4k^0m_N^2}{q_0-2p^0}C_s^9\nonumber\\
    &-\frac{m_Ng_A^2}{(2\pi)^4f^2}\int\frac{\dd |\vec p|\dd|\vec k|\dd\cos\theta|\vec p|^2|\vec k|^2}{p_0k_0}\frac{\Theta_{\vec p}^p\Theta_{\vec k}^n}{(p-k)^2-m_\pi^2}\frac{2q_0m_N^2-2q_0(k\cdot p)+4p^0(k\cdot p)-4k^0m_N^2}{q_0+2p^0}C_s^{10},
\end{align}
\begin{align}
    \Sigma^{\mathrm{2\text{L}}}_{\pi^0}(q_0)
    &=\frac{m_N^2g_A^2}{5(2\pi f)^4}\int \dd |\vec p|\,\dd |\vec k|\,\dd \cos \theta\frac{\vec p^2\vec k^2}{p_0k_0}\frac{p_0k_0-|\vec p||\vec k|\cos\theta-m_N^2}{((p-k)^2-m_\pi^2)^2}\nonumber\\
    &\times[(-8m_N^2+8p_0k_0-8|\vec p||\vec k|\cos\theta-4q_0^2-2m_\pi^2)(\Theta_{\boldsymbol{k}}^p\Theta^n_{\boldsymbol{p}}+\Theta^p_{\boldsymbol{p}}\Theta^n_{\boldsymbol{k}})\nonumber\\
    &+(-8p_0k_0+8|\vec p||\vec k|\cos\theta+8m_N^2+4q_0^2-3m_\pi^2)(\Theta_{\boldsymbol{k}}^p\Theta_{\boldsymbol{p}}^p+\Theta_{\boldsymbol{k}}^n\Theta_{\boldsymbol{p}}^n)]+\frac{m_N^2g_A^2}{5(2\pi f)^4}\int \dd |\vec p|\,\dd |\vec k|\,\dd \cos \theta\frac{\vec p^2\vec k^2}{p_0k_0}\nonumber\\
    &\times\frac{1}{(p-k)^2-m_\pi^2}[(4p_0k_0-4|\vec p||\vec k|\cos\theta-4m_N^2)(\Theta_{\boldsymbol{k}}^p\Theta^n_{\boldsymbol{p}}+\Theta^n_{\boldsymbol{k}}\Theta^p_{\boldsymbol{p}})\nonumber\\
    &+(-4p_0k_0+4|\vec p||\vec k|\cos\theta+4m_N^2)(\Theta^p_{\boldsymbol{p}}\Theta^p_{\boldsymbol{k}}+\Theta^n_{\boldsymbol{p}}\Theta^n_{\boldsymbol{k}})]\nonumber\\
    &-\frac{q_0^2}{(2\pi)^4f^4}\int \dd |\vec p|\,\dd |\vec k|\,\dd \cos \theta\frac{\vec p^2\vec k^2}{p_0k_0}\frac{p^0 k^0+|\vec p||\vec k|\cos\theta+m_N^2}{q_0^2+2q_0(p^0-k^0)+2m_N^2-2p_0k_0+2|\vec p||\vec k|\cos\theta-m_\pi^2}(\Theta^n_{\boldsymbol{p}}\Theta^p_{\boldsymbol{k}}+\Theta^n_{\boldsymbol{k}}\Theta^p_{\boldsymbol{p}})\nonumber\\
    &-\frac{2q_0^2}{(2\pi)^4}\int \dd |\vec p|\,\dd |\vec k|\,\dd \cos \theta\frac{\vec p^2\vec k^2}{p_0k_0}\frac{C_s^1C_s^2(p_0k_0-|\vec p||\vec k|\cos\theta+m_N^2)}{q_0^2+2q_0(p^0-k^0)+2m_N^2-2p_0k_0+2|\vec p||\vec k|\cos\theta-m_\pi^2}(\Theta^p_{\boldsymbol{p}}\Theta^p_{\boldsymbol{k}}+\Theta^n_{\boldsymbol{k}}\Theta^n_{\boldsymbol{p}})\nonumber\\
    &+\frac{q_0g_A^2}{(2\pi)^4f^4}\int\frac{\dd |\vec p|\dd|\vec k|\dd\cos\theta|\vec p|^2|\vec k|^2}{p_0k_0}\frac{\Theta^p_{\vec k}\Theta^n_{\vec p}+\Theta^n_{\vec k}\Theta^p_{\vec p}}{q_0^2-2q_0(p^0-k^0)+2m_N^2-2p_0k_0+2|\vec p||\vec k|\cos\theta-m_\pi^2}\nonumber\\
    &\times\frac{2k^0p^0q_0^2+q_0^2m_N^2-(q_0^2+2k^0q_0)(k\cdot p)-2p^0q_0m_N^2+4k_0^2p^0q_0+4k^0p^0m_N^2-4m_N^2(k\cdot p+k_0^2)+4m_N^4}{q_0+2k^0}\nonumber\\
    &-\frac{q_0g_A^2}{(2\pi)^4f^4}\int\frac{\dd |\vec p|\dd|\vec k|\dd\cos\theta|\vec p|^2|\vec k|^2}{p_0k_0}\frac{\Theta^p_{\vec k}\Theta^n_{\vec p}+\Theta^n_{\vec k}\Theta^p_{\vec p}}{q_0^2+2q_0(p^0-k^0)+2m_N^2-2p_0k_0+2|\vec p||\vec k|\cos\theta-m_\pi^2}\nonumber\\
    &\times\frac{-q_0^2m_N^2-2k^0p^0q_0^2+(q_0^2-2k^0q_0)(k\cdot p)+4k_0^2p^0q_0-2p^0q_0m_N^2-4k^0p^0m_N^2+4m_N^2(k\cdot p+k_0^2)-4m_N^4}{q_0-2k^0}\nonumber\\
    &-\frac{q_0m_N g_A^2}{2(2\pi)^4f^2}\int\frac{\dd |\vec p|\dd|\vec k|\dd\cos\theta|\vec p|^2|\vec k|^2}{p_0k_0}\frac{\Theta_{\vec k}^p\Theta_{\vec p}^p+\Theta_{\vec k}^n\Theta_{\vec p}^n}{q_0^2-2q_0(p^0-k^0)+2m_N^2-2p_0k_0+2|\vec p||\vec k|\cos\theta-m_\pi^2}C_s^3\nonumber\\
    &\times\frac{q_0^2p_0+q_0^2k^0+2p^0k^0q_0+2k_0^2q_0-2q_0m_N^2-(2q_0+4k^0)(k\cdot p)+4p^0m_N^2}{q_0+2k^0}\nonumber\\
    &-\frac{q_0m_N g_A^2}{2(2\pi)^4f^2}\int\frac{\dd |\vec p|\dd|\vec k|\dd\cos\theta|\vec p|^2|\vec k|^2}{p_0k_0}\frac{\Theta_{\vec k}^p\Theta_{\vec p}^p+\Theta_{\vec k}^n\Theta_{\vec p}^n}{q_0^2+2q_0(p^0-k^0)+2m_N^2-2p_0k_0+2|\vec p||\vec k|\cos\theta-m_\pi^2}C_s^4\nonumber\\
    &\times\frac{-q_0^2p^0-k^0q_0^2+2p^0k^0q_0+2q_0(k^0)^2-2q_0m_N^2-(2q_0-4k^0)(k\cdot p)-4p^0m_N^2}{q_0-2k^0}\nonumber\\
    &-\frac{q_0m_Ng_A^2}{2(2\pi)^4f^2}\int\frac{\dd |\vec p|\dd|\vec k|\dd\cos\theta|\vec p|^2|\vec k|^2}{p_0k_0}\frac{\Theta_{\vec k}^n\Theta_{\vec p}^n+\Theta_{\vec k}^p\Theta_{\vec p}^p}{q_0^2+2q_0(p^0-k^0)+2m_N^2-2p_0k_0+2|\vec p||\vec k|\cos\theta-m_\pi^2}C_s^5\nonumber\\
    &\times\frac{p^0q_0^2+k^0q_0^2+2k^0p^0q_0+2p_0^2q_0-2q_0m_N^2-2q_0(k\cdot p)-4p^0(k\cdot p)+4k^0m_N^2}{q_0+2p^0}\nonumber\\
    &-\frac{q_0m_Ng_A^2}{2(2\pi)^4f^2}\int\frac{\dd |\vec p|\dd|\vec k|\dd\cos\theta|\vec p|^2|\vec k|^2}{p_0k_0}\frac{\Theta_{\vec k}^p\Theta_{\vec p}^p+\Theta_{\vec k}^n\Theta_{\vec p}^n}{q_0^2-2q_0(p^0-k^0)+2m_N^2-2p_0k_0+2|\vec p||\vec k|\cos\theta-m_\pi^2}C_s^6\nonumber\\
    &\times\frac{-k^0q_0^2-p_0q_0^2+2k^0p^0q_0-2q_0(k\cdot p)+2p_0^2q_0-2q_0m_N^2+4p^0(k\cdot p)-4k_0m_N^2}{q_0-2p^0}\nonumber\\
    &-\frac{q_0m_N^2g_A^2}{(2\pi)^4f^4}\int\frac{\dd |\vec p|\dd|\vec k|\dd\cos\theta|\vec p|^2|\vec k|^2}{p_0k_0}\frac{\Theta^p_{\vec k}\Theta^n_{\vec p}+\Theta^n_{\vec k}\Theta^p_{\vec p}}{(p-k)^2-m_\pi^2}\frac{3(k\cdot p)-3m_N^2}{q_0-2k^0}\nonumber\\
    &-\frac{q_0m_N^2g_A^2}{(2\pi)^4f^4}\int\frac{\dd |\vec p|\dd|\vec k|\dd\cos\theta|\vec p|^2|\vec k|^2}{p_0k_0}\frac{\Theta^p_{\vec k}\Theta^n_{\vec p}+\Theta^n_{\vec k}\Theta^p_{\vec p}}{(p-k)^2-m_\pi^2}\frac{3(k\cdot p)-3m_N^2}{q_0+2k^0}\nonumber\\
    &-\frac{m_Ng_A^2}{(2\pi)^4f^2}\int\frac{\dd |\vec p|\dd|\vec k|\dd\cos\theta|\vec p|^2|\vec k|^2}{p_0k_0}\frac{\frac{1}{2}(\Theta_{\vec k}^p\Theta_{\vec p}^p+\Theta_{\vec k}^n\Theta_{\vec p}^n)}{(p-k)^2-m_\pi^2}\frac{2q_0m_N^2-2q_0(k\cdot p)-4k^0(k\cdot p)+4p^0m_N^2}{q_0-2k^0}C_s^7\nonumber\\
    &-\frac{m_Ng_A^2}{(2\pi)^4f^2}\int\frac{\dd |\vec p|\dd|\vec k|\dd\cos\theta|\vec p|^2|\vec k|^2}{p_0k_0}\frac{\frac{1}{2}(\Theta_{\vec k}^p\Theta_{\vec p}^p+\Theta_{\vec k}^n\Theta_{\vec p}^n)}{(p-k)^2-m_\pi^2}\frac{2q_0m_N^2-2q_0(k\cdot p)+4k^0(k\cdot p)-4p^0m_N^2}{q_0+2k^0}C_s^8\nonumber\\
    &-\frac{m_Ng_A^2}{(2\pi)^4f^2}\int\frac{\dd |\vec p|\dd|\vec k|\dd\cos\theta|\vec p|^2|\vec k|^2}{p_0k_0}\frac{\frac{1}{2}(\Theta_{\vec k}^p\Theta_{\vec p}^p+\Theta_{\vec k}^n\Theta_{\vec p}^n)}{(p-k)^2-m_\pi^2}\frac{2q_0m_N^2-2q_0(k\cdot p)-4p^0(k\cdot p)+4k^0m_N^2}{q_0-2p^0}C_s^9\nonumber\\
    &-\frac{m_Ng_A^2}{(2\pi)^4f^2}\int\frac{\dd |\vec p|\dd|\vec k|\dd\cos\theta|\vec p|^2|\vec k|^2}{p_0k_0}\frac{\frac{1}{2}(\Theta_{\vec k}^p\Theta_{\vec p}^p+\Theta_{\vec k}^n\Theta_{\vec p}^n)}{(p-k)^2-m_\pi^2}\frac{2q_0m_N^2-2q_0(k\cdot p)+4p^0(k\cdot p)-4k^0m_N^2}{q_0+2p^0}C_s^{10},
\end{align}
\end{subequations}
where
\begin{align}
    \MoveEqLeft C_s^1=\frac{8B_0c_1m_q}{q_0f^2}-\frac{2c_2}{f^2m_N^2}q_0p_0^2-\frac{2c_3}{f^2}q_0-\frac{2c_2}{f^2m_N^2}p_0(p_\mu p^\mu-k_\mu p^\mu)-\frac{2c_3}{f^2}(p_0-k_0),\nonumber\\
    \MoveEqLeft C_s^2=\frac{8B_0c_1m_q}{q_0f^2}-\frac{2c_2}{f^2m_N^2}q_0k_0^2 -\frac{2c_3}{f^2}q_0-\frac{2c_2}{f^2m_N^2}k_0(k_\mu p^\mu-k_\mu k^\mu)-\frac{2c_3}{f^2}(p_0-k_0),\nonumber\\
    \MoveEqLeft C_s^3=\frac{8B_0c_1m_q}{q_0f^2}-\frac{2c_2}{f^2m_N^2}q_0 k_0^2-\frac{2c_3}{f^2}q_0-\frac{2c_2}{f^2m_N^2}k^0(k^\mu k_\mu-k_\mu p^\mu)-\frac{2c_3}{f^2}(k_0-p_0),\nonumber
\end{align}
\begin{align}
     \MoveEqLeft C_s^4=\frac{8B_0c_1m_q}{q_0f^2}-\frac{2c_2}{f^2m_N^2}q_0 p_0^2-\frac{2c_3}{f^2}q_0-\frac{2c_2}{f^2m_N^2}p^0(p_\mu p^\mu-p_\mu k^\mu)-\frac{2c_3}{f^2}(p_0-k_0),\nonumber\\
     \MoveEqLeft C_s^5=\frac{8B_0c_1m_q}{q_0f^2}-\frac{2c_2}{f^2m_N^2}k_0^2 q_0 -\frac{2c_3}{f^2}q_0-\frac{2c_2}{f^2m_N^2}k_0(k_\mu p^\mu-k_\mu k^\mu)-\frac{2c_3}{f^2}(p_0-k_0),\nonumber\\
     \MoveEqLeft C_s^6=\frac{8B_0c_1m_q}{q_0f^2}-\frac{2c_2}{f^2m_N^2}p_0^2 q_0-\frac{2c_3}{f^2}q_0-\frac{2c_2}{f^2m_N^2}p_0(p_\mu k^\mu-p_\mu p^\mu)-\frac{2c_3}{f^2}(k^0-p^0),\nonumber\\
    \MoveEqLeft C_s^7=\frac{8B_0c_1m_q}{f^2}-\frac{2c_2}{f^2m_N^2}q_\nu (k^\nu-q^\nu)(k_\mu-p_\mu)( k^\mu-q^\mu))-\frac{2c_3}{f^2}q^\mu(k_\mu-p_\mu),\nonumber\\
    \MoveEqLeft C_s^8=\frac{8B_0c_1m_q}{f^2}-\frac{2c_2}{f^2m_N^2}q_\nu p^\nu p_\mu (p^\mu-k^\mu)-\frac{2c_3}{f^2}q^\mu(p_\mu-k_\mu),\nonumber\\
    \MoveEqLeft C_s^9=\frac{8B_0c_1m_q}{f^2}+\frac{2c_2}{f^2m_N^2}k^\mu q_\mu k_\nu(k^\nu-p^\nu)+\frac{2c_3}{f^2}q^\mu(k_\mu-p_\mu),\nonumber\\
    \MoveEqLeft C_s^{10}=\frac{8B_0c_1m_q}{f^2}+\frac{2c_2}{f^2m_N^2}q_\mu(p^\mu+q^\mu)(p_\nu-k_\nu)(p^\nu+q^\nu)+\frac{2c_3}{f^2}q^\mu(p_\mu-k_\mu).
\end{align}
The self-energies are evaluated in the pion rest frame, $q^\mu=(q^0,\vec 0)$, $p^\mu$ and $k^\mu$ are the vacuum on-shell nucleon four momenta, $\theta$ is the angle between the three nucleon momenta, and $E(\vec p)$ denotes the vacuum on-shell nucleon energy. 

The vertex corrections are now given by
\begin{subequations}\label{exdecayone}
\begin{align}
    \hat{\Gamma}_{0,\ \mathrm{1\text{L}}}^{\pi^+}(q_0)
    &=-\int \dd |\vec p| \frac{\vec p^2}{4\pi^2p_0f} \left[ \frac{q_0g_A^2p_0}{q_0+2p_0}\big[(\Theta^p_{\boldsymbol{p}}+\Theta^n_{\boldsymbol{p}})-(\Theta^p_{\boldsymbol{p}}-\Theta^n_{\boldsymbol{p}})\big]-\frac{q_0g_A^2p_0}{q_0-2p_0}\big[(\Theta^p_{\boldsymbol{p}}+\Theta^n_{\boldsymbol{p}})+(\Theta^p_{\boldsymbol{p}}-\Theta^n_{\boldsymbol{p}})\big]\right. \nonumber\\
    &\qquad +\left.  \frac{2g_A^2\vec p^2}{q_0+2p_0}\big[(\Theta^p_{\boldsymbol{p}}+\Theta^n_{\boldsymbol{p}})-(\Theta^p_{\boldsymbol{p}}-\Theta^n_{\boldsymbol{p}})\big]+\frac{2g_A^2\vec p^2}{q_0-2p_0}\big[(\Theta^p_{\boldsymbol{p}}+\Theta^n_{\boldsymbol{p}})+(\Theta^p_{\boldsymbol{p}}-\Theta^n_{\boldsymbol{p}})\big]\right.\nonumber\\
    &\qquad \left. -4q_0^2m_N\left(\frac{2c_2}{m_N^2}p_0^2+2c_3\right)(\Theta^p_{\boldsymbol{p}}+\Theta^n_{\boldsymbol{p}})+2q_0p_0(\Theta^p_{\boldsymbol{p}}-\Theta^n_{\boldsymbol{p}})\right] \nonumber\\
    &\approx-\frac{q_0m_Ng_A^2}{f(4m_N^2-q_0^2)}\rho+\frac{g_A^2q_0^2(r-1)}{2f(1+r)(4m_N^2-q_0^2)}\rho+\frac{r-1}{2f(r+1)}\rho+\frac{2q_0(c_2+c_3)}{f}\rho,
\end{align}
\begin{align}
    \hat{\Gamma}_{0,\ \mathrm{1\text{L}}}^{\pi^-}(q_0)
    &=-\int \dd |\vec p| \frac{\vec p^2}{4\pi^2p_0f} \left[ \frac{q_0g_A^2p_0}{q_0+2p_0}\big[(\Theta^p_{\boldsymbol{p}}+\Theta^n_{\boldsymbol{p}})+(\Theta^p_{\boldsymbol{p}}-\Theta^n_{\boldsymbol{p}})\big]-\frac{q_0g_A^2p_0}{q_0-2p_0}\big[(\Theta^p_{\boldsymbol{p}}+\Theta^n_{\boldsymbol{p}})-(\Theta^p_{\boldsymbol{p}}-\Theta^n_{\boldsymbol{p}})\big]\right. \nonumber\\
    &\qquad +\left.  \frac{2g_A^2\vec p^2}{q_0+2p_0}\big[(\Theta^p_{\boldsymbol{p}}+\Theta^n_{\boldsymbol{p}})+(\Theta^p_{\boldsymbol{p}}-\Theta^n_{\boldsymbol{p}})\big]+\frac{2g_A^2\vec p^2}{q_0-2p_0}\big[(\Theta^p_{\boldsymbol{p}}+\Theta^n_{\boldsymbol{p}})-(\Theta^p_{\boldsymbol{p}}-\Theta^n_{\boldsymbol{p}})\big]\right.\nonumber\\
    &\qquad \left. -4q_0m_N\left(\frac{2c_2}{m_N^2}p_0^2+2c_3\right)(\Theta^p_{\boldsymbol{p}}+\Theta^n_{\boldsymbol{p}})-2p_0(\Theta^p_{\boldsymbol{p}}-\Theta^n_{\boldsymbol{p}})\right]\nonumber\\
    &\approx-\frac{q_0m_Ng_A^2}{f(4m_N^2-q_0^2)}\rho+\frac{g_A^2q_0^2(1-r)}{2f(1+r)(4m_N^2-q_0^2)}\rho+\frac{1-r}{2f(1+r)}\rho+\frac{2q_0(c_2+c_3)}{f}\rho,
\end{align}
\begin{align}
    \hat{\Gamma}_{0,\ \mathrm{1\text{L}}}^{\pi^0} (q_0)
    &=-\int \dd |\vec p|\frac{q_0g_A^2\vec p^2}{4\pi^2fp_0}\frac{4p_0^2-4\vec p^2}{4p_0^2-q_0^2}(\Theta^p_{\boldsymbol{p}}+\Theta^n_{\boldsymbol{p}})+\int \dd |\vec p|\frac{2m_Nq_0\vec p^2}{\pi^2p_0f}\left(c_2\frac{p_0^2}{m_N^2}+c_3\right)(\Theta^p_{\boldsymbol{p}}+\Theta^n_{\boldsymbol{p}})\nonumber\\
    &\approx-\frac{q_0m_Ng_A^2}{f(4m_N^2-q_0^2)}\rho+\frac{2q_0(c_2+c_3)}{f}\rho,
\end{align}
\end{subequations}
\begin{subequations}\label{eq:exdecaytwo}
\begin{align}
    \hat{\Gamma}_{0,\ \mathrm{2\text{L}}}^{\pi^+}(q_0)
    &=-\frac{2g_A^2m_N^2}{5(2\pi)^4f^3}\int \dd |\vec p|\,\dd |\vec k|\,\dd \cos \theta\frac{\vec p^2\vec k^2}{p_0k_0}\frac{p_0k_0-|\vec p||\vec k|\cos{\theta}-m_N^2}{((p-k)^2-m_\pi^2)^2}\nonumber\\
    &\times\left((-2q_0+10p_0-10k_0)\Theta_{\boldsymbol{k}}^p\Theta^n_{\boldsymbol{p}}+(-2q_0-10p_0+10k_0)\Theta^p_{\boldsymbol{p}}\Theta^n_{\boldsymbol{k}}-3q_0(\Theta^p_{\boldsymbol{p}}\Theta^p_{\boldsymbol{k}}+\Theta^n_{\boldsymbol{p}}\Theta^n_{\boldsymbol{k}})\right)\nonumber\\
    &+\frac{g_A^2m_N^2}{2(2\pi)^4f^3}\int \dd |\vec p|\,\dd |\vec k|\,\dd \cos \theta\frac{\vec p^2\vec k^2}{p_0k_0}\frac{k_0-p_0}{(p-k)^2-m_\pi^2}\left(-4\Theta^p_{\boldsymbol{p}}\Theta^n_{\boldsymbol{k}}+2\Theta^p_{\boldsymbol{k}}\Theta^n_{\boldsymbol{p}}\right)\nonumber\\
    &+\frac{q_0}{2(2\pi)^4f^3}\int \dd |\vec p|\,\dd |\vec k|\,\dd \cos \theta\frac{\vec p^2\vec k^2}{p_0k_0}\frac{ p^0 k^0+|\vec p||\vec k|\cos\theta+m_N^2}{q_0^2+2q_0(p^0-k^0)+2m_N^2-2p_0k_0+2|\vec p||\vec k|\cos\theta-m_\pi^2}\nonumber\\
    &\times\big[(\Theta^p_{\boldsymbol{p}}\Theta^p_{\boldsymbol{k}}+\Theta^n_{\boldsymbol{k}}\Theta^n_{\boldsymbol{p}})+(\Theta^n_{\boldsymbol{p}}\Theta^p_{\boldsymbol{k}}+\Theta^p_{\boldsymbol{p}}\Theta^n_{\boldsymbol{k}})+(\Theta^n_{\boldsymbol{p}}\Theta^p_{\boldsymbol{k}}-\Theta^p_{\boldsymbol{p}}\Theta^n_{\boldsymbol{k}})\big]\nonumber\\
    &-\frac{2q_0}{(2\pi)^4}\int \dd |\vec p|\,\dd |\vec k|\,\dd \cos \theta\frac{\vec p^2\vec k^2}{p_0k_0}\frac{C_f^1C_f^2(p_0k_0-|\vec p||\vec k|\cos\theta+m_N^2)}{q_0^2+2q_0(p^0-k^0)+2m_N^2-2p_0k_0+2|\vec p||\vec k|\cos\theta-m_\pi^2}(\Theta^p_{\boldsymbol{p}}\Theta^p_{\boldsymbol{k}}+\Theta^n_{\boldsymbol{k}}\Theta^n_{\boldsymbol{p}})\nonumber\\
    &-\frac{q_0m_N}{(2\pi)^4f^2}\int \dd |\vec p|\,\dd |\vec k|\,\dd \cos \theta\frac{\vec p^2\vec k^2}{p_0k_0}\frac{(p_0+k_0)}{q_0^2+2q_0(p^0-k^0)+2m_N^2-2p_0k_0+2|\vec p||\vec k|\cos\theta-m_\pi^2}(\Theta^p_{\boldsymbol{p}}\Theta^p_{\boldsymbol{k}}-\Theta^n_{\boldsymbol{k}}\Theta^n_{\boldsymbol{p}})C_f^2\nonumber\\
    &+\frac{q_0m_N}{(2\pi)^4f}\int \dd |\vec p|\,\dd |\vec k|\,\dd \cos \theta\frac{\vec p^2\vec k^2}{p_0k_0}\frac{(p_0+k_0)}{q_0^2+2q_0(p^0-k^0)+2m_N^2-2p_0k_0+2|\vec p||\vec k|\cos\theta-m_\pi^2}(\Theta^p_{\boldsymbol{p}}\Theta^p_{\boldsymbol{k}}-\Theta^n_{\boldsymbol{p}}\Theta^n_{\boldsymbol{k}})C_f^1\nonumber\\
    &-\frac{ g_A^2}{(2\pi)^4f^3}\int\frac{\dd |\vec p|\dd|\vec k|\dd\cos\theta|\vec p|^2|\vec k|^2}{p_0k_0}\frac{\Theta^n_{\vec k}\Theta^p_{\vec p}+\Theta^n_{\vec k}\Theta^n_{\vec p}}{q_0^2-2q_0(p^0-k^0)+2m_N^2-2p_0k_0+2|\vec p||\vec k|\cos\theta-m_\pi^2}\nonumber\\
    &\times\frac{2k^0p^0q_0^2+q_0^2m_N^2-(q_0^2+2k^0q_0)(k\cdot p)-2p^0q_0m_N^2+4k_0^2p^0q_0+4k^0p^0m_N^2-4m_N^2(k\cdot p+k_0^2)+4m_N^4}{q_0+2k^0}\nonumber\\
    &+\frac{ g_A^2}{(2\pi)^4f^3}\int\frac{\dd |\vec p|\dd|\vec k|\dd\cos\theta|\vec p|^2|\vec k|^2}{p_0k_0}\frac{\Theta^p_{\vec k}\Theta^n_{\vec p}+\Theta^p_{\vec k}\Theta^p_{\vec p}}{q_0^2+2q_0(p^0-k^0)+2m_N^2-2p_0k_0+2|\vec p||\vec k|\cos\theta-m_\pi^2}\nonumber\\
    &\times\frac{-q_0^2m_N^2-2k^0p^0q_0^2+(q_0^2-2k^0q_0)(k\cdot p)+4k_0^2p^0q_0-2p^0q_0m_N^2-4k^0p^0m_N^2+4m_N^2(k\cdot p+k_0^2)-4m_N^4}{q_0-2k^0}\nonumber\\
    &+\frac{m_Ng_A^2}{(2\pi)^4f}\int\frac{\dd |\vec p|\dd|\vec k|\dd\cos\theta|\vec p|^2|\vec k|^2}{p_0k_0}\frac{\Theta_{\vec k}^n\Theta_{\vec p}^n}{q_0^2-2q_0(p^0-k^0)+2m_N^2-2p_0k_0+2|\vec p||\vec k|\cos\theta-m_\pi^2}C_f^3\nonumber\\
    &\times\frac{q_0^2p^0+q_0^2k^0+2p^0q_0k^0+2k_0^2q^0-2q_0m_N^2-(2q_0+4k^0)(k\cdot p)+4p^0m_N^2}{q_0+2k^0}\nonumber\\
    &+\frac{m_N g_A^2}{(2\pi)^4f}\int\frac{\dd |\vec p|\dd|\vec k|\dd\cos\theta|\vec p|^2|\vec k|^2}{p_0k_0}\frac{\Theta_{\vec k}^p\Theta_{\vec p}^p}{q_0^2+2q_0(p^0-k^0)+2m_N^2-2p_0k_0+2|\vec p||\vec k|\cos\theta-m_\pi^2}C_f^4\nonumber\\
    &\times\frac{-q_0^2p^0-k^0q_0^2+2p^0k^0q_0+2q_0(k^0)^2-2q_0m_N^2-(2q_0-4k^0)(k\cdot p)-4p^0m_N^2}{q_0-2k^0}\nonumber\\
    &+\frac{m_Ng_A^2}{(2\pi)^4f^2}\int\frac{\dd |\vec p|\dd|\vec k|\dd\cos\theta|\vec p|^2|\vec k|^2}{p_0k_0}\frac{\Theta_{\vec k}^n\Theta_{\vec p}^n}{q_0^2+2q_0(p^0-k^0)+2m_N^2-2p_0k_0+2|\vec p||\vec k|\cos\theta-m_\pi^2}C_f^5\nonumber\\
    &\times\frac{p^0q_0^2+k^0q_0^2+2k^0p^0q_0+2p_0^2q_0-2q_0m_N^2-2q_0(k\cdot p)-4p^0(k\cdot p)+4k^0m_N^2}{q_0+2 p^0}\nonumber\\
    &+\frac{m_Ng_A^2}{(2\pi)^4f^2}\int\frac{\dd |\vec p|\dd|\vec k|\dd\cos\theta|\vec p|^2|\vec k|^2}{p_0k_0}\frac{\Theta_{\vec k}^p\Theta_{\vec p}^p}{q_0^2-2q_0(p^0-k^0)+2m_N^2-2p_0k_0+2|\vec p||\vec k|\cos\theta-m_\pi^2}C_f^6\nonumber\\
    &\times\frac{-k^0q_0^2-p_0q_0^2+2k^0p^0q_0-2q_0(k\cdot p)+2p_0^2q_0-2q_0m_N^2+4p^0(k\cdot p)-4k_0m_N^2}{q_0-2p^0}\nonumber\\
    &+\frac{m_N^2g_A^2}{(2\pi)^4f^3}\int\frac{\dd |\vec p|\dd|\vec k|\dd\cos\theta|\vec p|^2|\vec k|^2}{p_0k_0}\frac{\Theta^p_{\vec k}\Theta^p_{\vec p}+\Theta^p_{\vec k}\Theta^n_{\vec p}}{(p-k)^2-m_\pi^2}\frac{3(k\cdot p)-3m_N^2}{q_0-2k^0}\nonumber\\
    &+\frac{m_N^2g_A^2}{(2\pi)^4f^3}\int\frac{\dd |\vec p|\dd|\vec k|\dd\cos\theta|\vec p|^2|\vec k|^2}{p_0k_0}\frac{\Theta^n_{\vec k}\Theta^p_{\vec p}+\Theta^n_{\vec k}\Theta^n_{\vec p}}{(p-k)^2-m_\pi^2}\frac{3(k\cdot p)-3m_N^2}{q_0+2k^0}\nonumber\\
    &+\frac{m_N^2g_A^2}{(2\pi)^4f^3}\int\frac{\dd |\vec p|\dd|\vec k|\dd\cos\theta|\vec p|^2|\vec k|^2}{p_0k_0}\frac{\Theta^n_{\vec k}\Theta^p_{\vec p}-\Theta^p_{\vec k}\Theta^n_{\vec p}}{(p-k)^2-m_\pi^2}(k^0-p^0)\nonumber\\
    &+\frac{m_Ng_A^2}{(2\pi)^4f}\int\frac{\dd |\vec p|\dd|\vec k|\dd\cos\theta|\vec p|^2|\vec k|^2}{p_0k_0}\frac{\Theta_{\vec k}^p\Theta_{\vec p}^n}{(p-k)^2-m_\pi^2}\frac{2q_0m_N^2-2q_0(k\cdot p)-4k^0(k\cdot p)+4p^0m_N^2}{q_0-2k^0}C_f^7\nonumber\\
    &+\frac{m_Ng_A^2}{(2\pi)^4f}\int\frac{\dd |\vec p|\dd|\vec k|\dd\cos\theta|\vec p|^2|\vec k|^2}{p_0k_0}\frac{\Theta_{\vec k}^n\Theta_{\vec p}^p}{(p-k)^2-m_\pi^2}\frac{2q_0m_N^2-2q_0(k\cdot p)+4k^0(k\cdot p)-4p^0m_N^2}{q_0+2k^0}C_f^8\nonumber\\
    &+\frac{m_Ng_A^2}{(2\pi)^4f^2}\int\frac{\dd |\vec p|\dd|\vec k|\dd\cos\theta|\vec p|^2|\vec k|^2}{p_0k_0}\frac{\Theta_{\vec k}^n\Theta_{\vec p}^p}{(p-k)^2-m_\pi^2}\frac{2q_0m_N^2-2q_0(k\cdot p)-4p^0(k\cdot p)+4k^0m_N^2}{q_0-2p^0}C_f^9\nonumber
\end{align}
\begin{align}
    \quad\qquad&+\frac{m_Ng_A^2}{(2\pi)^4f^2}\int\frac{\dd |\vec p|\dd|\vec k|\dd\cos\theta|\vec p|^2|\vec k|^2}{p_0k_0}\frac{\Theta_{\vec k}^p\Theta_{\vec p}^n}{(p-k)^2-m_\pi^2}\frac{2q_0m_N^2-2q_0(k\cdot p)+4p^0(k\cdot p)-4k^0m_N^2}{q_0+2p^0}C_f^{10},
\end{align}
\begin{align}
    \hat{\Gamma}_{0,\ \mathrm{2\text{L}}}^{\pi^-}(q_0)
    &=-\frac{2g_A^2m_N^2}{5(2\pi)^4f^3}\int \dd |\vec p|\,\dd |\vec k|\,\dd \cos \theta\frac{\vec p^2\vec k^2}{p_0k_0}\frac{p_0k_0-|\vec p||\vec k|\cos{\theta}-m_N^2}{((p-k)^2-m_\pi^2)^2}\nonumber\\
    &\times\left((-2q_0-10p_0+10k_0)\Theta_{\boldsymbol{k}}^p\Theta^n_{\boldsymbol{p}}+(-2q_0+10p_0-10k_0)\Theta^p_{\boldsymbol{p}}\Theta^n_{\boldsymbol{k}}-3q_0(\Theta^p_{\boldsymbol{p}}\Theta^p_{\boldsymbol{k}}+\Theta^n_{\boldsymbol{p}}\Theta^n_{\boldsymbol{k}})\right)\nonumber\\
    &+\frac{g_A^2m_N^2}{2(2\pi)^4f^3}\int \dd |\vec p|\,\dd |\vec k|\,\dd \cos \theta\frac{\vec p^2\vec k^2}{p_0k_0}\frac{k_0-p_0}{(p-k)^2-m_\pi^2}\left(2\Theta^p_{\boldsymbol{p}}\Theta^n_{\boldsymbol{k}}-4\Theta^p_{\boldsymbol{k}}\Theta^n_{\boldsymbol{p}}\right)\nonumber\\
    &+\frac{q_0}{2(2\pi)^4f^3}\int \dd |\vec p|\,\dd |\vec k|\,\dd \cos \theta\frac{\vec p^2\vec k^2}{p_0k_0}\frac{ p^0 k^0+|\vec p||\vec k|\cos\theta+m_N^2}{q_0^2+2q_0(p^0-k^0)+2m_N^2-2p_0k_0+2|\vec p||\vec k|\cos\theta-m_\pi^2}\nonumber\\
    &\times\big[(\Theta^p_{\boldsymbol{p}}\Theta^p_{\boldsymbol{k}}+\Theta^n_{\boldsymbol{k}}\Theta^n_{\boldsymbol{p}})+(\Theta^n_{\boldsymbol{p}}\Theta^p_{\boldsymbol{k}}+\Theta^p_{\boldsymbol{p}}\Theta^n_{\boldsymbol{k}})-(\Theta^n_{\boldsymbol{p}}\Theta^p_{\boldsymbol{k}}-\Theta^p_{\boldsymbol{p}}\Theta^n_{\boldsymbol{k}})\big]\nonumber\\
    &-\frac{2q_0}{(2\pi)^4}\int \dd |\vec p|\,\dd |\vec k|\,\dd \cos \theta\frac{\vec p^2\vec k^2}{p_0k_0}\frac{C_f^1C_f^2(p_0k_0-|\vec p||\vec k|\cos\theta+m_N^2)}{q_0^2+2q_0(p^0-k^0)+2m_N^2-2p_0k_0+2|\vec p||\vec k|\cos\theta-m_\pi^2}(\Theta^p_{\boldsymbol{p}}\Theta^p_{\boldsymbol{k}}+\Theta^n_{\boldsymbol{k}}\Theta^n_{\boldsymbol{p}})\nonumber\\
    &+\frac{q_0m_N}{(2\pi)^4f^2}\int \dd |\vec p|\,\dd |\vec k|\,\dd \cos \theta\frac{\vec p^2\vec k^2}{p_0k_0}\frac{(p_0+k_0)}{q_0^2+2q_0(p^0-k^0)+2m_N^2-2p_0k_0+2|\vec p||\vec k|\cos\theta-m_\pi^2}(\Theta^p_{\boldsymbol{p}}\Theta^p_{\boldsymbol{k}}-\Theta^n_{\boldsymbol{k}}\Theta^n_{\boldsymbol{p}})C_f^2\nonumber\\
    &-\frac{q_0m_N}{(2\pi)^4f}\int \dd |\vec p|\,\dd |\vec k|\,\dd \cos \theta\frac{\vec p^2\vec k^2}{p_0k_0}\frac{(p_0+k_0)}{q_0^2+2q_0(p^0-k^0)+2m_N^2-2p_0k_0+2|\vec p||\vec k|\cos\theta-m_\pi^2}(\Theta^p_{\boldsymbol{p}}\Theta^p_{\boldsymbol{k}}-\Theta^n_{\boldsymbol{k}}\Theta^n_{\boldsymbol{p}})C_f^1\nonumber\\
    &-\frac{g_A^2}{(2\pi)^4f^3}\int\frac{\dd |\vec p|\dd|\vec k|\dd\cos\theta|\vec p|^2|\vec k|^2}{p_0k_0}\frac{\Theta^p_{\vec k}\Theta^n_{\vec p}+\Theta^p_{\vec k}\Theta^p_{\vec p}}{q_0^2-2q_0(p^0-k^0)+2m_N^2-2p_0k_0+2|\vec p||\vec k|\cos\theta-m_\pi^2}\nonumber\\
    &\times\frac{2k^0p^0q_0^2+q_0^2m_N^2-(q_0^2+2k^0q_0)(k\cdot p)-2p^0q_0m_N^2+4k_0^2p^0q_0+4k^0p^0m_N^2-4m_N^2(k\cdot p+k_0^2)+4m_N^4}{q_0+2k^0}\nonumber\\
    &+\frac{g_A^2}{(2\pi)^4f^3}\int\frac{\dd |\vec p|\dd|\vec k|\dd\cos\theta|\vec p|^2|\vec k|^2}{p_0k_0}\frac{\Theta^n_{\vec k}\Theta^p_{\vec p}+\Theta^n_{\vec k}\Theta^n_{\vec p}}{q_0^2+2q_0(p^0-k^0)+2m_N^2-2p_0k_0+2|\vec p||\vec k|\cos\theta-m_\pi^2}\nonumber\\
    &\times\frac{-q_0^2m_N^2-2k^0p^0q_0^2+(q_0^2-2k^0q_0)(k\cdot p)+4k_0^2p^0q_0-2p^0q_0m_N^2-4k^0p^0m_N^2+4m_N^2(k\cdot p+k_0^2)-4m_N^4}{q_0-2k^0}\nonumber\\
    &+\frac{m_N g_A^2}{(2\pi)^4f}\int\frac{\dd |\vec p|\dd|\vec k|\dd\cos\theta|\vec p|^2|\vec k|^2}{p_0k_0}\frac{\Theta_{\vec p}^p\Theta_{\vec k}^p}{q_0^2-2q_0(p^0-k^0)+2m_N^2-2p_0k_0+2|\vec p||\vec k|\cos\theta-m_\pi^2}C_f^3\nonumber\\
    &\times\frac{q_0^2p^0+q_0^2k^0+2p^0q_0k^0+2k^2_0q^0-2q_0m_N^2-(2q_0+4k^0)(k\cdot p)+4p^0m_N^2}{q_0+2k^0}\nonumber\\
    &+\frac{ m_Ng_A^2}{(2\pi)^4f}\int\frac{\dd |\vec p|\dd|\vec k|\dd\cos\theta|\vec p|^2|\vec k|^2}{p_0k_0}\frac{\Theta_{\vec p}^n\Theta_{\vec k}^n}{q_0^2+2q_0(p^0-k^0)+2m_N^2-2p_0k_0+2|\vec p||\vec k|\cos\theta-m_\pi^2}C_f^4\nonumber\\
    &\times\frac{-q_0^2p^0-k^0q_0^2+2p^0k^0q_0+2q_0(k^0)^2-2q_0m_N^2-(2q_0-4k^0)(k\cdot p)-4p^0m_N^2}{q_0-2k^0}\nonumber\\
    &+\frac{m_Ng_A^2}{(2\pi)^4f^2}\int\frac{\dd |\vec p|\dd|\vec k|\dd\cos\theta|\vec p|^2|\vec k|^2}{p_0k_0}\frac{\Theta_{\vec p}^p\Theta_{\vec k}^p}{q_0^2+2q_0(p^0-k^0)+2m_N^2-2p_0k_0+2|\vec p||\vec k|\cos\theta-m_\pi^2}C_f^5\nonumber\\
    &\times\frac{p^0q_0^2+k^0q_0^2+2k^0p^0q_0+2p_0^2q_0-2q_0m_N^2-2q_0(k\cdot p)-4p^0(k\cdot p)+4k^0m_N^2}{q_0+2 p^0}\nonumber\\
    &+\frac{m_Ng_A^2}{(2\pi)^4f^2}\int\frac{\dd |\vec p|\dd|\vec k|\dd\cos\theta|\vec p|^2|\vec k|^2}{p_0k_0}\frac{\Theta_{\vec p}^n\Theta_{\vec k}^n}{q_0^2-2q_0(p^0-k^0)+2m_N^2-2p_0k_0+2|\vec p||\vec k|\cos\theta-m_\pi^2}C_f^6\nonumber\\
    &\times\frac{-k^0q_0^2-p_0q_0^2+2k^0p^0q_0-2q_0(k\cdot p)+2p_0^2q_0-2q_0m_N^2+4p^0(k\cdot p)-4k_0m_N^2}{q_0-2 p^0}\nonumber\\
    &+\frac{m_N^2g_A^2}{(2\pi)^4f^3}\int\frac{\dd |\vec p|\dd|\vec k|\dd\cos\theta|\vec p|^2|\vec k|^2}{p_0k_0}\frac{\Theta^n_{\vec k}\Theta^n_{\vec p}+\Theta^n_{\vec k}\Theta^p_{\vec p}}{(p-k)^2-m_\pi^2}\frac{3(k\cdot p)-3m_N^2}{q_0-2k^0}\nonumber\\
    &+\frac{m_N^2g_A^2}{(2\pi)^4f^3}\int\frac{\dd |\vec p|\dd|\vec k|\dd\cos\theta|\vec p|^2|\vec k|^2}{p_0k_0}\frac{\Theta^p_{\vec k}\Theta^n_{\vec p}+\Theta^p_{\vec k}\Theta^p_{\vec p}}{(p-k)^2-m_\pi^2}\frac{3(k\cdot p)-3m_N^2}{q_0+2k^0}\nonumber\\
    &+\frac{m_N^2g_A^2}{(2\pi)^4f^3}\int\frac{\dd |\vec p|\dd|\vec k|\dd\cos\theta|\vec p|^2|\vec k|^2}{p_0k_0}\frac{\Theta_{\vec k}^p\Theta_{\vec p}^n-\Theta_{\vec k}^n\Theta_{\vec p}^p}{(p-k)^2-m_\pi^2}(k^0-p^0)
\end{align}
\begin{align}
    \quad\qquad&+\frac{m_Ng_A^2}{(2\pi)^4f}\int\frac{\dd |\vec p|\dd|\vec k|\dd\cos\theta|\vec p|^2|\vec k|^2}{p_0k_0}\frac{\Theta_{\vec k}^n\Theta_{\vec p}^p}{(p-k)^2-m_\pi^2}\frac{2q_0m_N^2-2q_0(k\cdot p)-4k^0(k\cdot p)+4p^0m_N^2}{q_0-2k^0}C_f^7\nonumber\\
    &+\frac{ m_Ng_A^2}{(2\pi)^4f}\int\frac{\dd |\vec p|\dd|\vec k|\dd\cos\theta|\vec p|^2|\vec k|^2}{p_0k_0}\frac{\Theta_{\vec k}^p\Theta_{\vec p}^n}{(p-k)^2-m_\pi^2}\frac{2q_0m_N^2-2q_0(k\cdot p)+4k^0(k\cdot p)-4p^0m_N^2}{q_0+2k^0}C_f^8\nonumber\\
    &+\frac{m_Ng_A^2}{(2\pi)^4f^2}\int\frac{\dd |\vec p|\dd|\vec k|\dd\cos\theta|\vec p|^2|\vec k|^2}{p_0k_0}\frac{\Theta_{\vec k}^p\Theta_{\vec p}^n}{(p-k)^2-m_\pi^2}\frac{2q_0m_N^2-2q_0(k\cdot p)-4p^0(k\cdot p)+4k^0m_N^2}{q_0-2p^0}C_f^9\nonumber\\
    &+\frac{m_Ng_A^2}{(2\pi)^4f^2}\int\frac{\dd |\vec p|\dd|\vec k|\dd\cos\theta|\vec p|^2|\vec k|^2}{p_0k_0}\frac{\Theta_{\vec p}^p\Theta_{\vec k}^n}{(p-k)^2-m_\pi^2}\frac{2q_0m_N^2-2q_0(k\cdot p)+4p^0(k\cdot p)-4k^0m_N^2}{q_0+2p^0}C_f^{10},
\end{align}
\begin{align}
    \hat{\Gamma}_{0,\ \mathrm{2\text{L}}}^{\pi^0}(q_0)
    &=-\frac{2g_A^2m_N^2}{5(2\pi)^4f^3}\int \dd |\vec p|\,\dd |\vec k|\,\dd \cos \theta\frac{\vec p^2\vec k^2}{p_0k_0}\frac{p_0k_0-|\vec p||\vec k|\cos{\theta}-m_N^2}{((p-k)^2-m_\pi^2)^2}\nonumber\left(-6q_0\Theta_{\boldsymbol{k}}^p\Theta^n_{\boldsymbol{p}}-6q_0\Theta^p_{\boldsymbol{p}}\Theta^n_{\boldsymbol{k}}+q_0(\Theta^p_{\boldsymbol{p}}\Theta^p_{\boldsymbol{k}}+\Theta^n_{\boldsymbol{p}}\Theta^n_{\boldsymbol{k}})\right)\nonumber\\
    &+\frac{q_0}{(2\pi)^4f^3}\int \dd |\vec p|\,\dd |\vec k|\,\dd \cos \theta\frac{\vec p^2\vec k^2}{p_0k_0}\frac{p^0k^0+|\vec p||\vec k|\cos\theta+m_N^2}{q_0^2+2q_0(p^0-k^0)+2m_N^2-2p_0k_0+2|\vec p||\vec k|\cos\theta-m_\pi^2}(\Theta^p_{\boldsymbol{p}}\Theta^n_{\boldsymbol{k}}+\Theta^n_{\boldsymbol{p}}\Theta^p_{\boldsymbol{k}})\nonumber\\
    &-\frac{2q_0}{(2\pi)^4}\int \dd |\vec p|\,\dd |\vec k|\,\dd \cos \theta\frac{\vec p^2\vec k^2}{p_0k_0}\frac{C_f^1C_f^2(p_0k_0-|\vec p||\vec k|\cos\theta+m_N^2)}{q_0^2+2q_0(p^0-k^0)+2m_N^2-2p_0k_0+2|\vec p||\vec k|\cos\theta-m_\pi^2}(\Theta^p_{\boldsymbol{p}}\Theta^p_{\boldsymbol{k}}+\Theta^n_{\boldsymbol{k}}\Theta^n_{\boldsymbol{p}})\nonumber\\
    &-\frac{ g_A^2}{(2\pi)^4f^3}\int\frac{\dd |\vec p|\dd|\vec k|\dd\cos\theta|\vec p|^2|\vec k|^2}{p_0k_0}\frac{\Theta^p_{\vec k}\Theta^n_{\vec p}+\Theta^n_{\vec k}\Theta^p_{\vec p}}{q_0^2-2q_0(p^0-k^0)+2m_N^2-2p_0k_0+2|\vec p||\vec k|\cos\theta-m_\pi^2}\nonumber\\
    &\times\frac{2k^0p^0q_0^2+q_0^2m_N^2-(q_0^2+2k^0q_0)(k\cdot p)-2p^0q_0m_N^2+4k_0^2p^0q_0+4k^0p^0m_N^2-4m_N^2(k\cdot p+k_0^2)+4m_N^4}{q_0+2k^0}\nonumber\\
    &+\frac{g_A^2}{(2\pi)^4f^3}\int\frac{\dd |\vec p|\dd|\vec k|\dd\cos\theta|\vec p|^2|\vec k|^2}{p_0k_0}\frac{\Theta^p_{\vec k}\Theta^n_{\vec p}+\Theta^n_{\vec k}\Theta^p_{\vec p}}{q_0^2+2q_0(p^0-k^0)+2m_N^2-2p_0k_0+2|\vec p||\vec k|\cos\theta-m_\pi^2}\nonumber\\
    &\times\frac{-q_0^2m_N^2-2k^0p^0q_0^2+(q_0^2-2k^0q_0)(k\cdot p)+4k_0^2p^0q_0-2p^0q_0m_N^2-4k^0p^0m_N^2+4m_N^2(k\cdot p+k_0^2)-4m_N^4}{q_0-2k^0}\nonumber\\
    &+\frac{ m_Ng_A^2}{2(2\pi)^4f}\int\frac{\dd |\vec p|\dd|\vec k|\dd\cos\theta|\vec p|^2|\vec k|^2}{p_0k_0}\frac{\Theta_{\vec k}^p\Theta_{\vec p}^p+\Theta_{\vec k}^n\Theta_{\vec p}^n}{q_0^2-2q_0(p^0-k^0)+2m_N^2-2p_0k_0+2|\vec p||\vec k|\cos\theta-m_\pi^2}C_f^3\nonumber\\
    &\times\frac{q_0^2p^0+q_0^2k^0+2p^0q_0k^0+2k_0^2q^0-2q_0m_N^2-(2q_0+4k^0)(k\cdot p)+4p^0m_N^2}{q_0+2k^0}\nonumber\\
    &+\frac{ m_Ng_A^2}{2(2\pi)^4f}\int\frac{\dd |\vec p|\dd|\vec k|\dd\cos\theta|\vec p|^2|\vec k|^2}{p_0k_0}\frac{\Theta_{\vec k}^p\Theta_{\vec p}^p+\Theta_{\vec k}^n\Theta_{\vec p}^n}{q_0^2+2q_0(p^0-k^0)+2m_N^2-2p_0k_0+2|\vec p||\vec k|\cos\theta-m_\pi^2}C_f^4\nonumber\\
    &\times\frac{-q_0^2p^0-k^0q_0^2+2p^0k^0q_0+2q_0(k^0)^2-2q_0m_N^2-(2q_0-4k^0)(k\cdot p)-4p^0m_N^2}{q_0-2k^0}\nonumber\\
    &+\frac{m_Ng_A^2}{2(2\pi)^4f^2}\int\frac{\dd |\vec p|\dd|\vec k|\dd\cos\theta|\vec p|^2|\vec k|^2}{p_0k_0}\frac{\Theta_{\vec k}^n\Theta_{\vec p}^n+\Theta_{\vec k}^p\Theta_{\vec p}^p}{q_0^2+2q_0(p^0-k^0)+2m_N^2-2p_0k_0+2|\vec p||\vec k|\cos\theta-m_\pi^2}C_f^5\nonumber\\
    &\times\frac{p^0q_0^2+k^0q_0^2+2k^0p^0q_0+2p_0^2q_0-2q_0^2m_N^2-2q_0(k\cdot p)-4p^0(k\cdot p)+4k^0m_N^2}{q_0+2p^0}\nonumber\\
    &+\frac{m_Ng_A^2}{2(2\pi)^4f^2}\int\frac{\dd |\vec p|\dd|\vec k|\dd\cos\theta|\vec p|^2|\vec k|^2}{p_0k_0}\frac{\Theta_{\vec k}^p\Theta_{\vec p}^p+\Theta_{\vec k}^n\Theta_{\vec p}^n}{q_0^2-2q_0(p^0-k^0)+2m_N^2-2p_0k_0+2|\vec p||\vec k|\cos\theta-m_\pi^2}C_f^6\nonumber\\
    &\times\frac{-k^0q_0^2-p_0q_0^2+2k^0p^0q_0-2q_0(k\cdot p)+2p_0^2q_0-2q_0m_N^2+4p^0(k\cdot p)-4k_0m_N^2}{q_0-2 p^0}\nonumber\\
    &+\frac{m_N^2g_A^2}{(2\pi)^4f^3}\int\frac{\dd |\vec p|\dd|\vec k|\dd\cos\theta|\vec p|^2|\vec k|^2}{p_0k_0}\frac{\Theta^p_{\vec k}\Theta^n_{\vec p}+\Theta^n_{\vec k}\Theta^p_{\vec p}}{(p-k)^2-m_\pi^2}\frac{3(k\cdot p)-3m_N^2}{q_0-2k^0}\nonumber\\
    &+\frac{m_N^2g_A^2}{(2\pi)^4f^3}\int\frac{\dd |\vec p|\dd|\vec k|\dd\cos\theta|\vec p|^2|\vec k|^2}{p_0k_0}\frac{\Theta^p_{\vec k}\Theta^n_{\vec p}+\Theta^n_{\vec k}\Theta^p_{\vec p}}{(p-k)^2-m_\pi^2}\frac{3(k\cdot p)-3m_N^2}{q_0+2k^0}\nonumber
\end{align}
\begin{align}
    \quad\qquad&+\frac{m_Ng_A^2}{(2\pi)^4f}\int\frac{\dd |\vec p|\dd|\vec k|\dd\cos\theta|\vec p|^2|\vec k|^2}{p_0k_0}\frac{\frac{1}{2}(\Theta_{\vec k}^p\Theta_{\vec p}^p+\Theta_{\vec k}^n\Theta_{\vec p}^n)}{(p-k)^2-m_\pi^2}\frac{2q_0m_N^2-2q_0(k\cdot p)-4k^0(k\cdot p)+4p^0m_N^2}{q_0-2k^0}C_f^7\nonumber\\
    &+\frac{m_Ng_A^2}{(2\pi)^4f}\int\frac{\dd |\vec p|\dd|\vec k|\dd\cos\theta|\vec p|^2|\vec k|^2}{p_0k_0}\frac{\frac{1}{2}(\Theta_{\vec k}^p\Theta_{\vec p}^p+\Theta_{\vec k}^n\Theta_{\vec p}^n)}{(p-k)^2-m_\pi^2}\frac{2q_0m_N^2-2q_0(k\cdot p)+4k^0(k\cdot p)-4p^0m_N^2}{q_0+2k^0}C_f^8\nonumber\\
    &+\frac{m_N g_A^2}{(2\pi)^4f^2}\int\frac{\dd |\vec p|\dd|\vec k|\dd\cos\theta|\vec p|^2|\vec k|^2}{p_0k_0}\frac{\frac{1}{2}(\Theta_{\vec k}^p\Theta_{\vec p}^p+\Theta_{\vec k}^n\Theta_{\vec p}^n)}{(p-k)^2-m_\pi^2}\frac{2q_0m_N^2-2q_0(k\cdot p)-4p^0(k\cdot p)+4k^0m_N^2}{q_0-2p^0}C_f^9\nonumber\\
    &+\frac{m_Ng_A^2}{(2\pi)^4f^2}\int\frac{\dd |\vec p|\dd|\vec k|\dd\cos\theta|\vec p|^2|\vec k|^2}{p_0k_0}\frac{\frac{1}{2}(\Theta_{\vec k}^p\Theta_{\vec p}^p+\Theta_{\vec k}^n\Theta_{\vec p}^n)}{(p-k)^2-m_\pi^2}\frac{2q_0m_N^2-2q_0(k\cdot p)+4p^0(k\cdot p)-4k^0m_N^2}{q_0+2p^0}C_f^{10},
\end{align}
\end{subequations}
where
\begin{align}
    \MoveEqLeft C_f^1=\frac{8B_0m_qc_1}{q_0f^2}-\frac{2c_2}{f^2m_N^2}q_0p_0^2 -\frac{2c_3}{f^2}q_0-\frac{2c_2}{f^2m_N^2}p_0(p^\mu p_\mu-k_\mu p^\mu)-\frac{2c_3}{f^2}(p_0-k_0), \nonumber\\
    \MoveEqLeft C_f^2=\frac{2c_2}{fm_N^2}q_0k_0^2+\frac{2c_3}{f}q_0+\frac{2c_2}{fm_N^2}k_0(k_\mu p^\mu-k^\mu k_\mu)+\frac{2c_3}{f}(p_0-k_0),\nonumber\\
    \MoveEqLeft C_f^3=\frac{8B_0m_qc_1}{q_0f^2}-\frac{2c_2}{f^2m_N^2}q_0 k_0^2-\frac{2c_3}{f^2}q_0-\frac{2c_2}{f^2m_N^2}k^0(k^\mu k_\mu-k_\mu p^\mu)-\frac{2c_3}{f^2}(k_0-p_0),\nonumber\\
     \MoveEqLeft C_f^4=\frac{8B_0m_qc_1}{q_0f^2}-\frac{2c_2}{f^2m_N^2}q_0 p_0^2-\frac{2c_3}{f^2}q_0-\frac{2c_2}{f^2m_N^2}p^0(p_\mu p^\mu-p_\mu k^\mu)-\frac{2c_3}{f^2}(p_0-k_0),\nonumber\\
     \MoveEqLeft C_f^5=-\frac{2c_2}{fm_N^2}k_0^2 q_0 -\frac{2c_3}{f}q_0-\frac{2c_2}{fm_N^2}k_0(k_\mu p^\mu-k_\mu k^\mu)-\frac{2c_3}{f}(p_0-k_0)\nonumber\\
     \MoveEqLeft C_f^6=-\frac{2c_2}{fm_N^2}p_0^2 q_0-\frac{2c_3}{f}q_0-\frac{2c_2}{fm_N^2}p_0(p_\mu k^\mu-p_\mu p^\mu)-\frac{2c_3}{f}(k^0-p^0),\nonumber\\
    \MoveEqLeft C_f^7=\frac{8B_0c_1m_q}{q_0f^2}-\frac{2c_2}{f^2m_N^2}(k^0-q^0)(k_\mu-p_\mu)( k^\mu-q^\mu)-\frac{2c_3}{f^2}(k_0-p_0),\nonumber\\
    \MoveEqLeft C_f^8=\frac{8B_0c_1m_q}{q_0f^2}-\frac{2c_2}{f^2m_N^2}p^0 p_\mu (p^\mu-k^\mu)-\frac{2c_3}{f^2}(p_0-k_0),\nonumber\\
    \MoveEqLeft C_f^9=\frac{2c_2}{fm_N^2}k_0 k_\mu (k^\mu-p^\mu)+\frac{2c_3}{f}(k^0-p^0),\nonumber\\
    \MoveEqLeft C_f^{10}=\frac{2c_2}{fm_N^2}(p_0+q_0) (p_\mu-k_\mu) (p^\mu+q^\mu)+\frac{2c_3}{f}(p^0-k^0).
\end{align}
The axial-vector form factors are obtained with the vertex corrections and wave function renormalizations using \cref{eq:axial-vector form,eq:defwavefunction}. Some of the Feynman diagrams have been evaluated with a Mathematica program, FeynArts \cite{hahn2001generating}.
\end{widetext}

\bibliographystyle{apsrev4-2}
\bibliography{references.bib}

\end{document}